\documentclass[structabstract]{aa}  
\usepackage{graphicx}
\usepackage{txfonts}
\usepackage{longtable}

\begin{document}
   \title{What causes the large extensions of red-supergiant atmospheres?}
   
   \titlerunning{Large extensions of RSG atmospheres}

   \subtitle{Comparisons of interferometric observations with 1-D hydrostatic, 3-D convection, and 1-D pulsating model atmospheres \thanks{Based on observations made with the VLT Interferometer (VLTI) at Paranal Observatory under programme ID 091.D-0275}}

   \author{B.~Arroyo-Torres \inst{1} \and
          M.~Wittkowski \inst{2} \and 
          A.~Chiavassa \inst{3} \and
          M.~Scholz \inst{4,5} \and
          B.~Freytag \inst{6} \and
          J.~M.~Marcaide \inst{1,7} \and 
          P.~H.~Hauschildt \inst{8} \and
          P.~R.~Wood \inst{9} \and
          F.~J.~Abellan \inst{1}
          }

   \institute{Dpt. Astronomia i Astrof\' isica, Universitat de Val\`encia, 
C/Dr. Moliner 50, 46100, Burjassot, Spain \email{belen.arroyo@uv.es}
         \and ESO, Karl-Schwarzschild-St. 2, 85748, Garching bei M\" unchen, 
Germany
		 \and Laboratoire Lagrange, UMR 7293, Universit\'e de Nice Sophia-Antipolis, CNRS,
Observatoire de la C\^ote d’Azur, BP. 4229, 06304 Nice Cedex 4, France
         \and Zentrum f\"ur Astronomie der Universit\"at Heidelberg (ZAH), Institut f\"ur Theoretische Astrophysik, Albert-Ueberle-Str. 2, 69120 Heidelberg, Germany
         \and Sydney Institute for Astronomy, School of Physics, University of Sydney, Sydney NSW 2006, Australia 
         \and Department of Physics and Astronomy at Uppsala University, Regementsv\"agen 1, Box 516, SE-75120 Uppsala, Sweden         
		 \and Donostia International Physics Center, Paseo de Manuel Lardizabal 4, 20018 Donostia-San Sebasti\'an, Spain
         \and Hamburger Sternwarte, Gojenbergsweg 112, 21029, Hamburg, Germany
         \and Research School of Astronomy and Astrophysics, Australian National University, Cotter Road, Weston Creek ACT 2611, Australia }

   \date{Received 24 October 2014; Accepted 23 December 2014}

  \abstract
   {}
   {This research has two main goals. First, we present the atmospheric structure and the fundamental parameters of three red supergiants (RSGs), increasing the sample of RSGs observed by near-infrared spectro-interferometry. Additionally, we test possible mechanisms that may explain the large observed atmospheric extensions of RSGs.}
   {We carried out spectro-interferometric observations of the RSGs V602~Car, HD~95687, and HD~183589 in the near-infrared $K$-band (1.92-2.47\,$\mu$m) with the VLTI/AMBER instrument at medium spectral resolution ($R\sim$1500). To categorize and comprehend the extended atmospheres, we compared our observational results to predictions by available hydrostatic \protect{\tt PHOENIX}, available 3-D convection, and new 1-D self-excited pulsation models of RSGs.}
   {Our near-infrared flux spectra of V602~Car, HD~95687, and HD~183589 are well reproduced by the \protect{\tt PHOENIX} model atmospheres. The continuum visibility values are consistent with a limb-darkened disk as predicted by the \protect{\tt PHOENIX} models, allowing us to determine the angular diameter and the fundamental parameters of our sources. Nonetheless, in the case of V602~Car and HD~95686, the \protect{\tt PHOENIX} model visibilities do not predict the large observed extensions of molecular layers, most remarkably in the CO bands. Likewise, the 3-D convection models and the 1-D pulsation models with typical parameters of RSGs lead to compact atmospheric structures as well, which are similar to the structure of the hydrostatic \protect{\tt PHOENIX} models. They can also not explain the observed decreases in the visibilities and thus the large atmospheric molecular extensions. The full sample of our RSGs indicates increasing observed atmospheric extensions with increasing luminosity and decreasing surface gravity, and no correlation with effective temperature or variability amplitude.}
   {The location of our RSG sources in the Hertzsprung-Russell diagram is confirmed to be consistent with the red limits of recent evolutionary tracks. The observed extensions of the atmospheric layers of our sample of RSGs are comparable to those of Mira stars. This phenomenon is not predicted by any of the considered model atmospheres including available 3-D convection and new 1-D pulsation models of RSGs. This confirms that neither convection nor pulsation alone can levitate the molecular atmospheres of RSGs. Our observed correlation of atmospheric extension with luminosity supports a scenario of radiative acceleration on  Doppler-shifted molecular lines.}


\maketitle

\section{Introduction}

Red supergiant (RSG) stars are known to lose mass with mass-loss rates of $2\times10^{-7}$\,M$_{\odot}$/yr -- $3\times10^{-4}$\,M$_{\odot}$/yr (de Beck et al.  \cite{DeBeck2010}), and they are one of the major sources of the chemical enrichment of galaxies and of dust in the universe, along with asymptotic giant branch (AGB) stars and supernovae. Currently, the mechanism of mass loss of RSG stars and semi-regular or irregular AGBs is not known in detail. Nevertheless, in the case of Mira-variable AGB stars (mass-loss rates of $10^{-6}$\,M$_{\odot}$/yr -- $10^{-4}$\,M$_{\odot}$/yr, Wood et al. \cite{Wood1983,Wood1992}), this mechanism is better understood. The theoretical models that explain the Mira mass-loss process are based on pulsations that extend the atmospheres to radii where dust can form, and subsequently on radiative pressure on dust grains that drives the wind (e.g., Bladh et al. \cite{Bladh2013}). In the case of variable RSGs, the amplitude of the light curves is about one-third of that of Miras (e.g., Wood et al. \cite{Wood1983}), so that pulsation is expected to play a less dominant role (cf. Josselin \& Plez \cite{Josselin2007}). Other mechanisms that might give rise to mass loss in the RSGs are convection and rotation (e.g., Langer \& Heger \cite{Langer1999}). Constraints on the mechanisms that levitate the atmospheres of RSGs are thus fundamental for our understanding of the mass-loss process of RSG stars.

\begin{table*}
\caption{VLTI/AMBER observations}
\centering
\begin{tabular}{ccccccc}
\hline
\hline
Target (Sp. type)  & Date   & Mode        & Baseline & Projected baseline & PA  & Calibrator   \\
                   & 2013-  & K- ($\mu$m) &          & m                  & deg &              \\
\hline
                   & 04-04 & 2.1 & A1-G1-K0 & 75.86/80.02/127.8 & 91/21/55 & \\
V602 Car (M3-M4 I) & 04-04 & 2.3 & A1-G1-K0 & 79.89/68.52/112.6 & 130/48/93 & HR 4164 - z Car\\
                   & 04-26 & 2.3 & D0-H0-G1 & 63.22/58.73/64.46 & 71/-172/125 & \\
                 & 04-04 & 2.1 & A1-G1-K0 & 73.06/80.57/128.6 & 80/14/45 & \\
HD95687 (M3 Iab) & 04-04 & 2.3 & A1-G1-K0 & 79.94/71.35/117.5 & 121/43/85 & HR 4164 - z Car\\
                 & 04-25 & 2.3 & D0-H0-G1 & 56.88/53.89/69.30 & 104/-153/153 & \\
                  & 05-04  & 2.1 & D0-I1-G1 & 78.89/45.56/63.26 & 99/-134/134 & \\
HD 183589 (K5 Ib) & 07-29 & 2.3 & D0-I1-G1 & 69.75/46.61/55.98 & 99/-134/141 & 38 Aql - HR 7404\\
                  & 08-04 & 2.1 &  A1-G1-K0 & 88.87/67.14/126.2 & -145/-72/-114 & \\ 
\hline
\end{tabular}
\tablefoot{Details of our observations. The projected baseline is the projected baseline length for the AT VLTI baseline used, and PA is the position angle of the baseline (north through east).}
\label{Log_obs}
\end{table*}

Observations of Mira variable stars using the IOTA or VLTI interferometers show evidence of molecular layers lying above the photospheric layers (e.g., Perrin et al. \cite{Perrin2004}; Wittkowski et al. \cite{Witt2008}, \cite{Witt2011}). Theoretical dynamic model atmospheres (Ireland et al. \cite{Ireland2004a}, \cite{Ireland2004b}, \cite{Ireland2008}, \cite{Ireland2011}; Scholz et al. \cite{Scholz2014}) can explain reasonably well these molecular layers for Miras. On the other hand, interferometric observations of RSGs also indicate the presence of extended molecular layers (CO and water), which cannot be explained by hydrostatic model atmospheres (Perrin et al. \cite{Perrin2005}; Ohnaka et al. \cite{Ohnaka2011}, \cite{Ohnaka2013}; Wittkowski et al. \cite{Witt2012}). The red supergiant VX~Sgr (Chiavassa et al. \cite{Chiavassa2010}) showed a good agreement with Mira models, although they have very different stellar parameters than expected for this source.

This paper is conceived as part of a series of three previous papers (Wittkowski et al. \cite{Witt2012}; Arroyo-Torres et al. \cite{Arroyo2013}, \cite{Arroyo2014}). In these previous works, we presented the atmosphere structure and the fundamental parameters of a sample of four RSG stars and five cool giant stars. One goal of the current paper is to add three red supergiants and thus to increase this sample. On the other hand, our previous works showed that the observed visibility data of the RSGs and of one of the red giants, $\beta$~Peg, indicate large extensions of the molecular layers, similar as those previously observed for Mira variable stars (Wittkowski et al. \cite{Witt2008}, \cite{Witt2011}). This was not predicted by hydrostatic {\tt PHOENIX} model atmospheres. However, the spectra of all our stars were reproduced well by the {\tt PHOENIX} models. This indicates that the molecular opacities were adequately included in these model atmospheres, but that they were too compact compared to observations. In order to understand the processes that may explain the extended molecular layers, the second goal of this paper is to investigate the effects of realistic three-dimensional (3-D) radiative hydrodynamical (RHD) simulations of stellar convection as well as of one-dimensional (1-D) self-excited pulsation models on the extensions of RSG atmospheres. These processes were previously discussed as possible mechanisms to levitate RSG atmospheres (Chiavassa et al. \cite{Chiavassa2010}, \cite{Chiavassa2011}). 

In this paper, in addition to studying RSGs, we also refer to asymptotic giant branch stars (AGBs), Mira variable stars, and red giant stars. AGB stars are low and intermediate mass evolved stars before they evolve toward hotter temperatures in the HR diagram. Mira stars are long-period large-amplitude variable AGB stars. With red giants we refer to giants on the first red giant branch.

Our work is structured as follows: In Sect.~\ref{sec:obs}, we describe our AMBER observations and the data reduction. In Sect.~\ref{sec:results} we present the results obtained from the PHOENIX model fitting and the fundamental parameters. In Sect.~\ref{sec:charac}, we characterize the extensions of the atmospheres. In Sect.~\ref{sec:comp}, we show the results obtained from the comparison with the convection and pulsation models and we discuss alternative mechanisms. Finally, in Sect.~\ref{sec:summ}, we summarize our results and conclusions. 

\section{Observations and data reduction}
\label{sec:obs}

\begin{figure*}
\centering
\includegraphics[width=0.49\hsize]{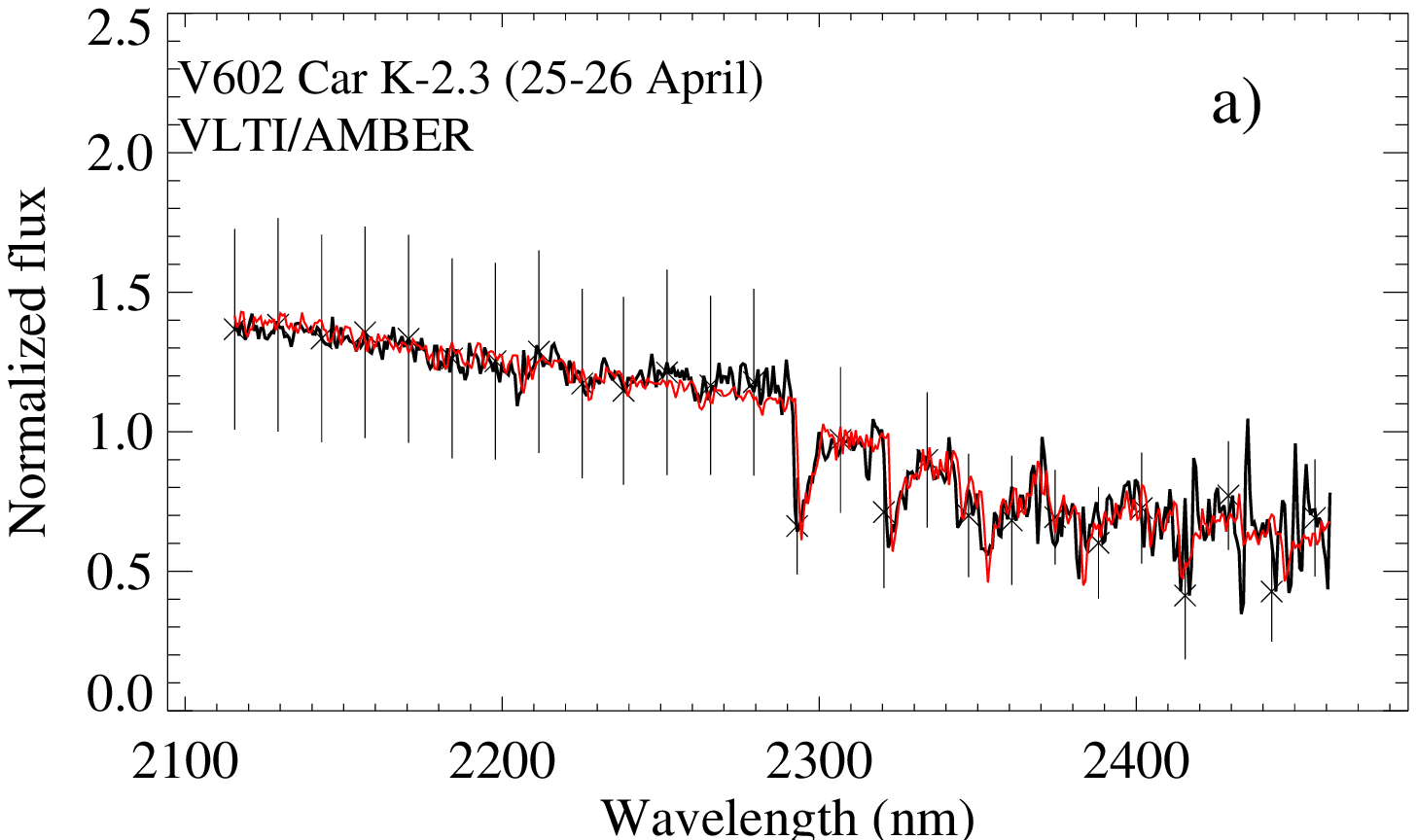}
\includegraphics[width=0.49\hsize]{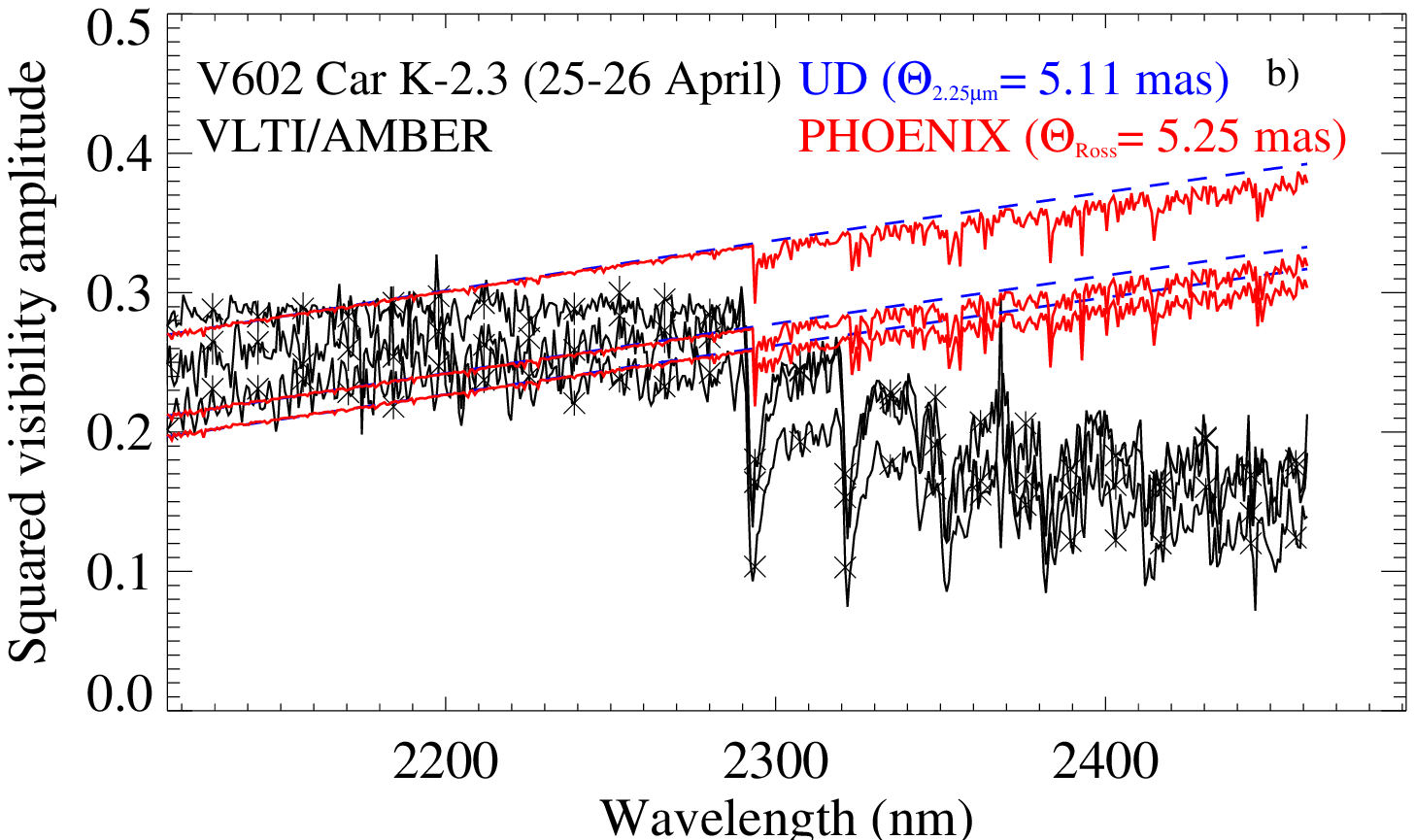}
\includegraphics[width=0.49\hsize]{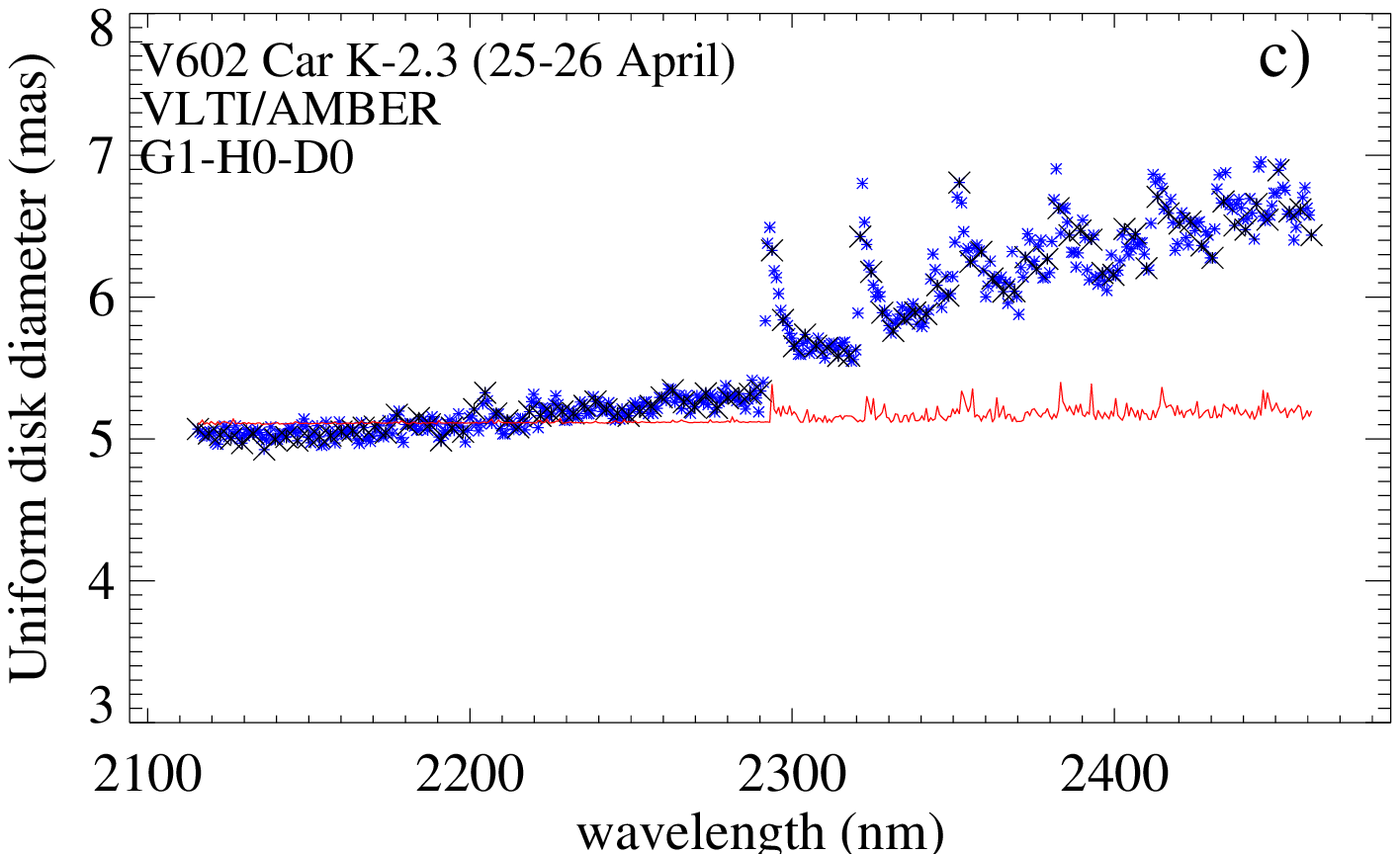}
\includegraphics[width=0.49\hsize]{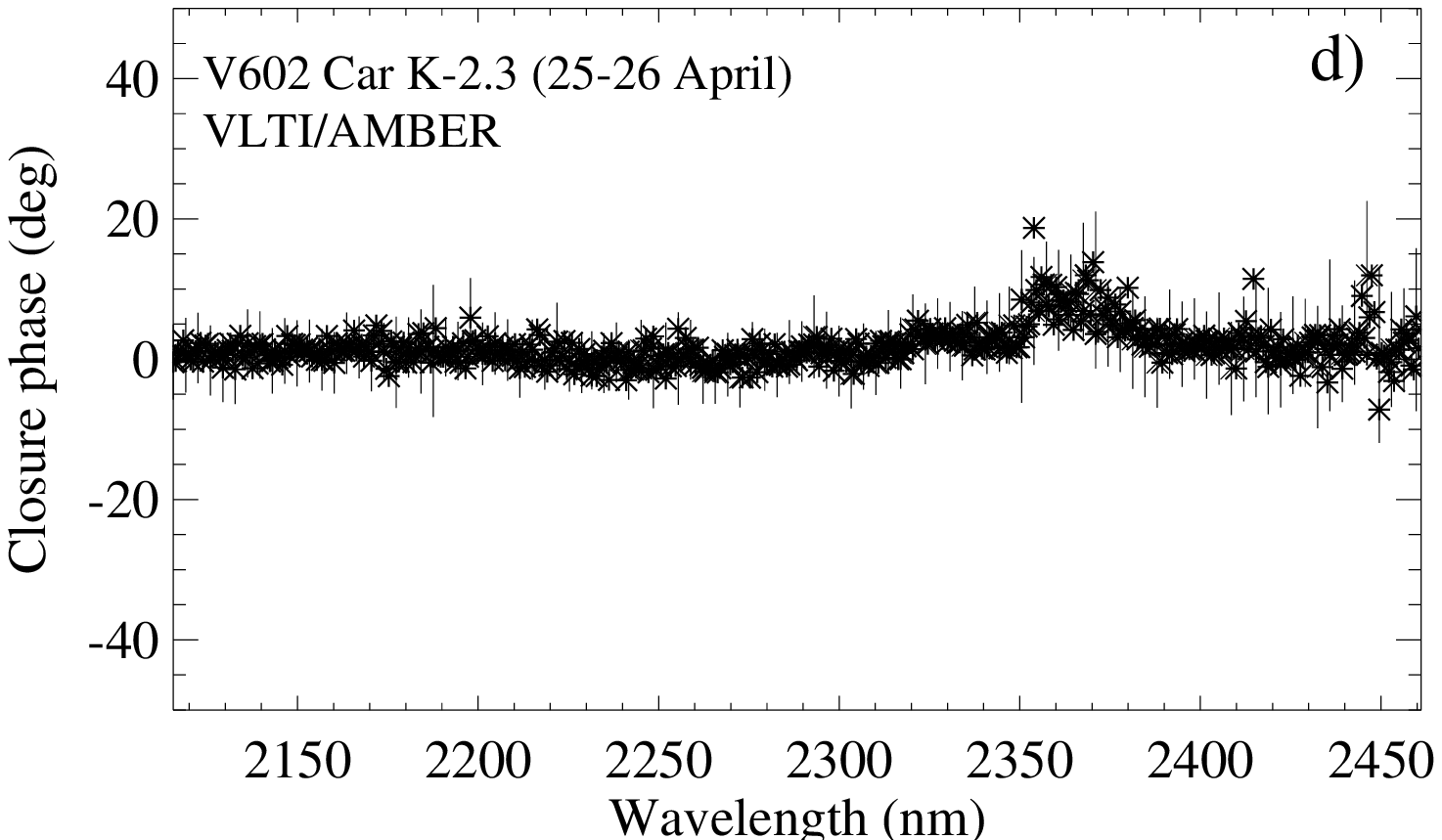}
\caption{Normalized flux (a), squared visibility amplitudes (b), UD diameters (c), and closure phases (d) for the example of V602~Car obtained with the MR-K 2.3\,$\mu$m setting, on 26 April 2013. In black the observed data, in blue the UD model, and in red the {\tt PHOENIX} model.}
\label{resul_V602Car_2604}
\end{figure*}

We observed V602~Car (Simbad spectral type M3-M4 I), HD~95687 (M3 Iab), and HD~183589 (K5 Ib) with the ESO Very Large Telescope Interferometer (VLTI), utilizing three of the Auxiliary Telescopes of 1.8\,m diameter, and the AMBER instrument (Astronomical Multi-BEam combineR) with the external fringe tracker FINITO (Petrov et al. \cite{Petrov2007}). We worked in medium-resolution mode ($R\sim$~1500) in the K-2.1\,$\mu$m and K-2.3\,$\mu$m bands. The integration time (DIT) of each frame was 20 ms. Our data were observed as sequences of cal-sci-cal (cal is calibrator and sci is our target), with 5 scans for each of them.  Table \ref{Log_obs} lists the detailed information about our observations and the calibrator used for each target. Table \ref{calibrator} shows the calibrators used for our observations together with their angular diameters. We selected them from the ESO Calibration Selector CalVin, in turn based on the catalog of Lafrasse et al. (\cite{Lafrasse2010}).
 
During the acquisition of one AMBER frame, there are optical path fluctuations (jitter) that produce fringe motions. These motions reduce the squared visibility by a factor $e^{-\sigma ^2_{\phi}}$, FINITO factor, where $\sigma _{\phi}$ is the fringe phase standard deviation over the frame acquisition time. This attenuation is corrected in the science data by the calibration provided the FINITO factors are similar in science and calibrator (more information in the AMBER User Manual \footnote{http://www.eso.org/sci/facilities/paranal/instruments/amber/doc.html}).

As a first step we selected the scans such that the FINITO factors were similar between cal and sci data. After that, we obtained the visibility data from our selected AMBER observations using the 3.0.7 version of the \textit{amdlib} data reduction package (Tatulli et al. \cite{Tatulli2007}; Chelli et al. \cite{Chelli2009}). This included the removal of the bad pixel map and the correction for the flat contribution. Afterwards, we calculated the pixel-to-visibility matrix (P2VM) to calibrate our data for the instrumental dispersive effects, and obtained the interferometric observables (visibility and closure phase). Next, we appended all scans of the same source taken consecutively, selected and averaged the resulting  visibilities of each frame using appropriate criteria. In our case, the criteria were based on the flux (we selected all frames having flux densities three times higher than the noise) and on the signal-to-noise ratio (S/N). We only used 80\% of the remaining frames with best S/N \footnote{see AMBER Data Reduction Software User Manual; http://www.jmmc.fr/doc/approved/JMMC-MAN-2720-0001.pdf}. 

Using scripts of IDL (Interactive Data Language), which have been developed by us, we performed the absolute wavelength calibration by correlating the AMBER flux spectra  with a reference spectrum, that of the star BS~4432 (spectral type K4.5 III, similar to our calibrators; Lan\c con \& Wood \cite{Lancon2000}). A relative flux calibration of the targets was performed by using the instrumental response, estimated by the calibrators and the BS~4432 spectrum. Finally, calibrated visibility spectra were obtained by using the average of two transfer function measurements taken before and after each science  target observation.  In the case of V602~Car (25-26 April), HD~95687 (24-25 April) and HD~183589 (03-04 August), we only used the first calibrator, because the other calibrator had very different FINITO factors and their visibilities were not of sufficient quality. The error of the transfer function was calculated as in our previous work (Arroyo-Torres et al. \cite{Arroyo2013}, \cite{Arroyo2014}).

\onlfig{2}{
\begin{figure*}
\centering
\includegraphics[width=0.49\hsize]{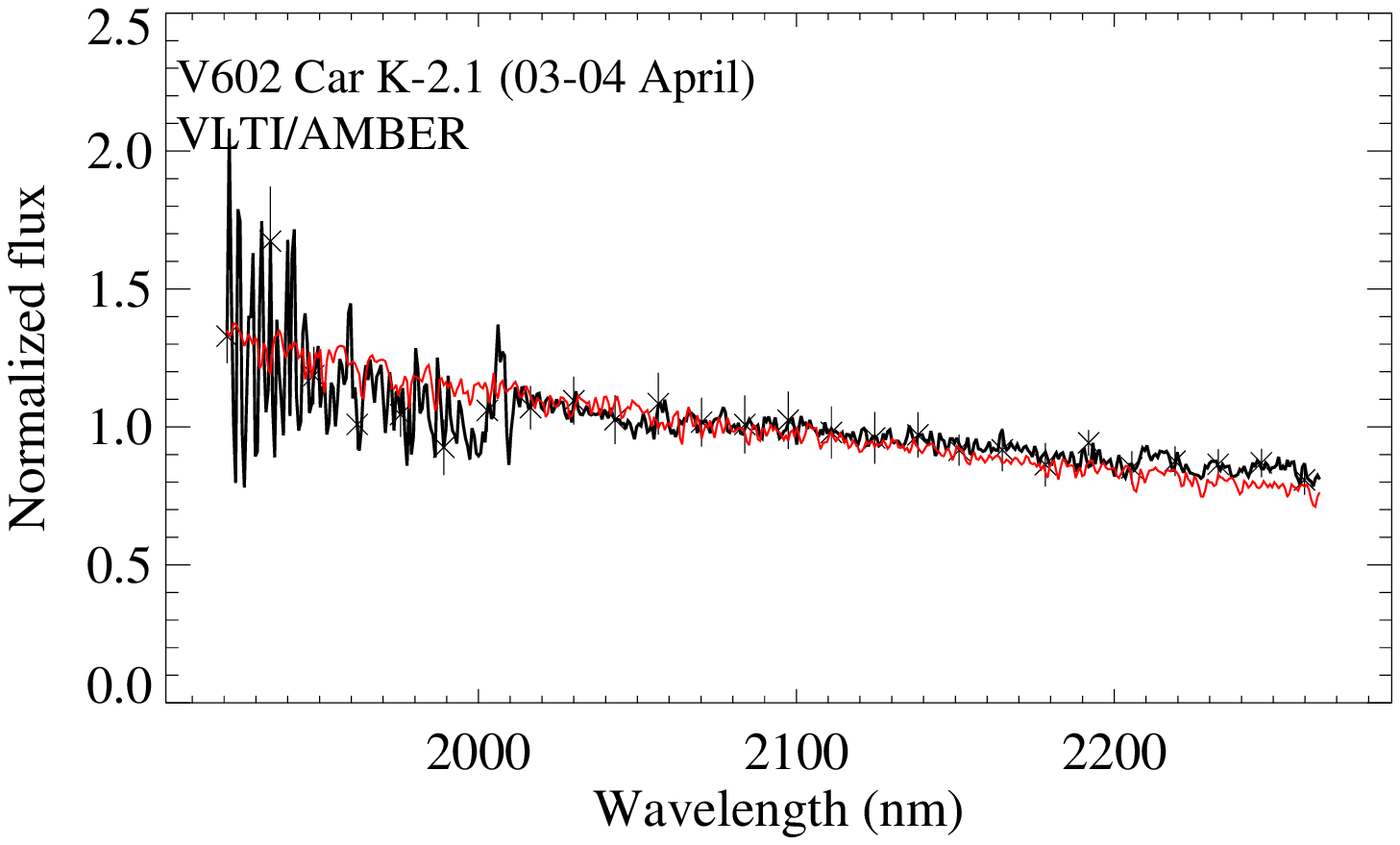}
\includegraphics[width=0.49\hsize]{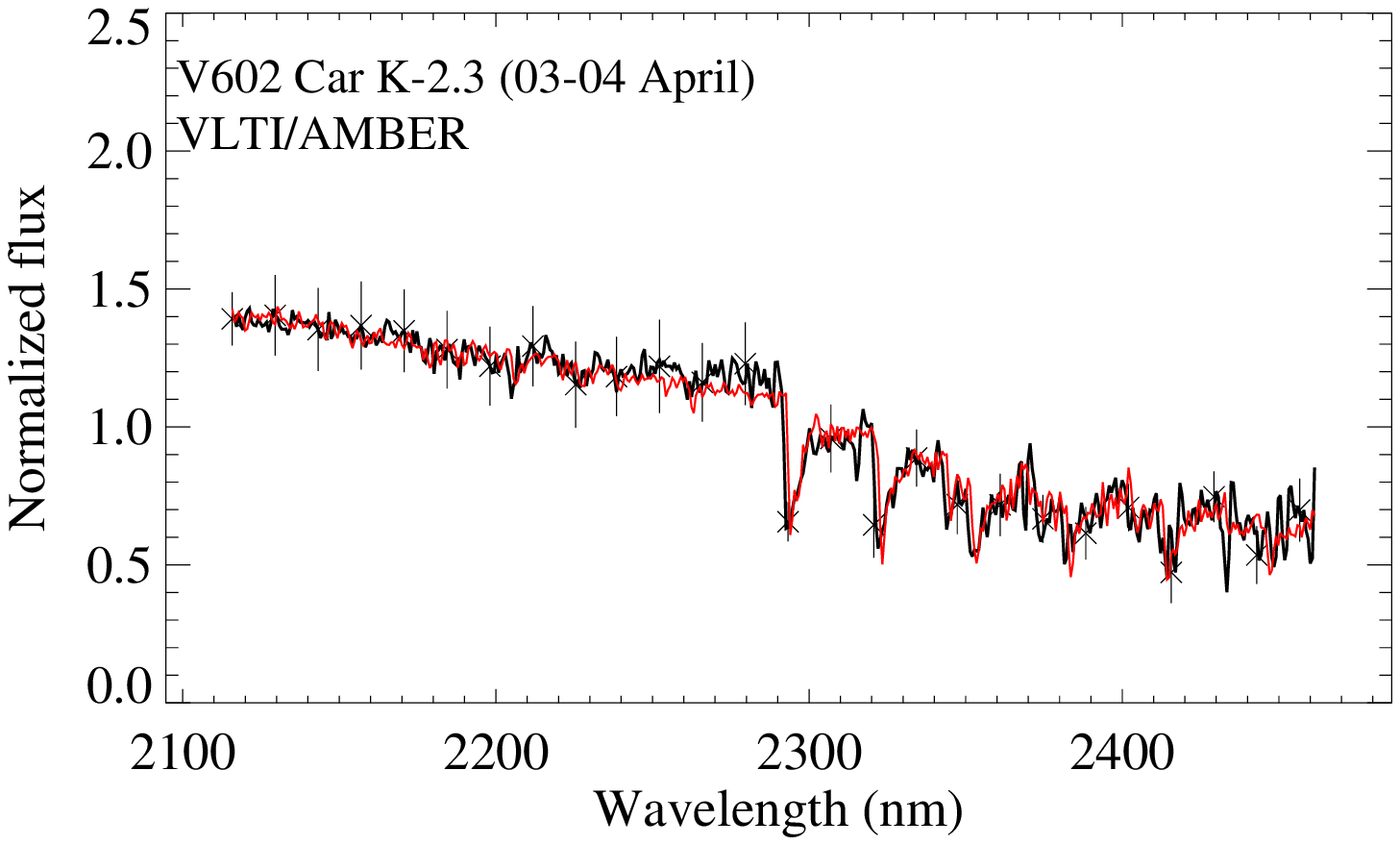}
\includegraphics[width=0.49\hsize]{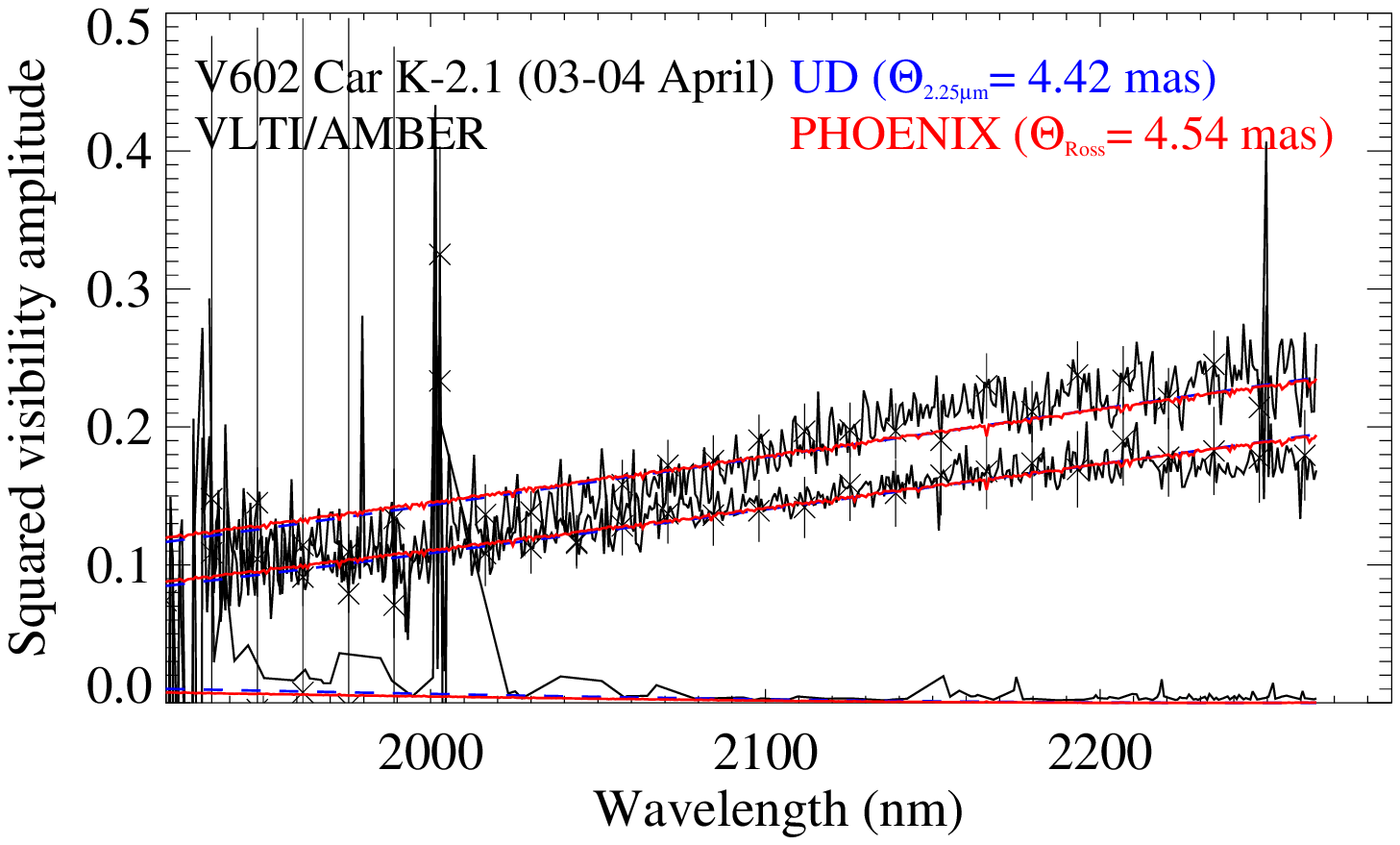}
\includegraphics[width=0.49\hsize]{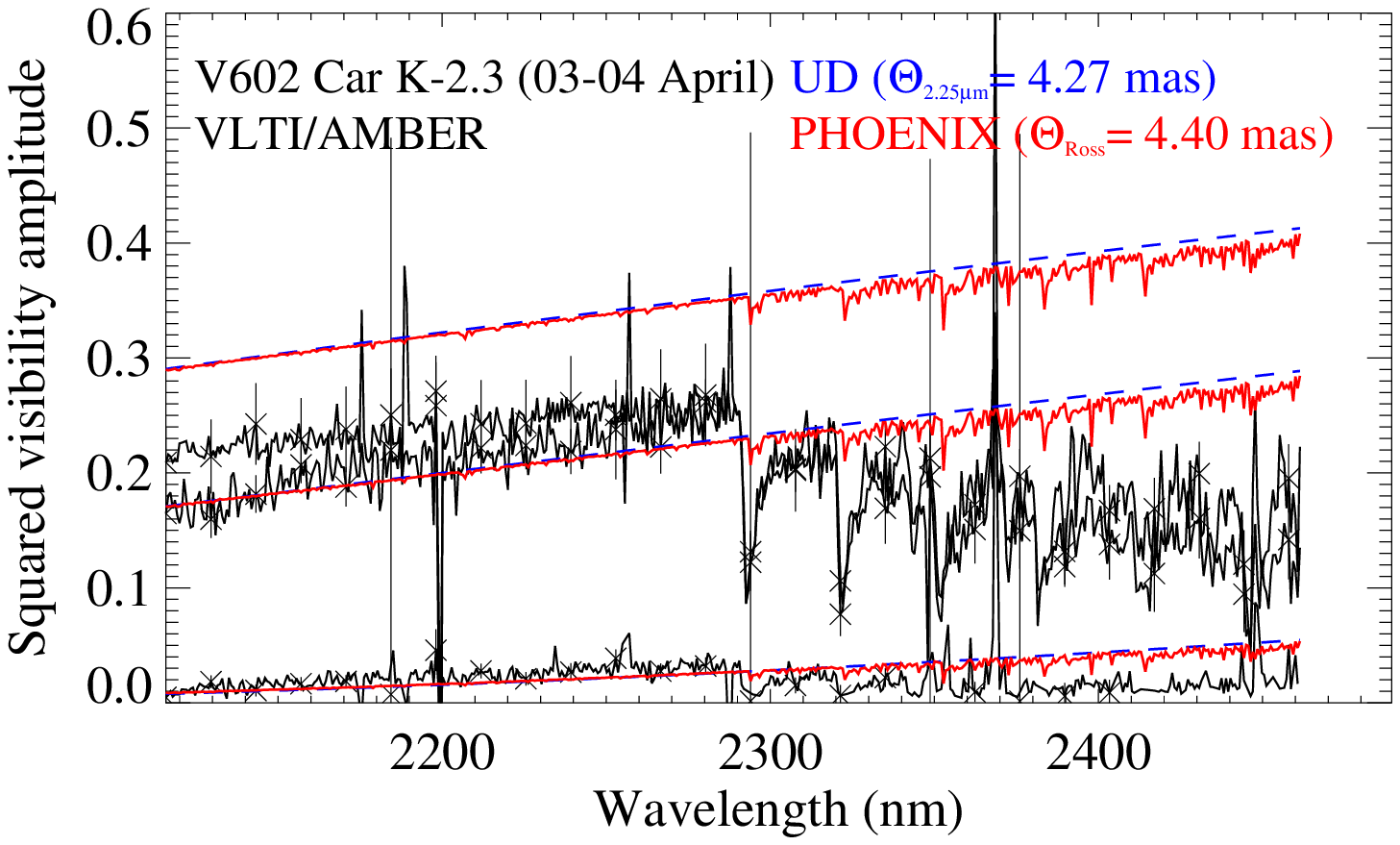}
\includegraphics[width=0.49\hsize]{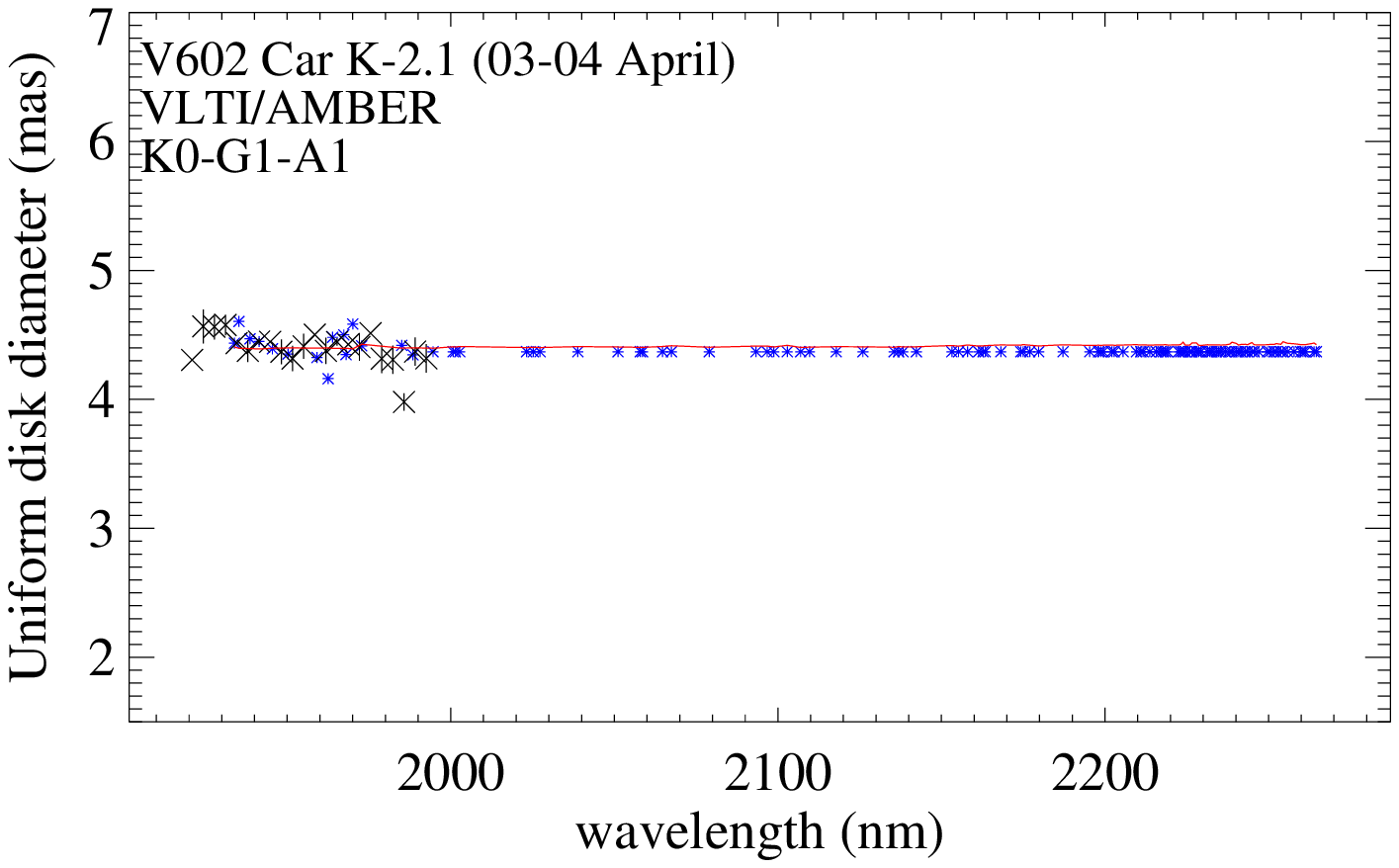}
\includegraphics[width=0.49\hsize]{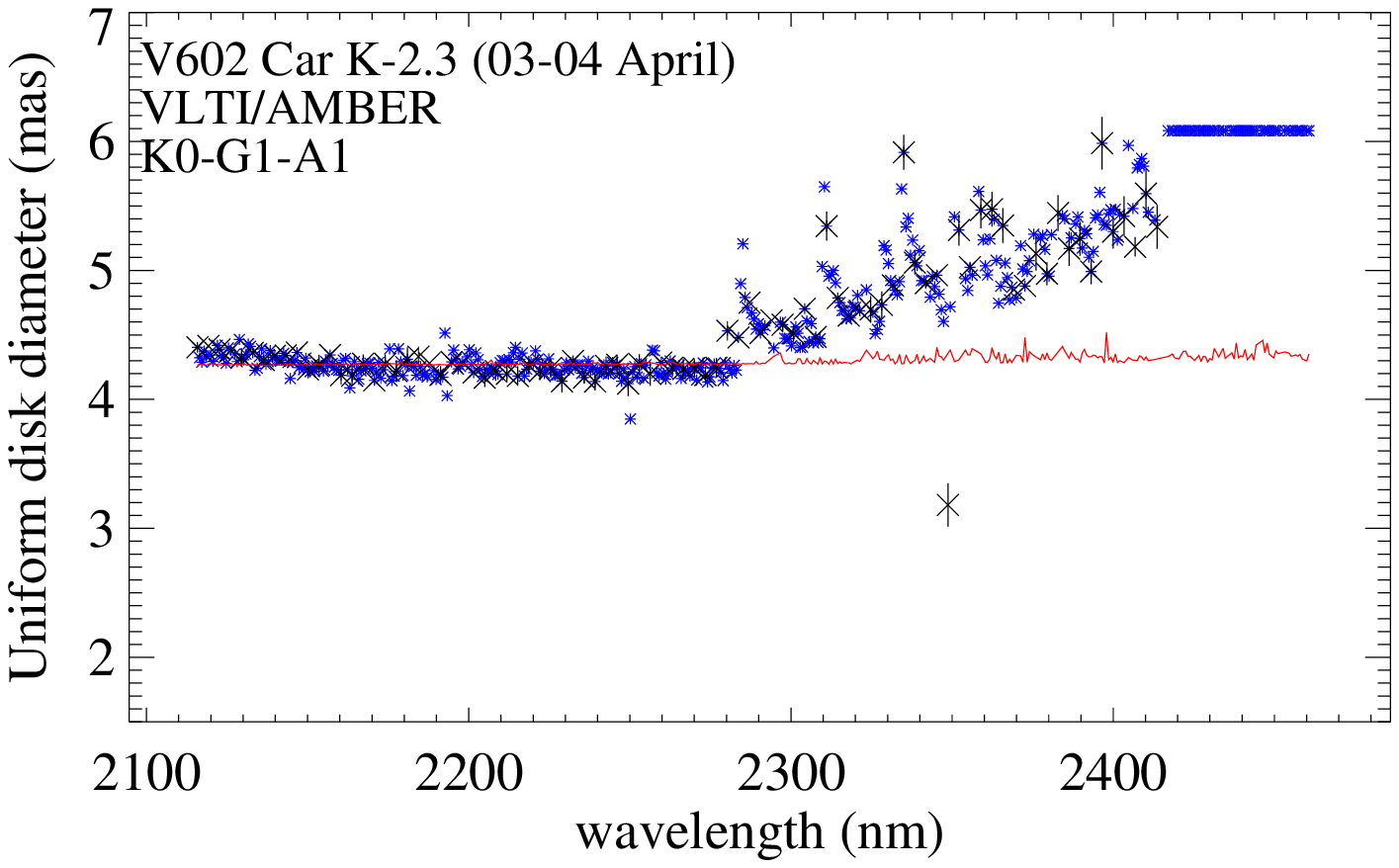}
\includegraphics[width=0.49\hsize]{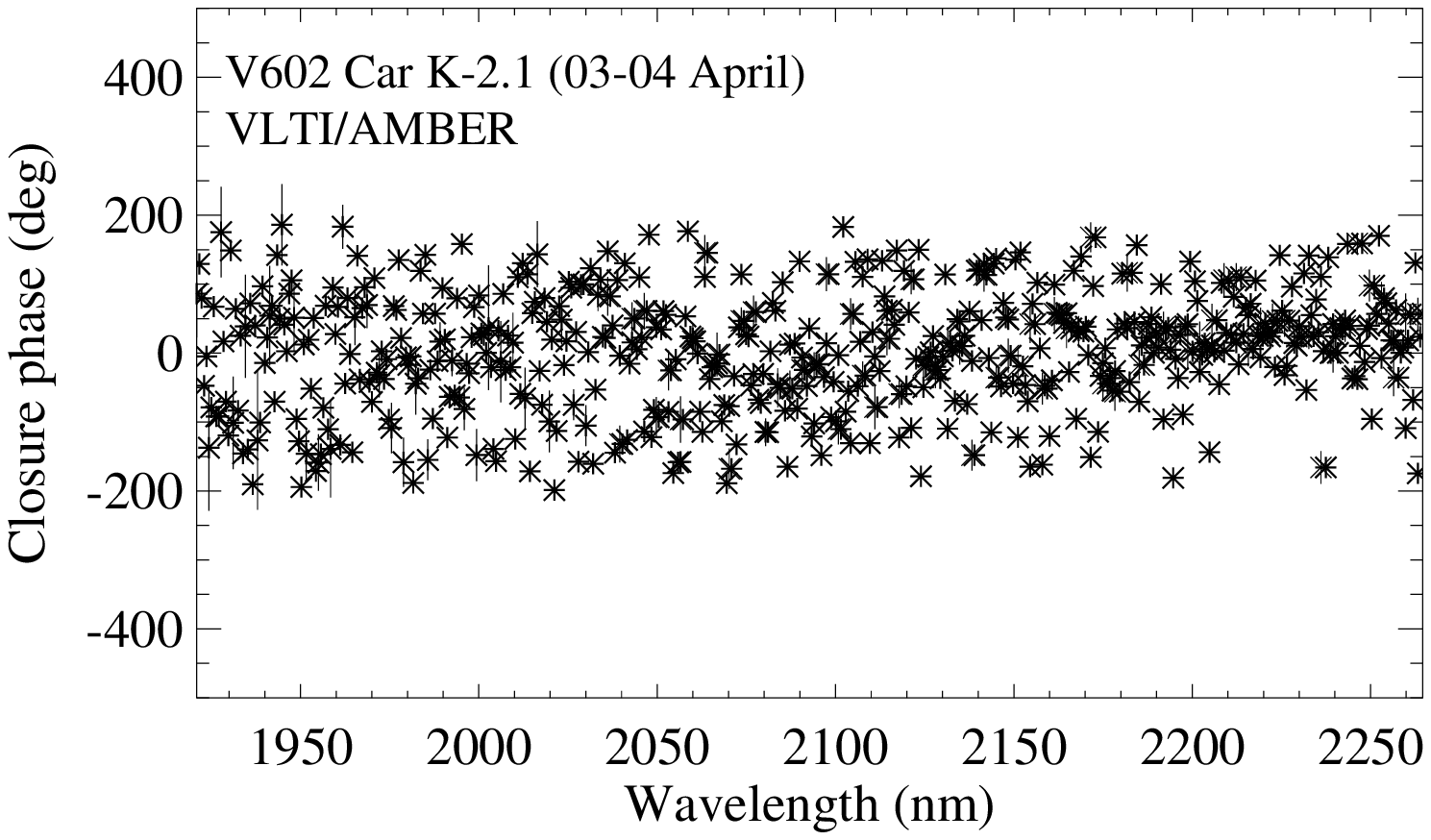}
\includegraphics[width=0.49\hsize]{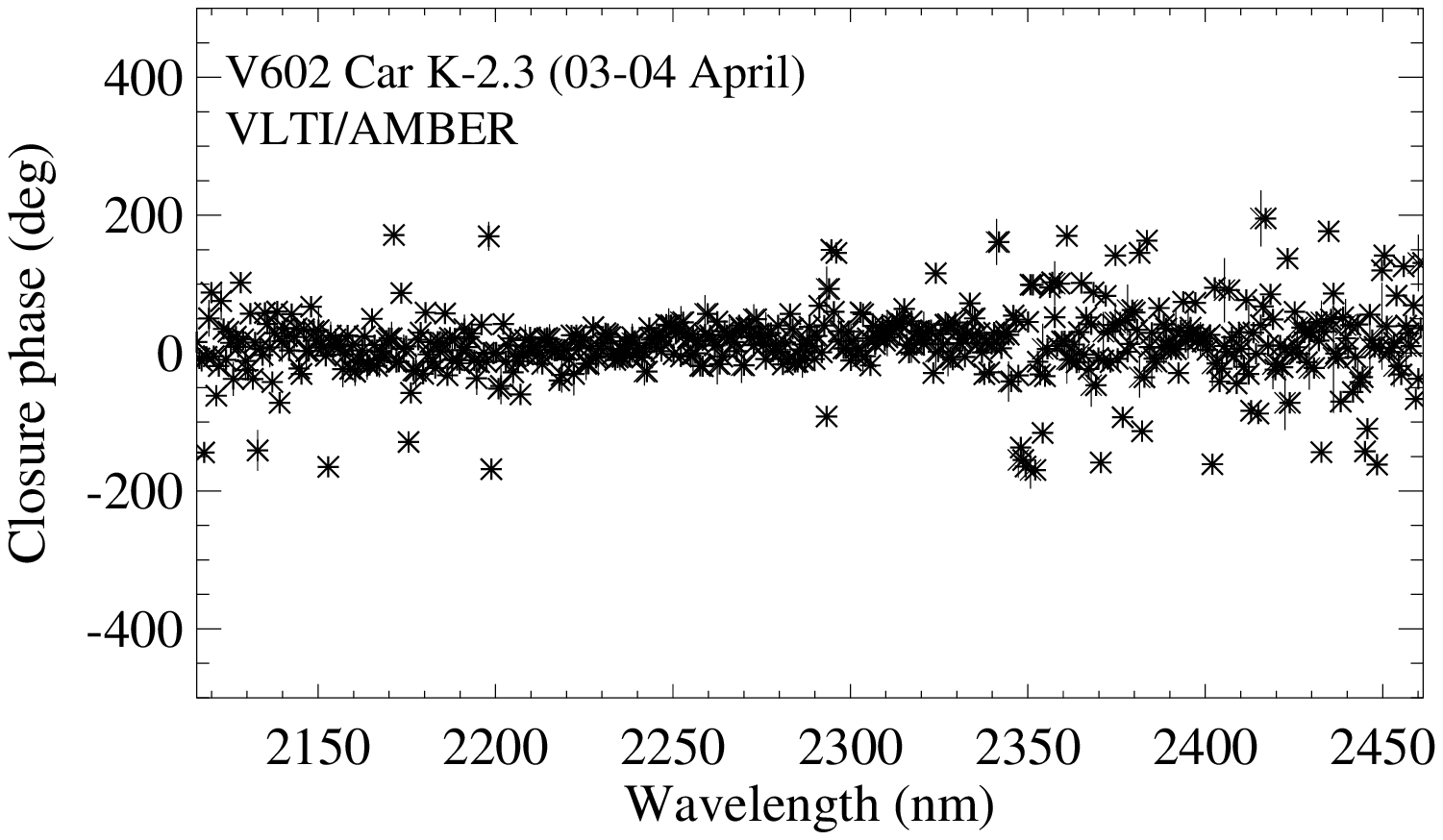}
\caption{As Fig.~\ref{resul_V602Car_2604}, but for data of V602~Car obtained with the MR-K 2.1\,$\mu$m setting on 4 April 2013 (left) and with the MR-K 2.3\,$\mu$m setting obtained on 4 April 2013 (right).}
\label{resul_V602Car_0404}
\end{figure*} }

\onlfig{3}{
\begin{figure*}
\centering
\includegraphics[width=0.49\hsize]{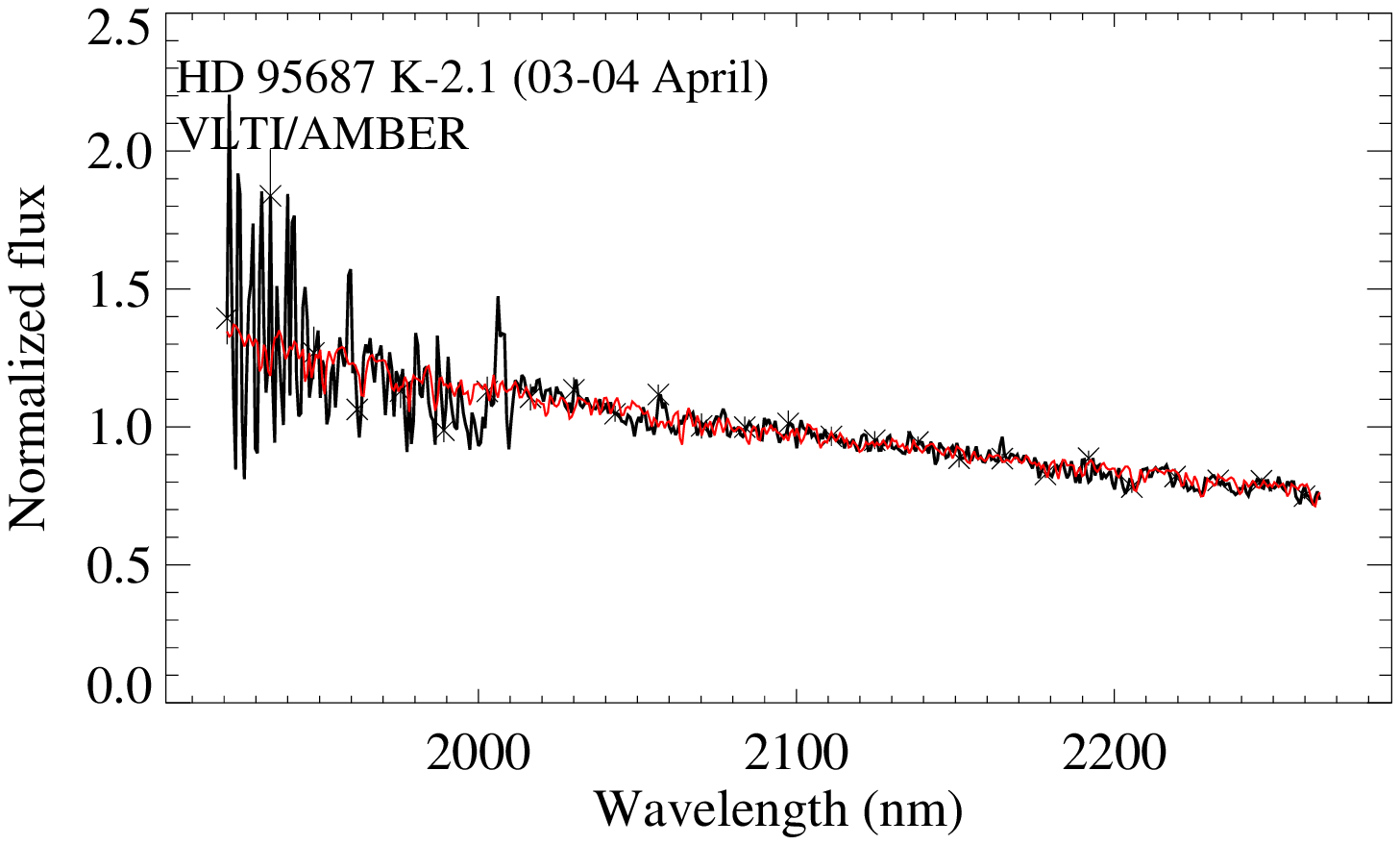}
\includegraphics[width=0.49\hsize]{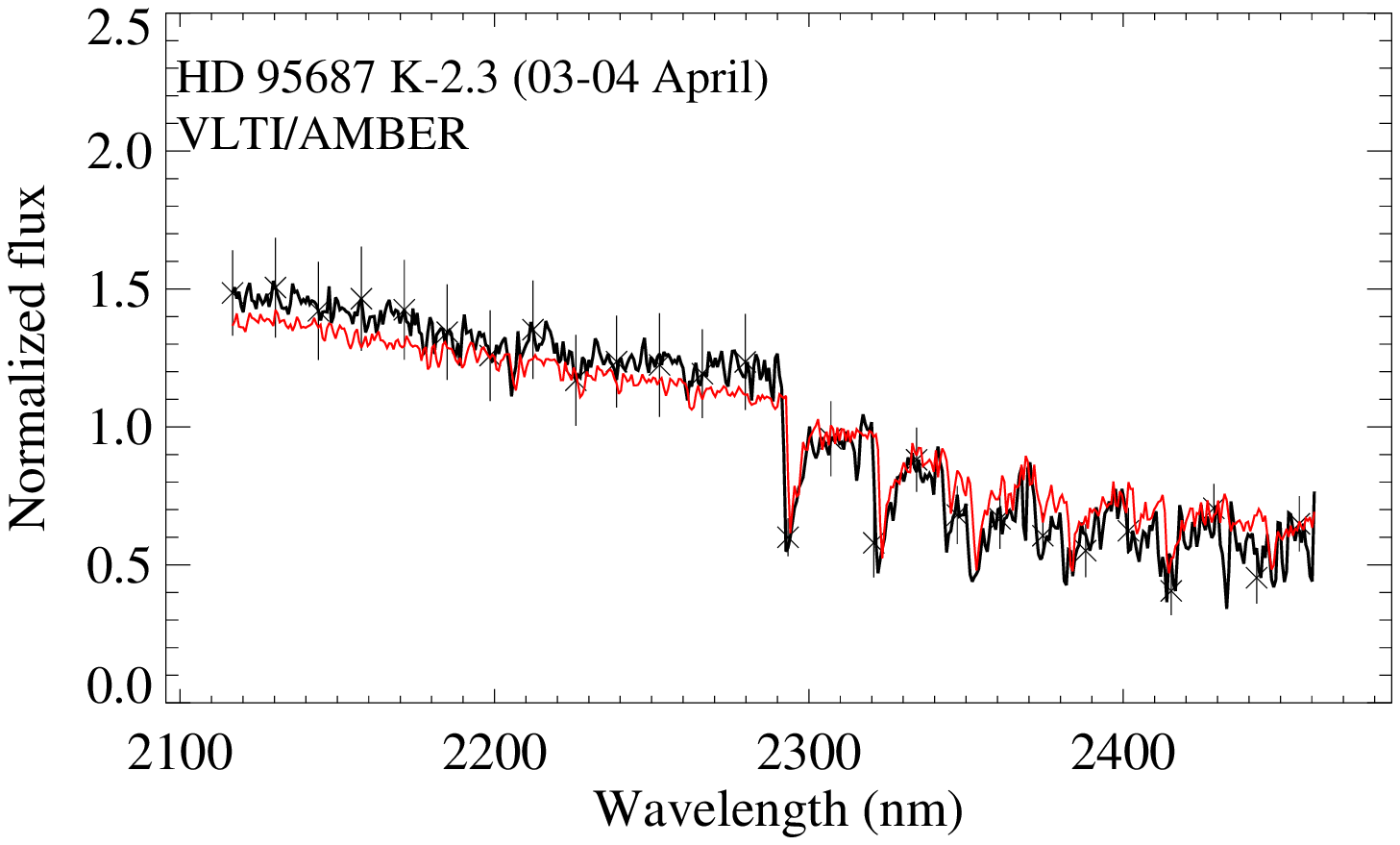}
\includegraphics[width=0.49\hsize]{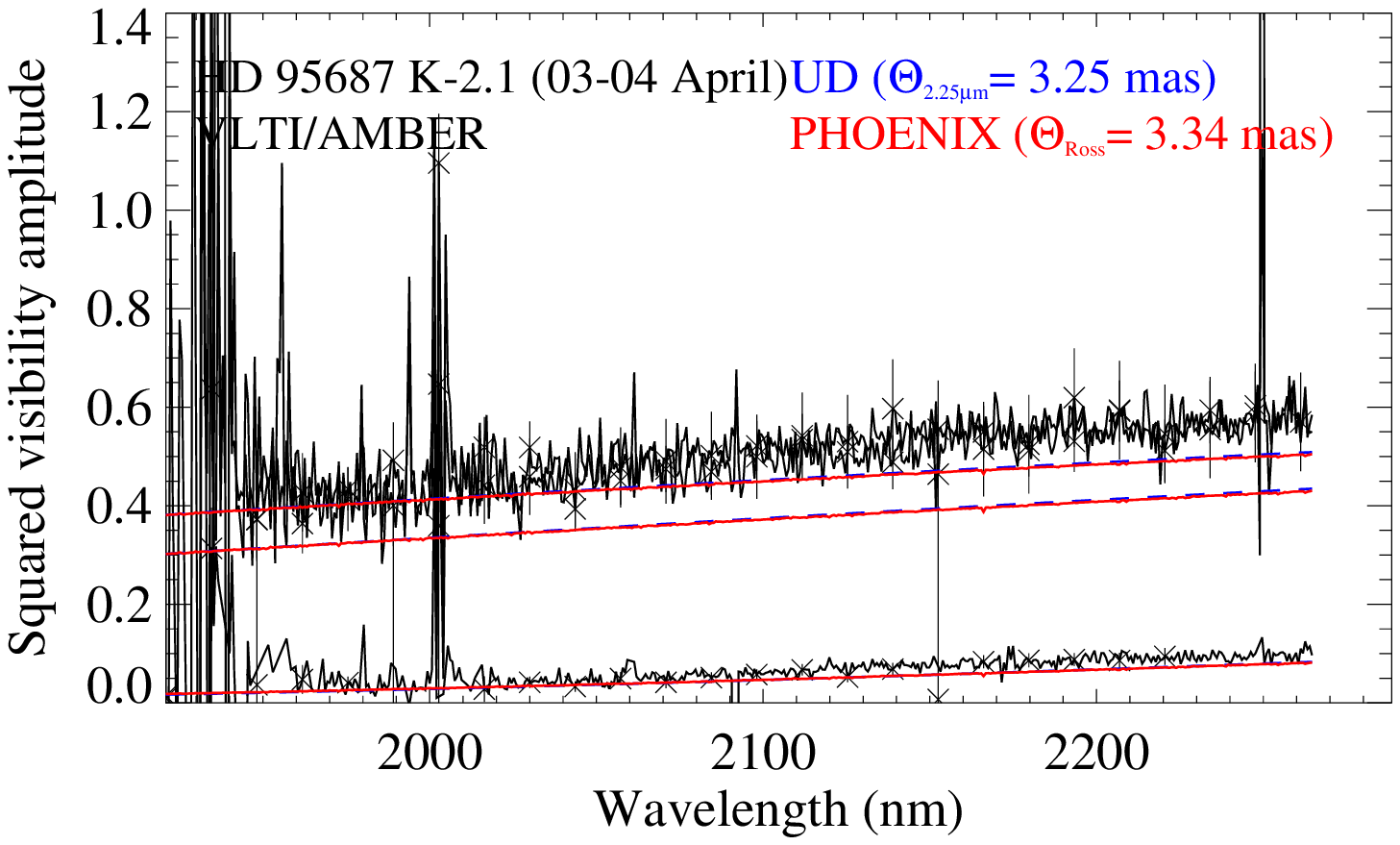}
\includegraphics[width=0.49\hsize]{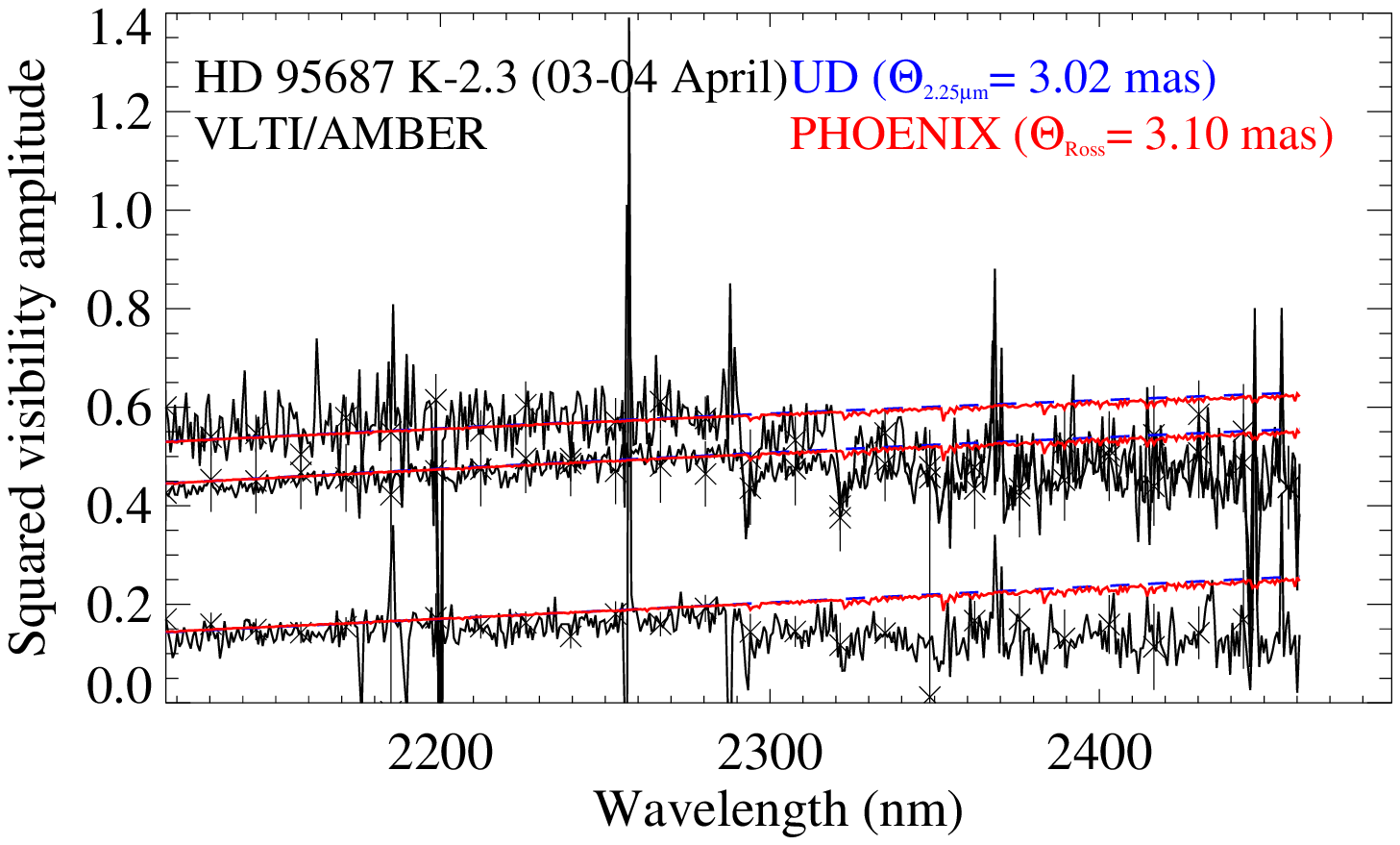}
\includegraphics[width=0.49\hsize]{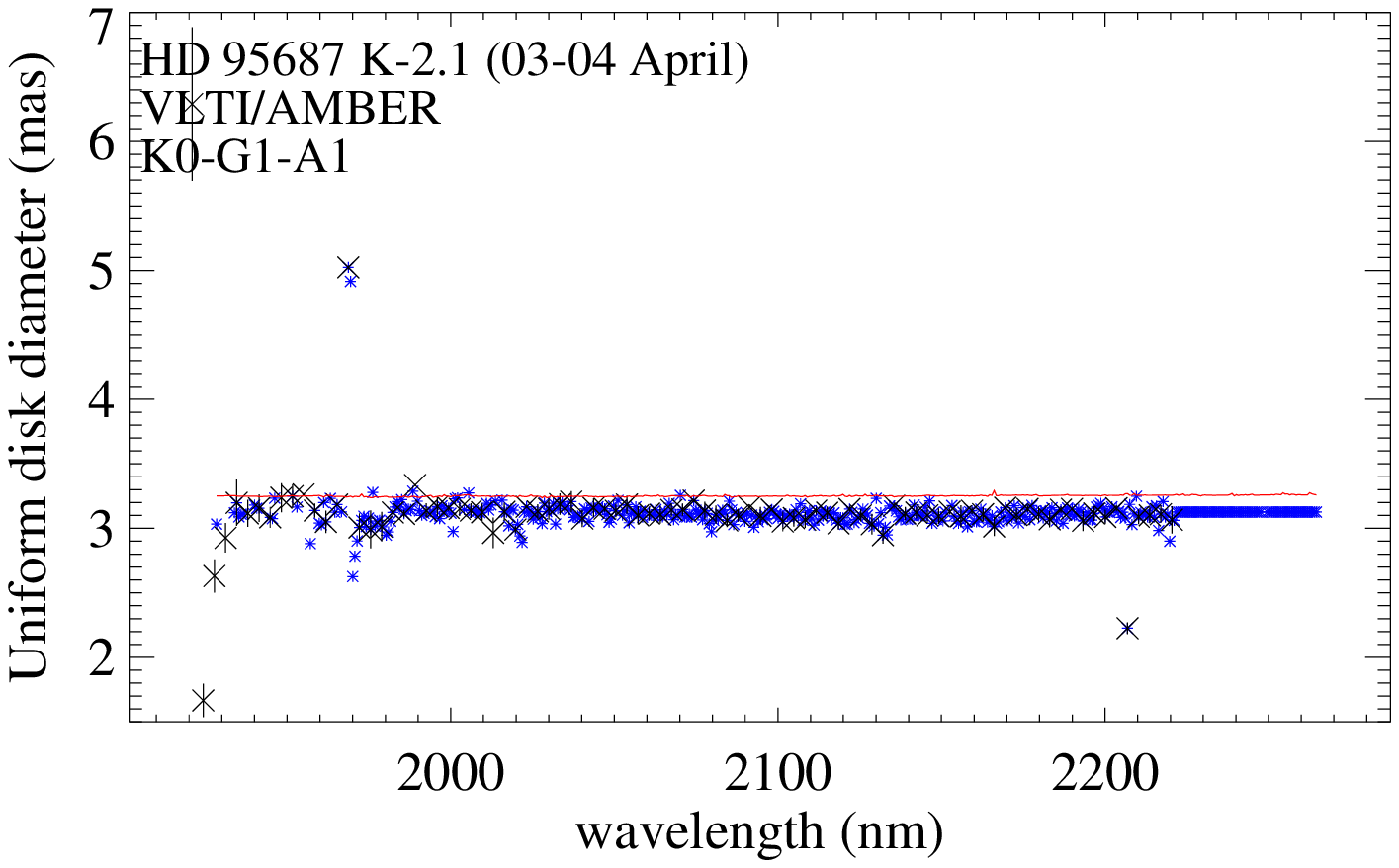}
\includegraphics[width=0.49\hsize]{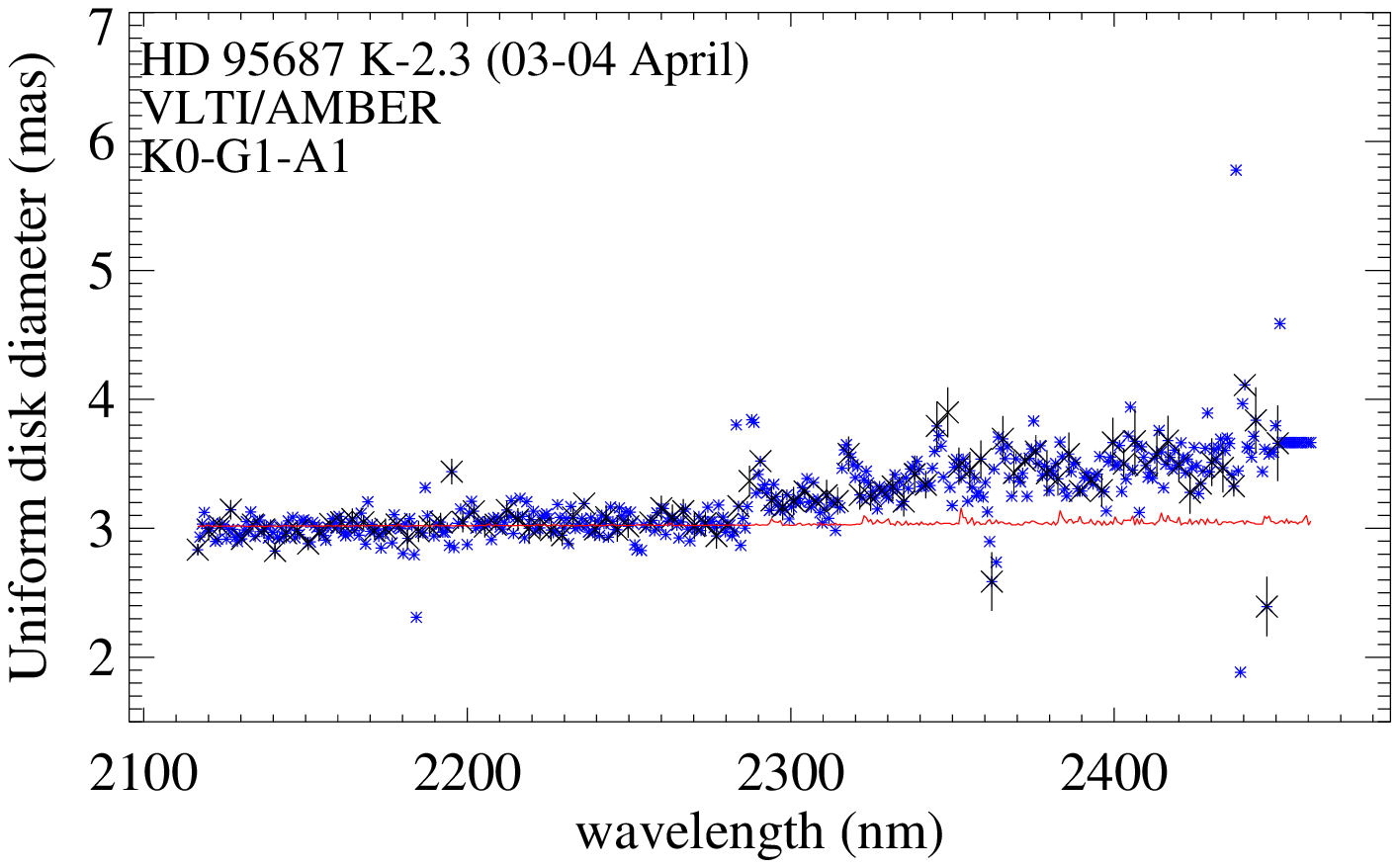}
\includegraphics[width=0.49\hsize]{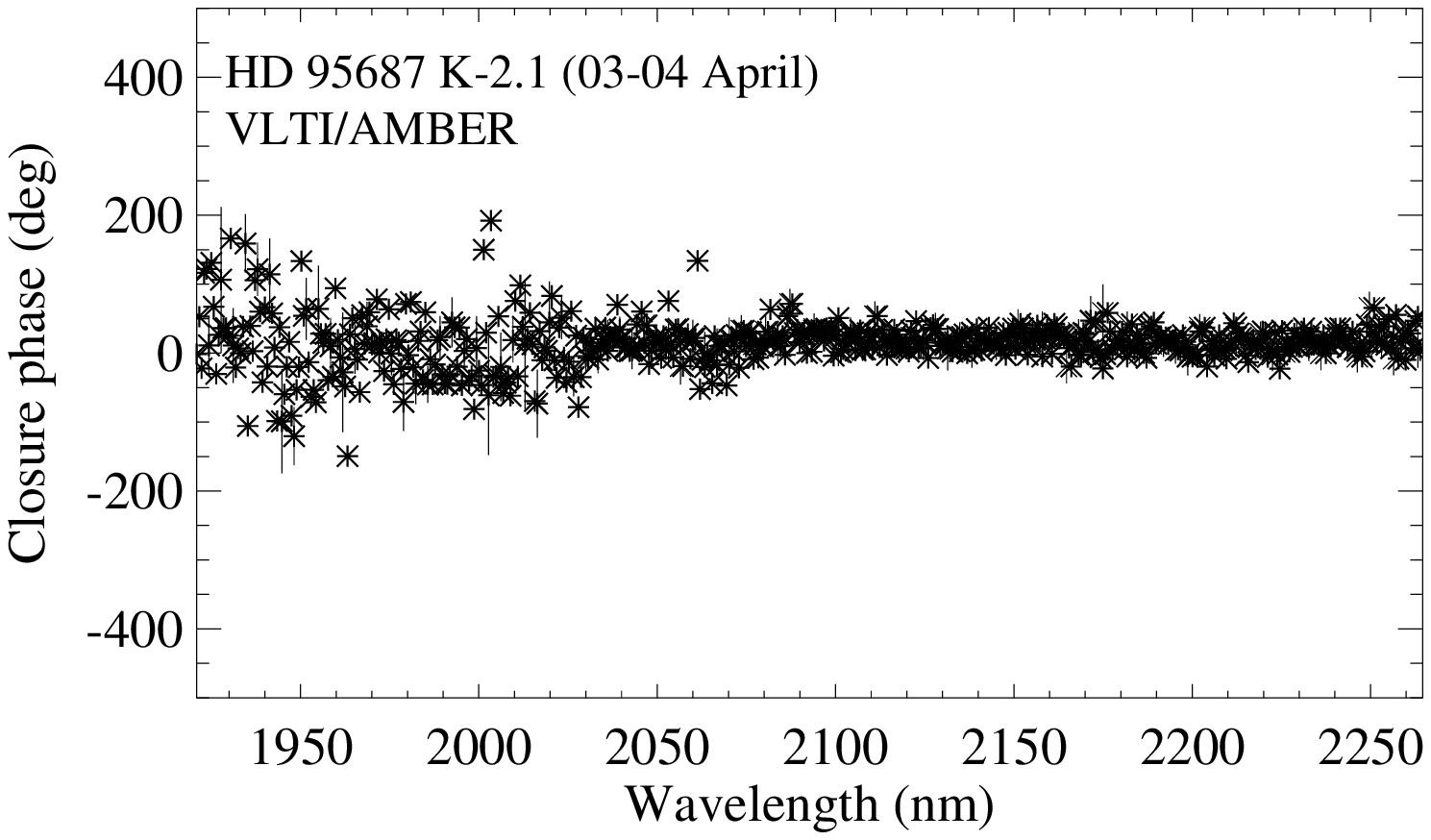}
\includegraphics[width=0.49\hsize]{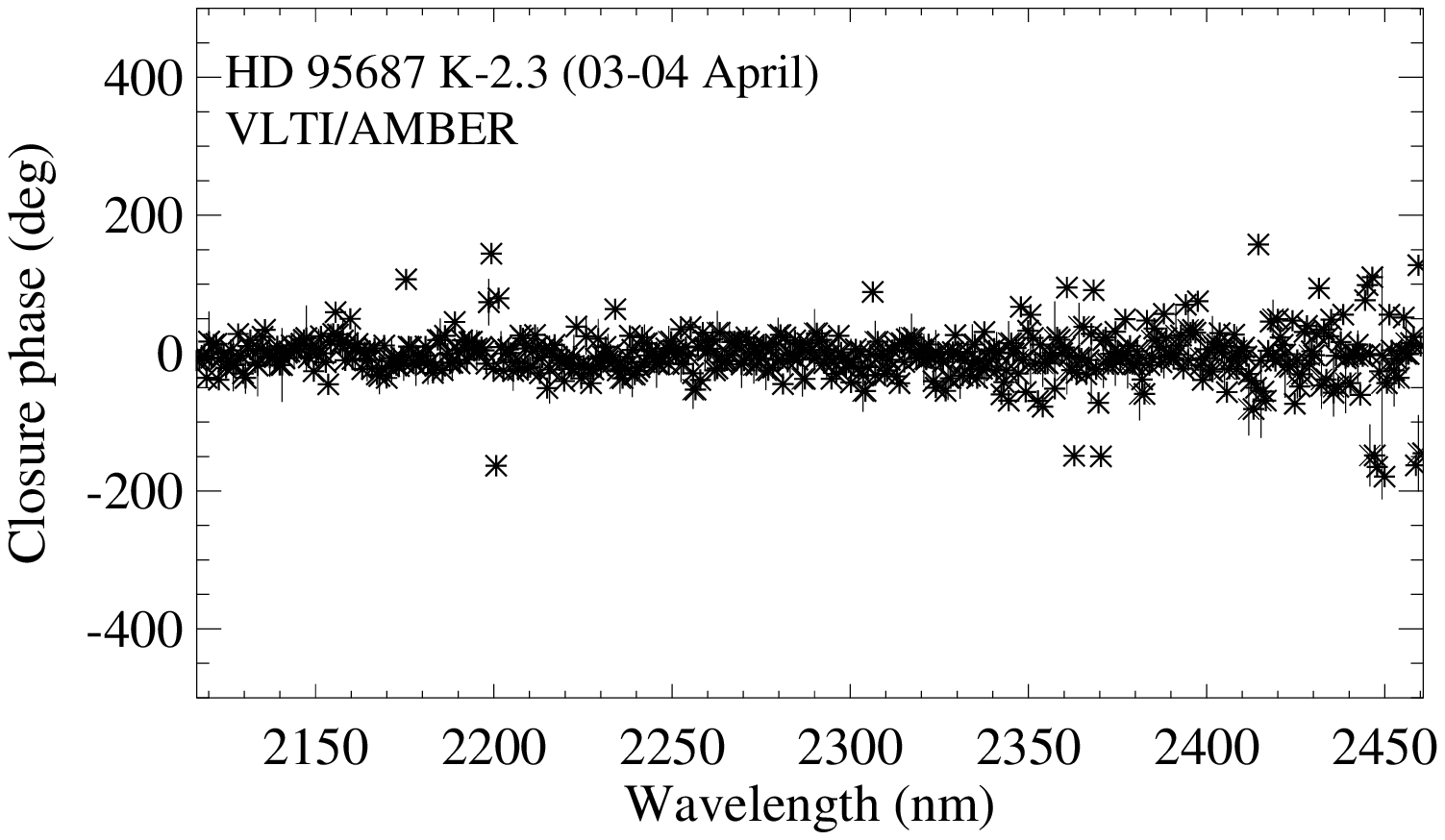}
\caption{As Fig.~\ref{resul_V602Car_2604}, but for data of HD~95687 obtained with the MR-K 2.1\,$\mu$m setting on 4 April 2013 (left) and with the MR-K 2.3\ $\mu$m setting on 4 April 2013 (right).}
\label{resul_HD95687_0404}
\end{figure*} }

\onlfig{4}{
\begin{figure*}
\centering
\includegraphics[width=0.49\hsize]{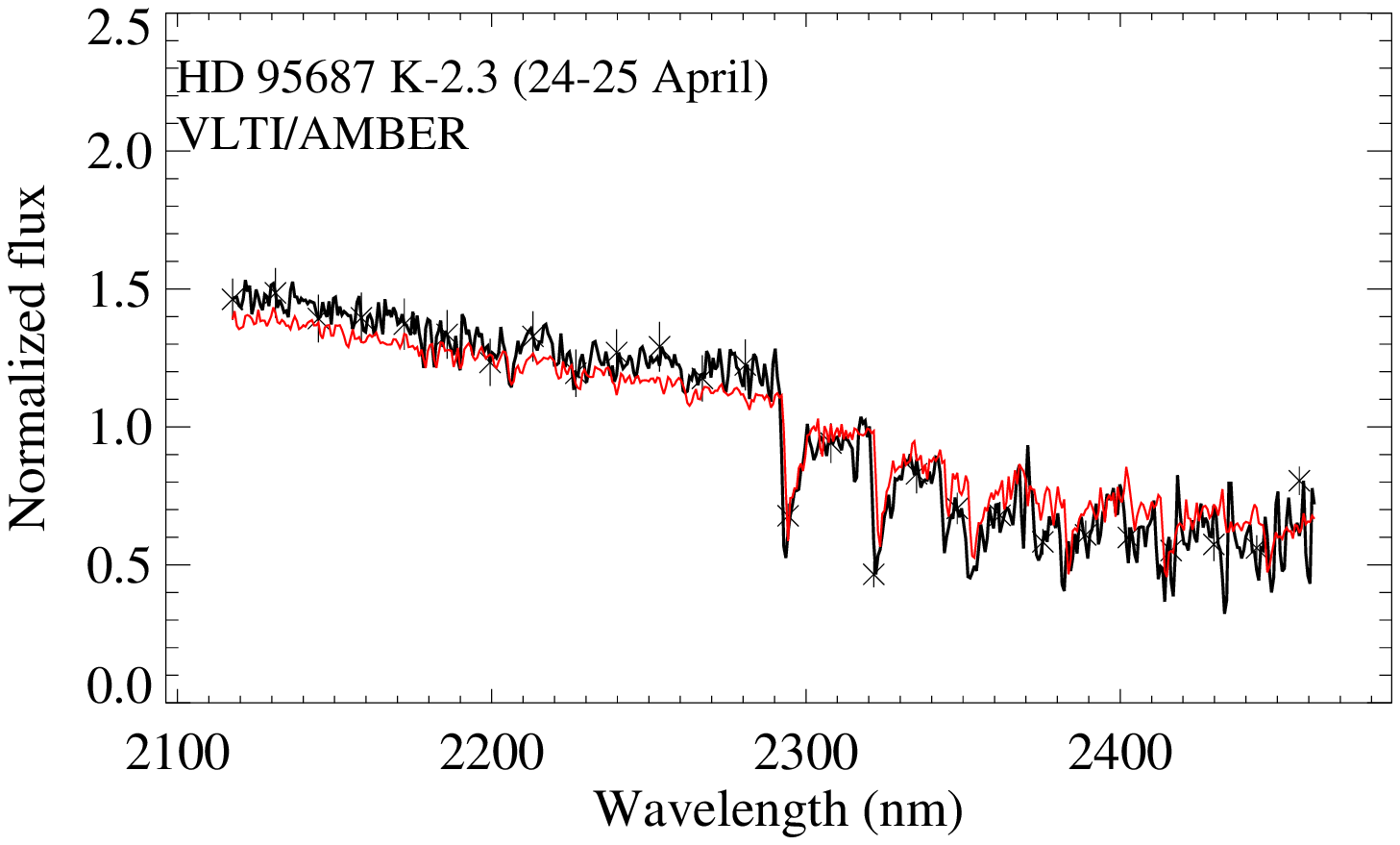}
\includegraphics[width=0.49\hsize]{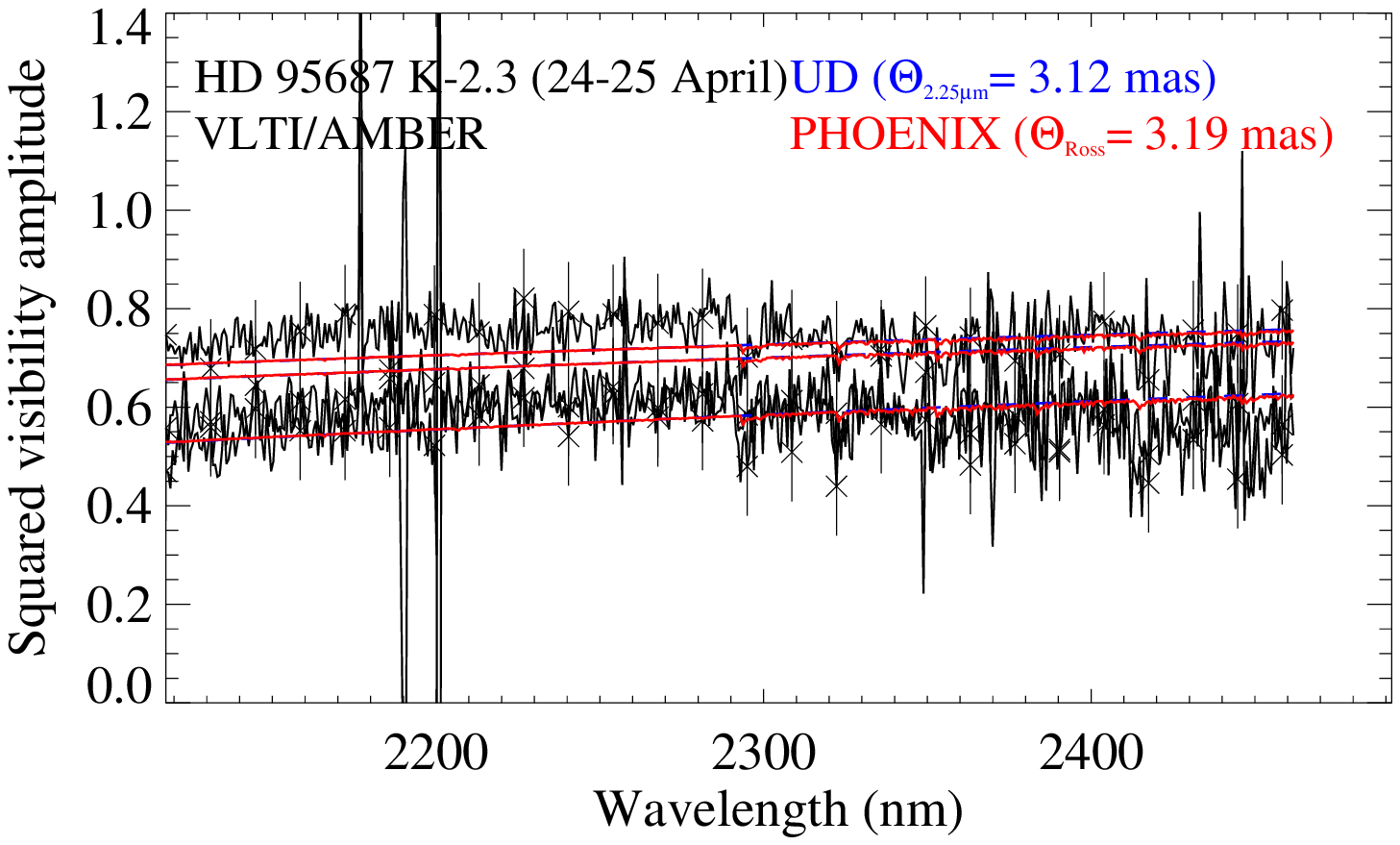}
\includegraphics[width=0.49\hsize]{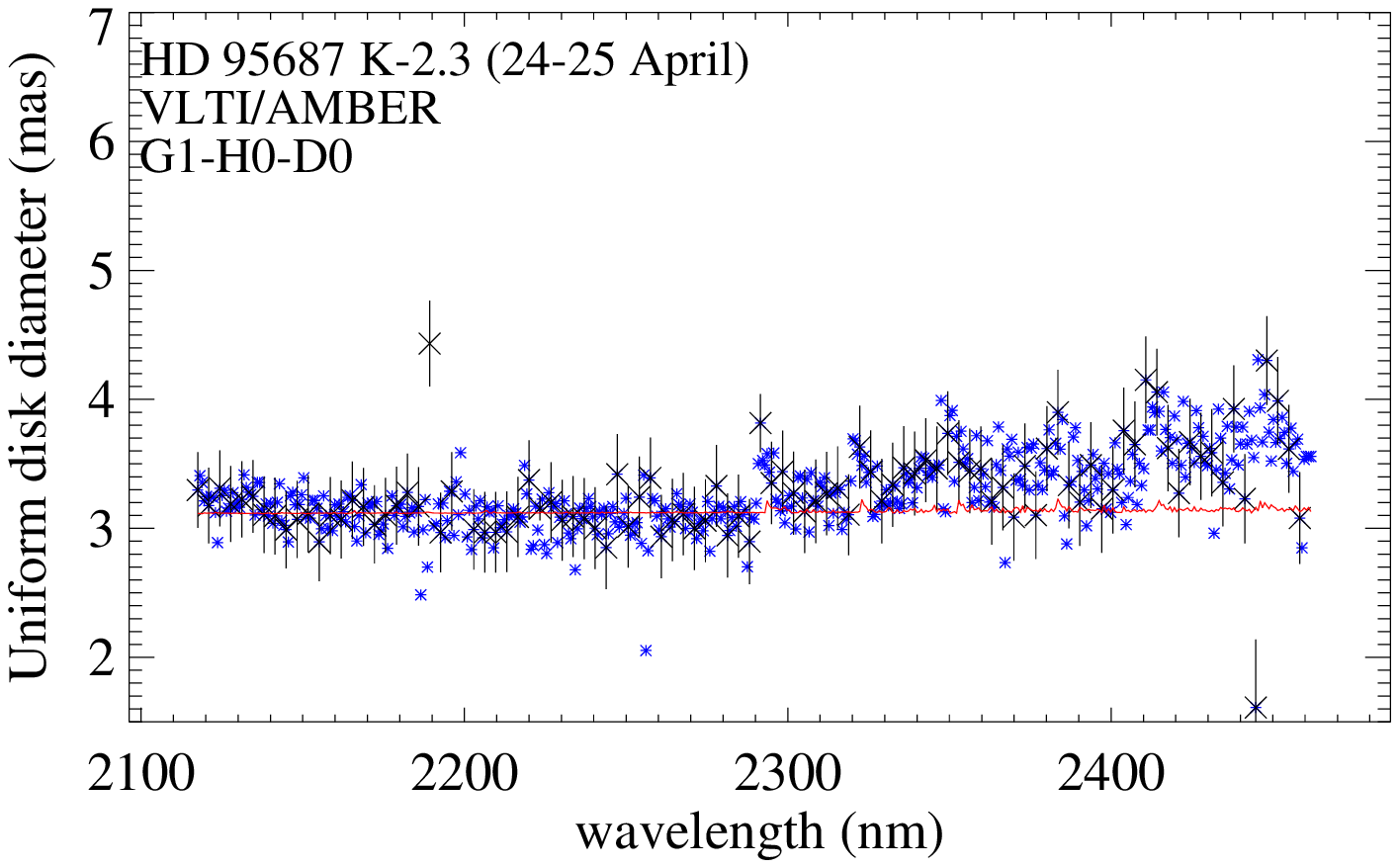}
\includegraphics[width=0.49\hsize]{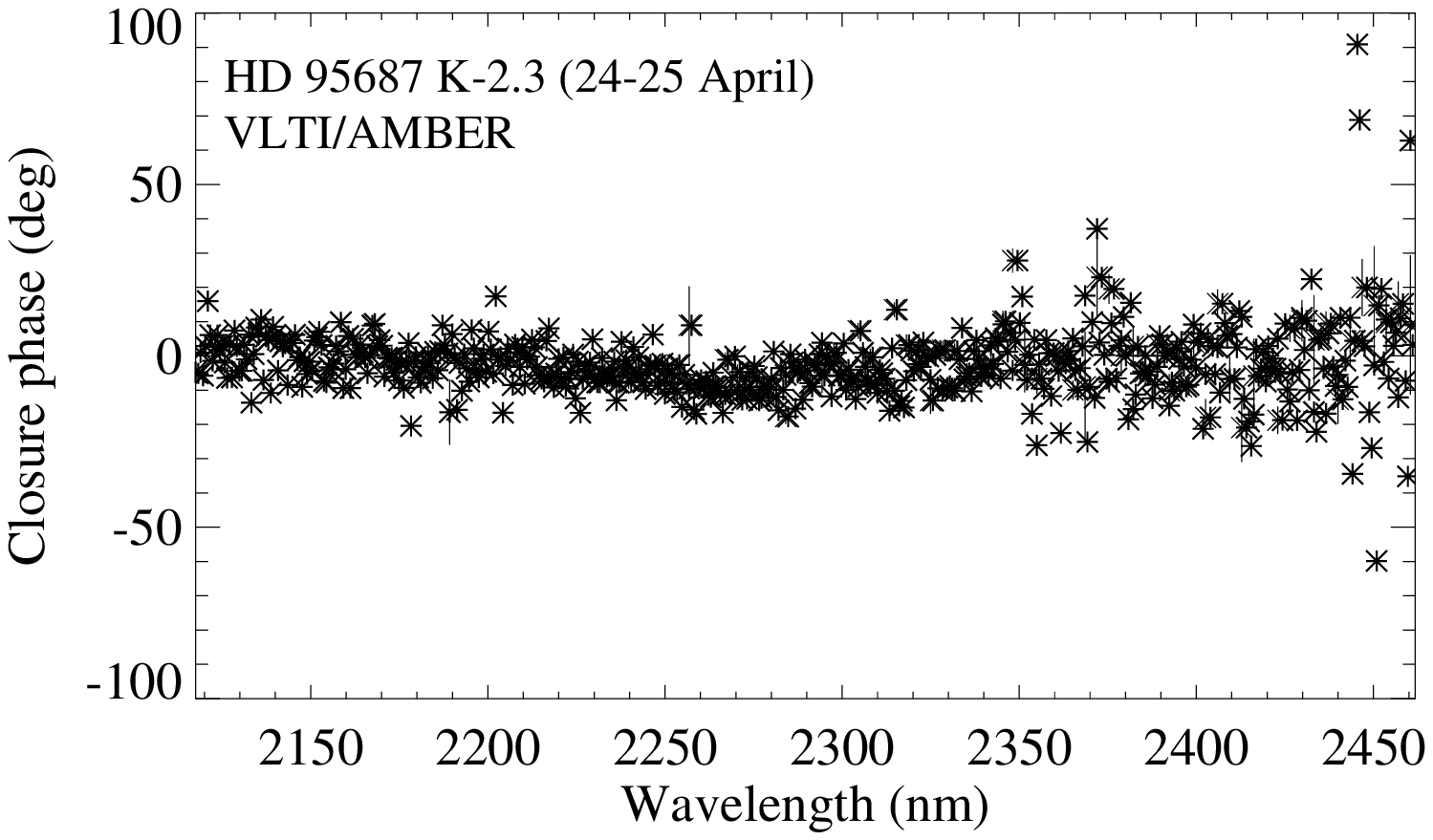}
\caption{As Fig.~\ref{resul_V602Car_2604}, but for data of HD~95687 obtained with the MR-K 2.3\,$\mu$m setting on 25 April 2013.}
\label{resul_HD95687_2504}
\end{figure*}}

\onlfig{5}{
\begin{figure*}
\centering
\includegraphics[width=0.49\hsize]{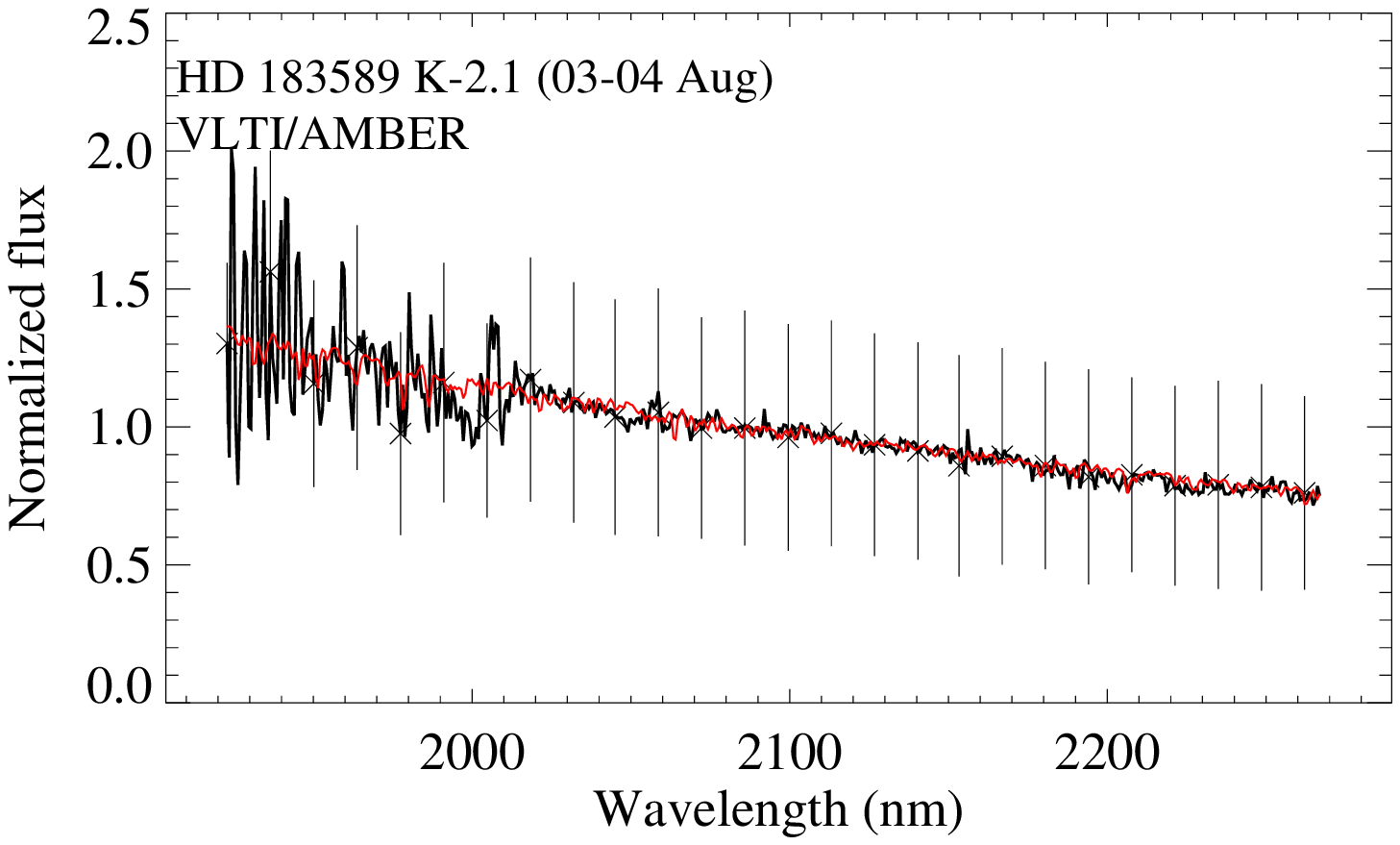}
\includegraphics[width=0.49\hsize]{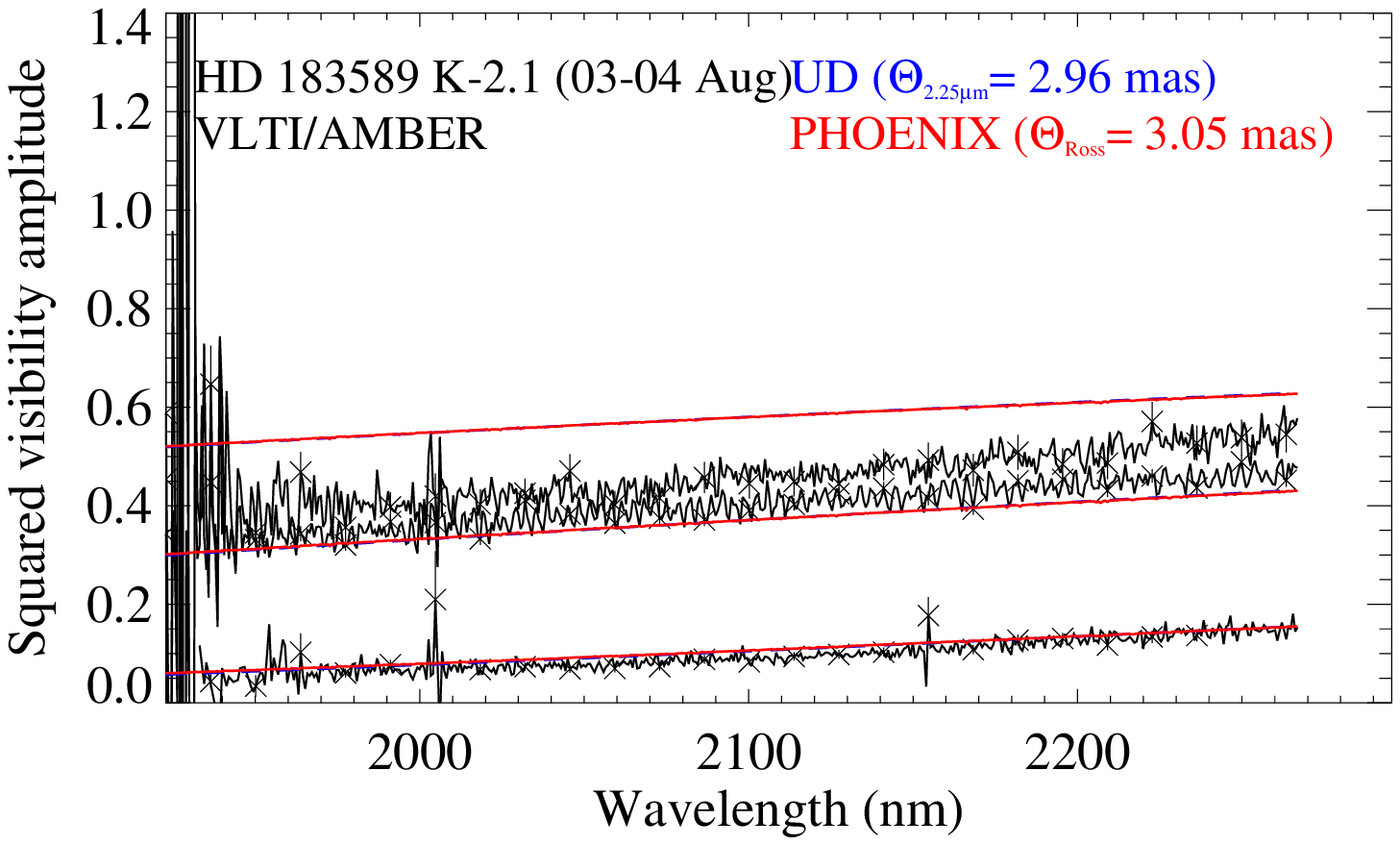}
\includegraphics[width=0.49\hsize]{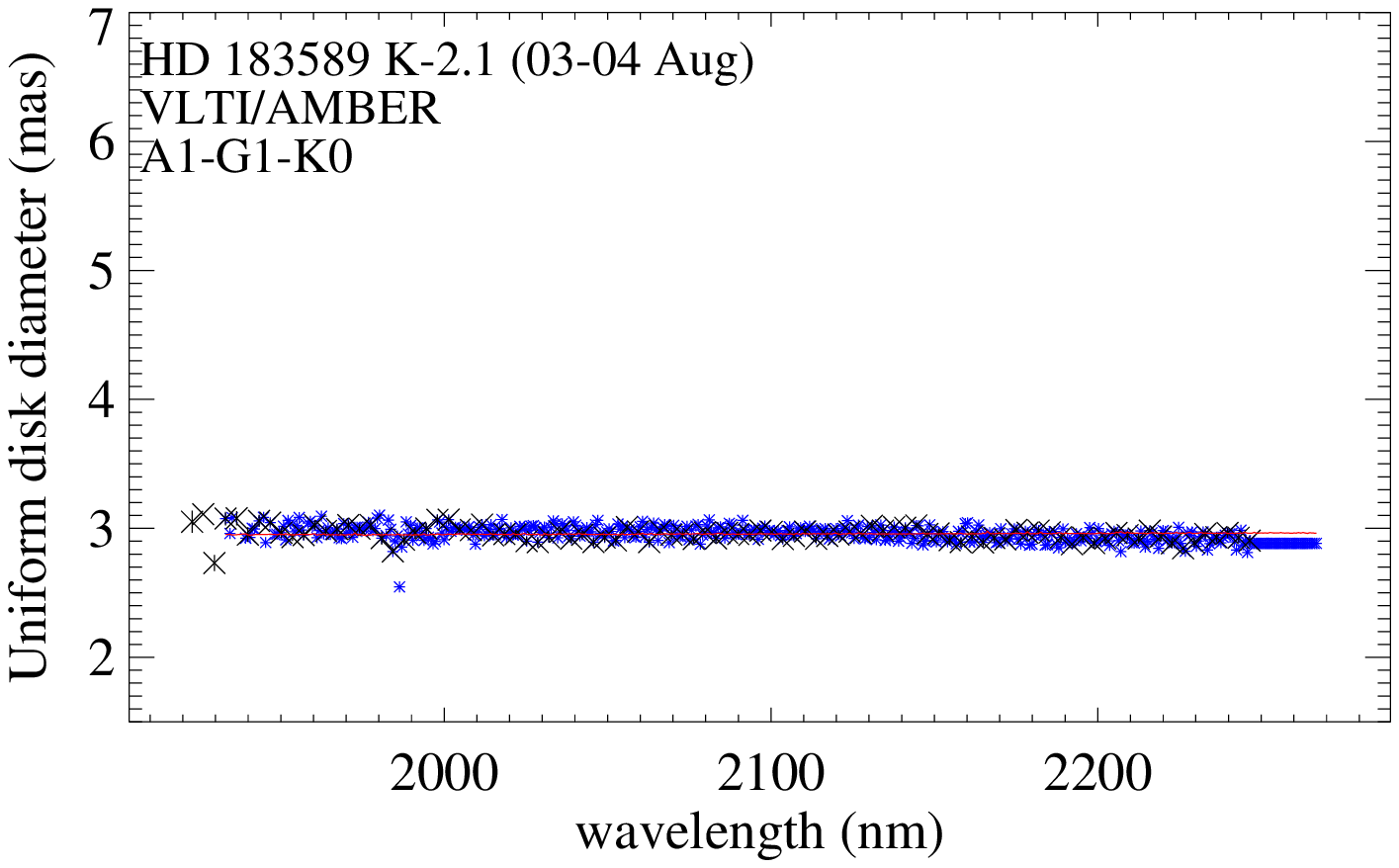}
\includegraphics[width=0.49\hsize]{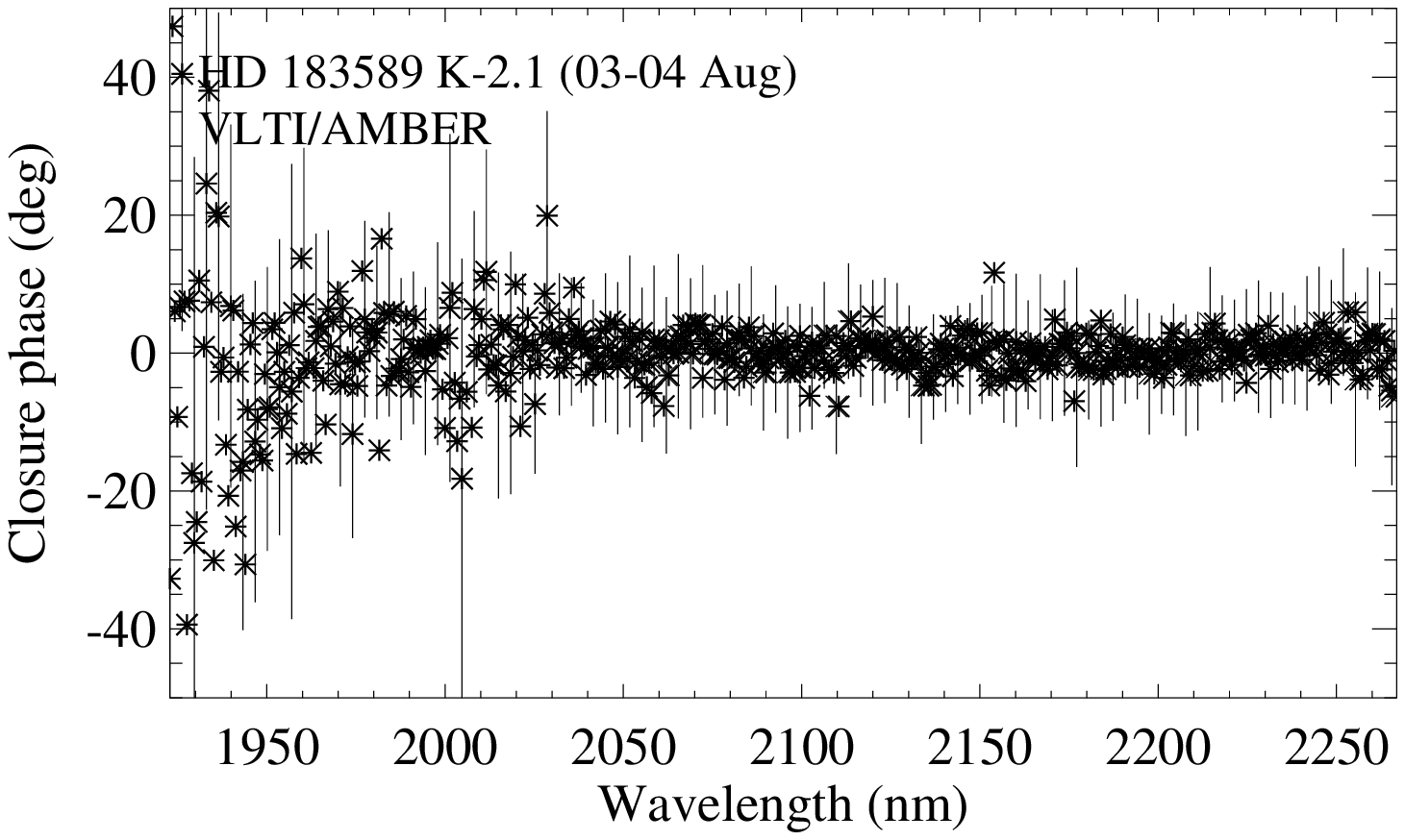}
\caption{As Fig.~\ref{resul_V602Car_2604}, but for data of HD~183889 obtained with the MR-K 2.1\,$\mu$m setting on 4 August 2013.}
\label{resul_HD183589_0408}
\end{figure*}}

\onlfig{6}{
\begin{figure*}
\centering
\includegraphics[width=0.49\hsize]{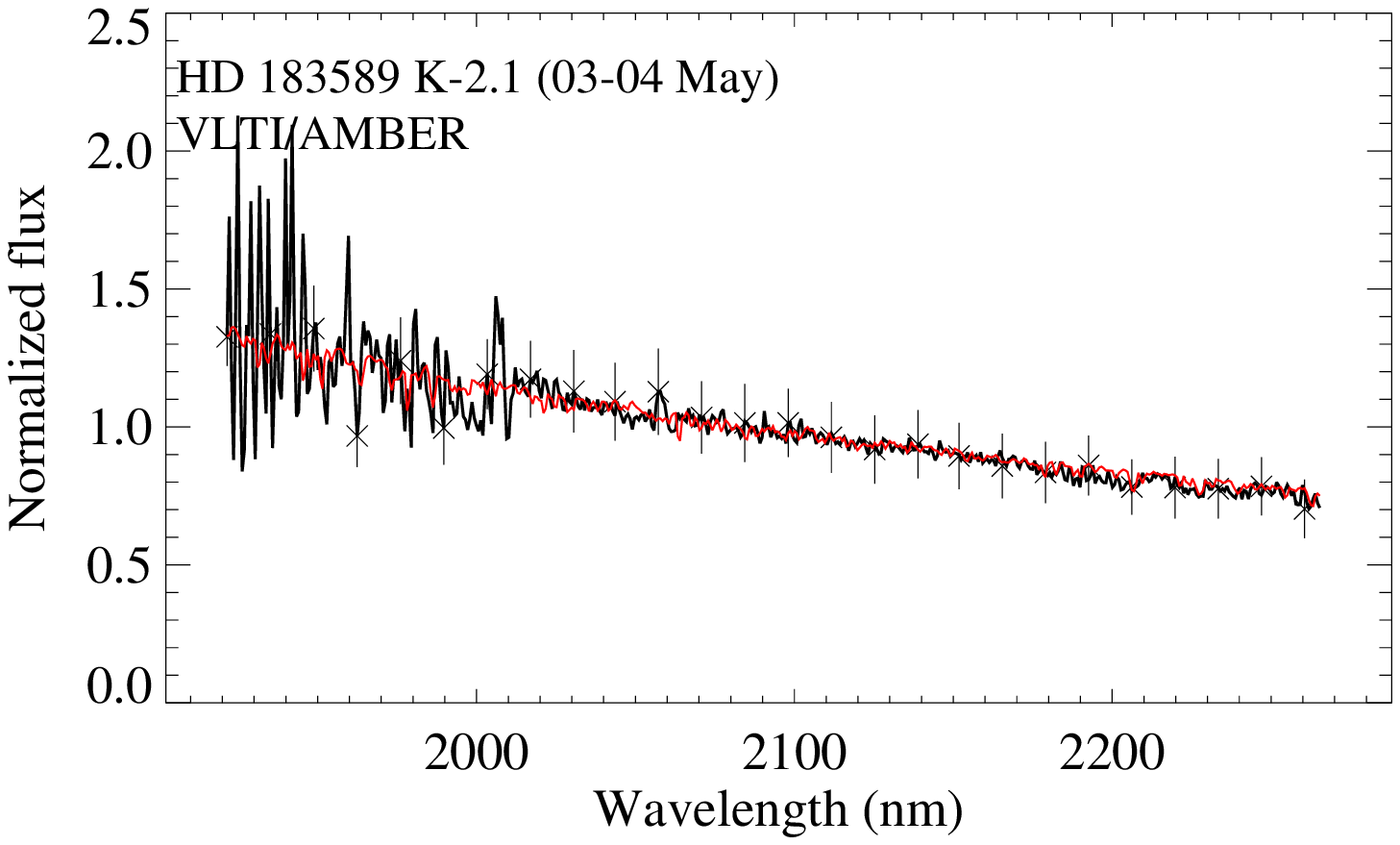}
\includegraphics[width=0.49\hsize]{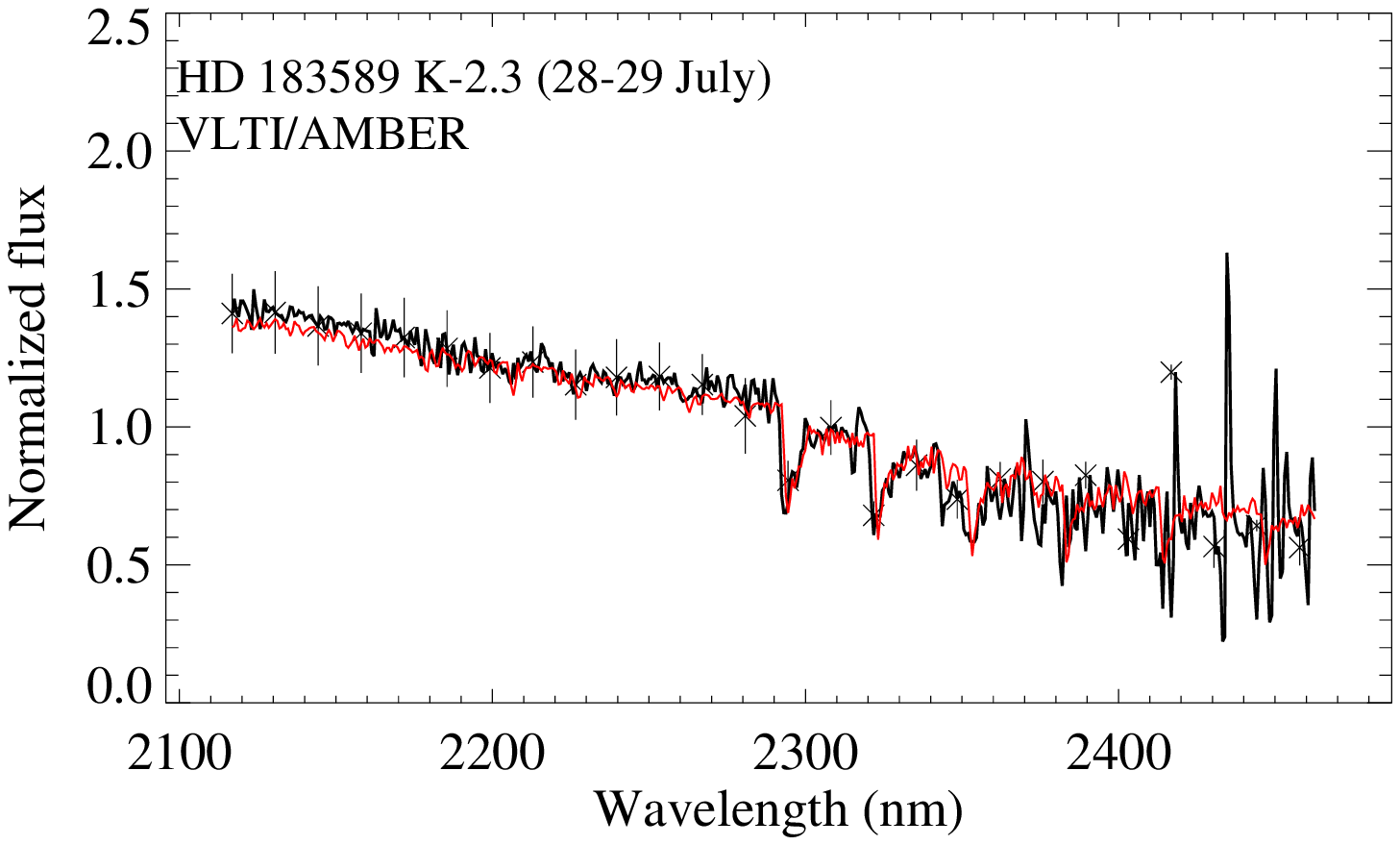}
\includegraphics[width=0.49\hsize]{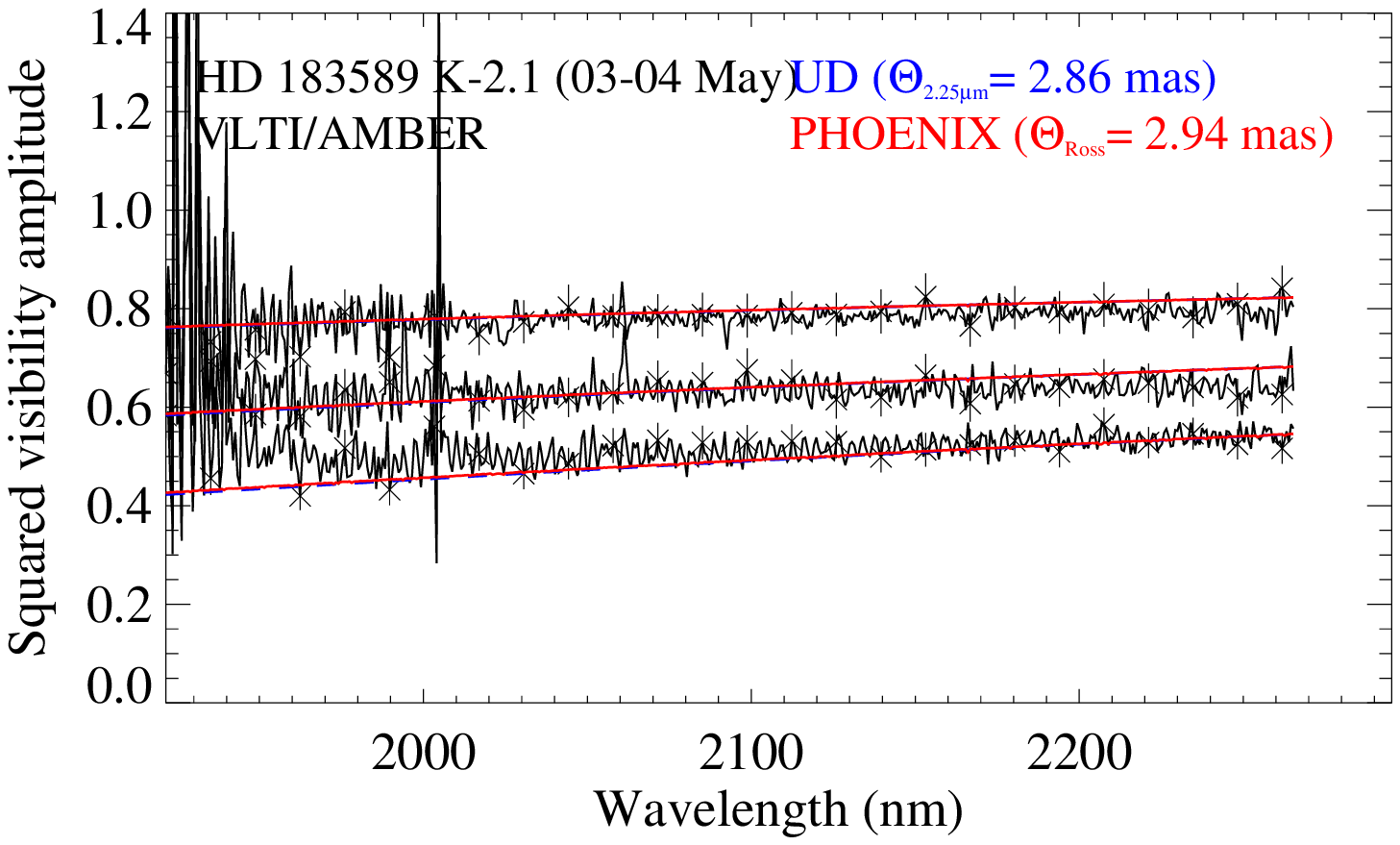}
\includegraphics[width=0.49\hsize]{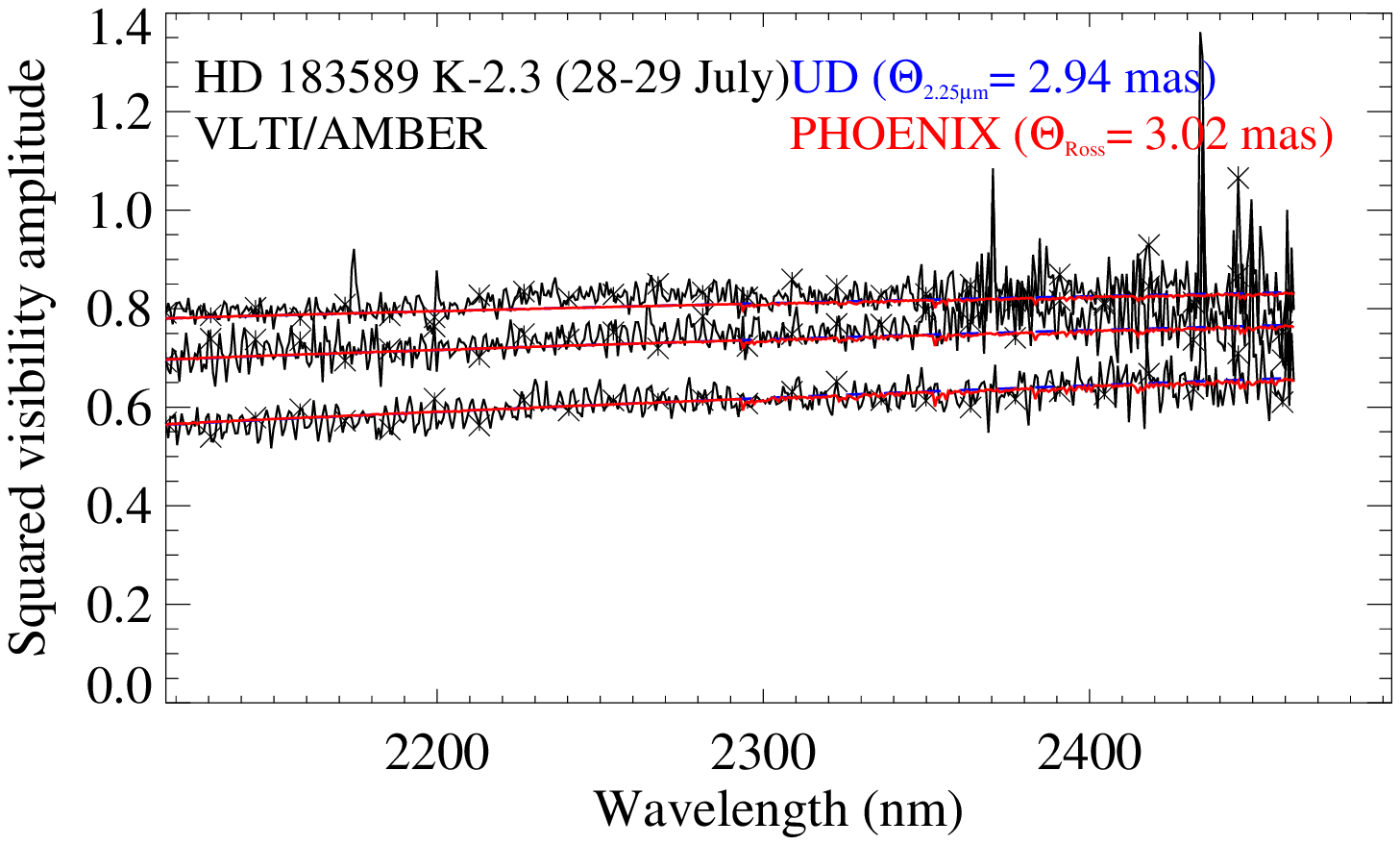}
\includegraphics[width=0.49\hsize]{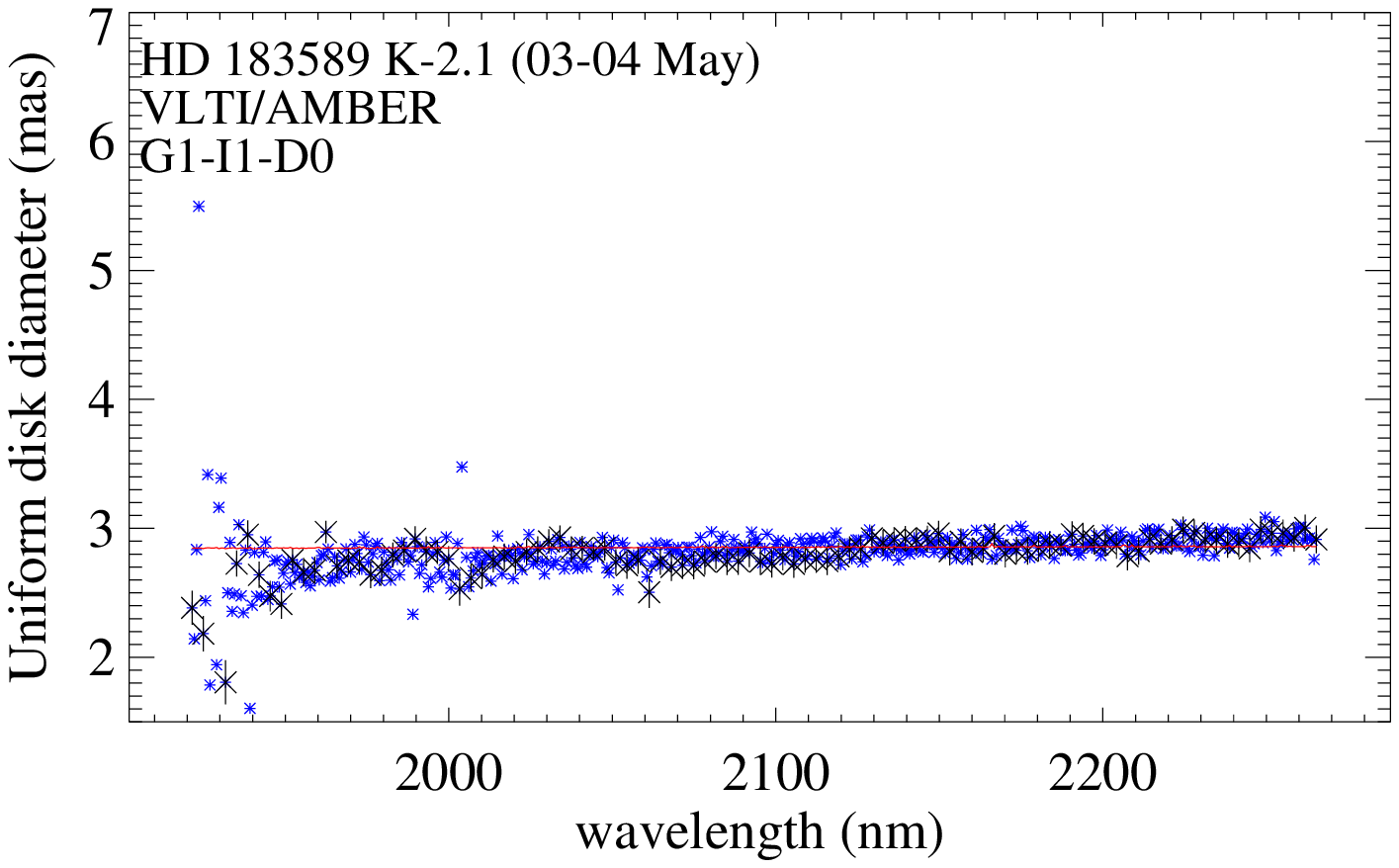}
\includegraphics[width=0.49\hsize]{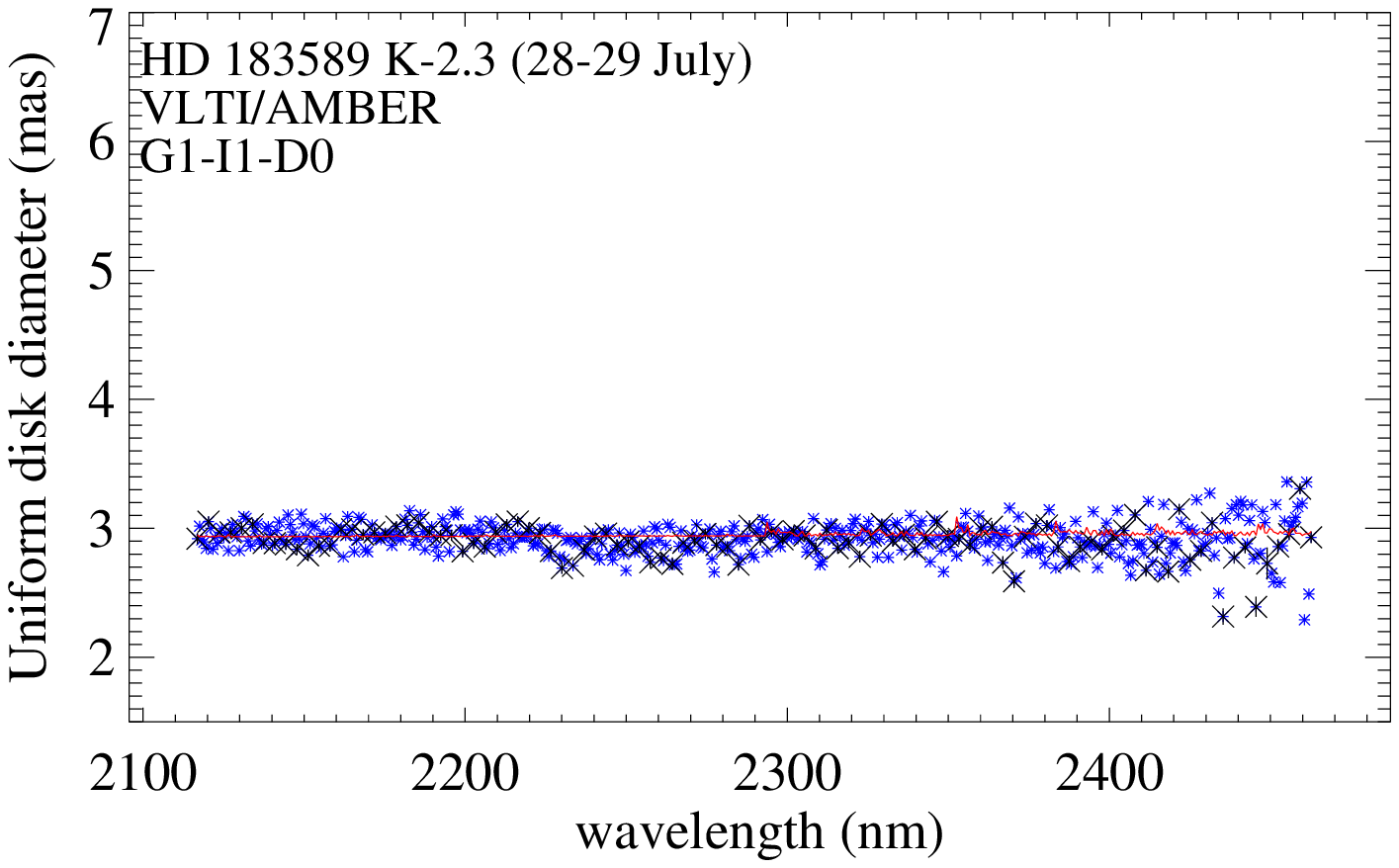}
\includegraphics[width=0.49\hsize]{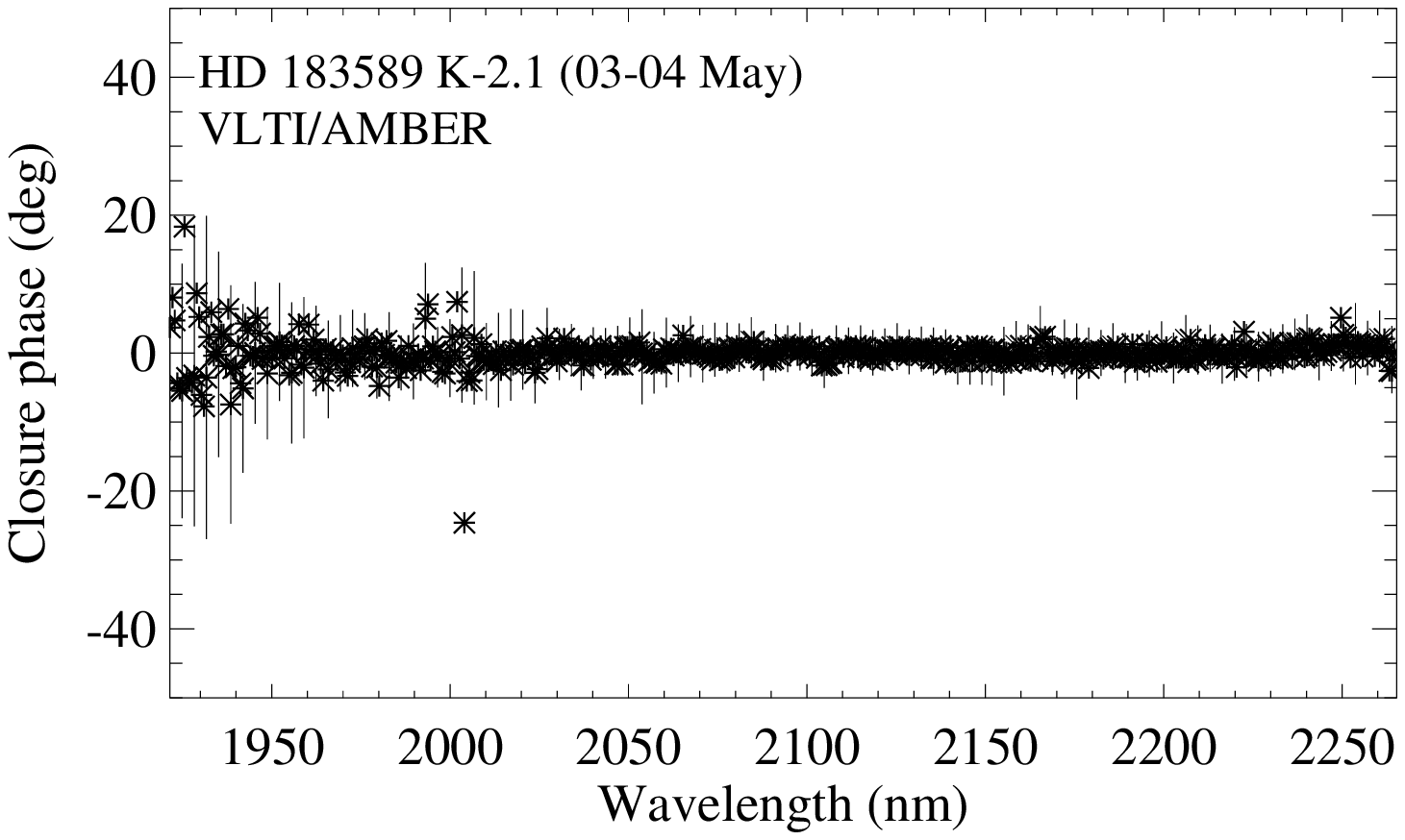}
\includegraphics[width=0.49\hsize]{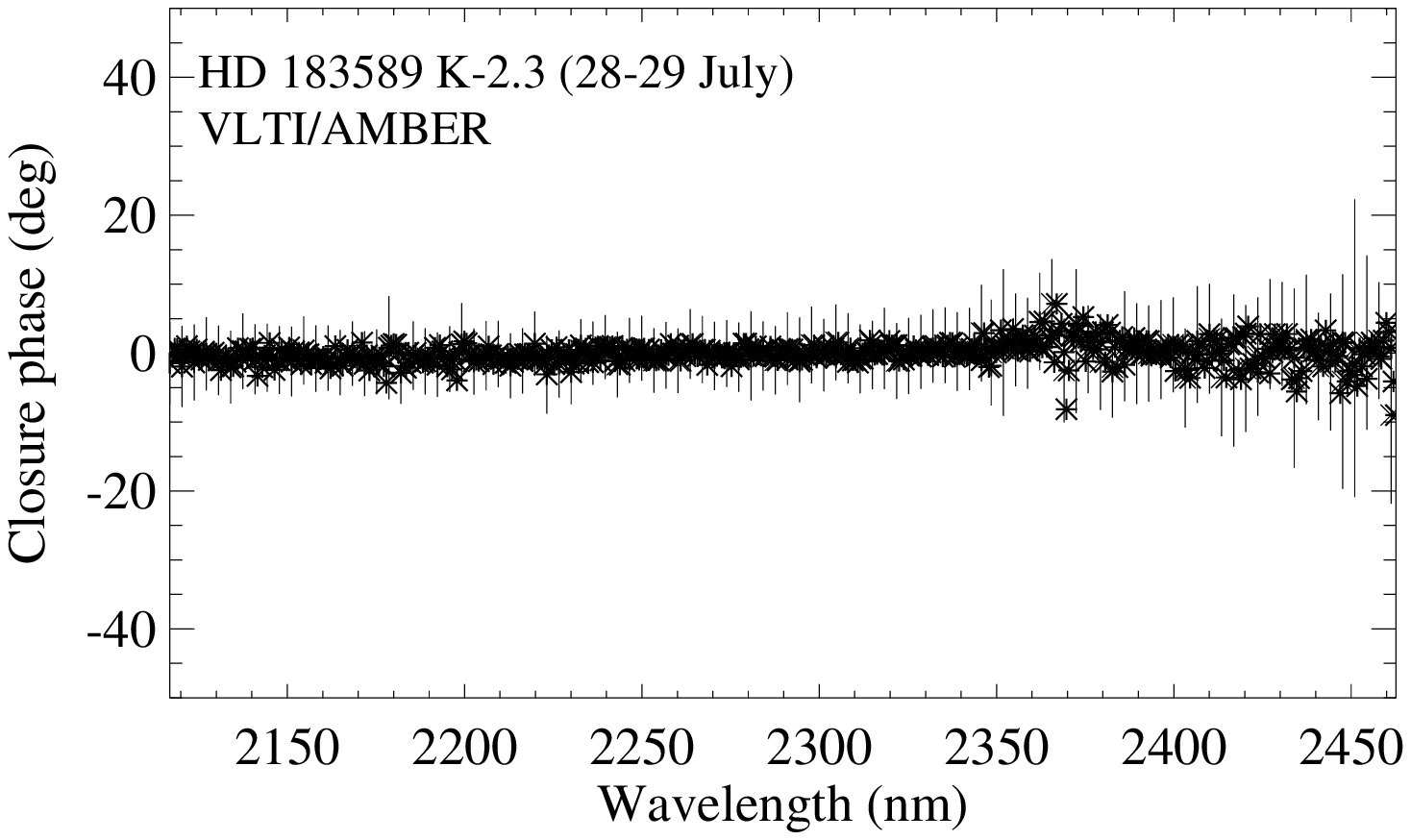}
\caption{As Fig.~\ref{resul_V602Car_2604}, but for data of HD~183589 obtained with the MR-K 2.1\,$\mu$m setting on 4 May 2013 (left) and with the MR-K 2.3\ $\mu$m setting on 29 July 2013 (right).}
\label{resul_HD183589_2907}
\end{figure*} }

\section{Results}
\label{sec:results}

We compared our observational data to synthetic data provided by a grid of {\tt PHOENIX} model atmospheres (version 16.03, Hauschildt \& Baron \cite{Hausch1999} from Arroyo-Torres et al. \cite{Arroyo2013}). These models are based on a hydrostatic atmosphere, local thermodynamic equilibrium and spherical geometry. The best-fit {\tt PHOENIX} model is obtained from an iterative process using the continuum band around 2.25\,$\mu$m, as explained by Arroyo-Torres et al. (\cite{Arroyo2013}). The final values for the used {\tt PHOENIX} model atmosphere are: For V602~Car, T$_\mathrm{eff}$=3400\,K, log(g)=-0.5; for HD~95687, T$_\mathrm{eff}$=3400\,K, log(g)=0.0; and for HD~183589, T$_\mathrm{eff}$=3700\,K, log(g)=1.0. For all cases, we used a model with solar metallicity and a micro-turbulent velocity of 2\,km$/$s. We chose a mass of 20\,$M_{\odot}$ for V602~Car and HD~95687 and of 1\,$M_{\odot}$ for HD~183589. We chose a low mass of 1\,$M_{\odot}$ for the latter target, because the final parameters indicate that it is a source with lower luminosity and thus lower mass compared to the other RSG sources. We note that the structure of the atmosphere is not very sensitive to variations of the mass (Hauschildt et al. \cite{Hausch1999-2}). Certainly, any of those structure variations are below the level of the detectability of our interferometer.

Fig.~\ref{resul_V602Car_2604} shows as an example the resulting normalized flux, squared visibility amplitude, uniform disk diameter, and closure phase data for one of our sources, V602~Car, obtained on 26 April 2013. Also shown are the best-fit uniform disk model (blue curve) and the best-fit {\tt PHOENIX} model atmosphere (red curve). The data and best-fit models for the remaining data are shown in the online appendix  (Figs.~\ref{resul_V602Car_0404}--\ref{resul_HD183589_2907}).

The normalized flux spectra show typical spectra of red supergiants in the K band (cf. Lan\c con et al. \cite{Lancon2007};  Arroyo-Torres et al. \cite{Arroyo2013}). The flux variations at wavelengths below about 2.0\,$\mu$m are due to a higher noise level, possibly caused by the lower atmospheric transmission. In the K-2.3 band, we observe a decreasing flux longwards of 2.25\,$\mu$m and strong absorption lines of CO. The synthetic spectra of the {\tt PHOENIX} model atmosphere are in a good agreement with our flux spectra including the CO bandheads. This indicates that the opacities of CO are well reproduced by the {\tt PHOENIX} model atmosphere.

The continuum visibility values near 2.25\,$\mu$m are consistent with the predictions by the {\tt PHOENIX} model atmospheres for all our sources. In the case of HD~183589, the visibility spectrum is featureless and  consistent with the {\tt PHOENIX} model atmosphere prediction across the whole observed wavelength range. In particular, the visibility spectrum of this source does not show features at the locations of the CO bandheads, which are visible in the flux spectrum, indicating a compact atmospheric structure where the CO layers are located close to the continuum-forming layers. Nonetheless, V602~Car and HD~95687 show large drops of the visibility in the CO bandheads between 2.3\,$\mu$m and 2.5\,$\mu$m that are not reproduced by the {\tt PHOENIX} model atmosphere. The synthetic {\tt PHOENIX} visibility spectra show features in the CO lines, but these are much weaker than the observed features. This effect is also reflected in the panels showing the uniform disk diameter. The size increases of UD fits at the CO bandheads are about 40\% for V602~Car and 20\% for HD~95687, while the {\tt PHOENIX} models predict UD size increases below 5\%. These results indicate that these sources exhibit a large contribution from extended atmospheric layers in the CO bands. The {\tt PHOENIX} model structures are too compact compared to our observations for these two sources. We also observe a monotonic decrease beyond 2.3\,$\mu$m, which may be caused by pseudo-continuum contributions from CO or by contributions from water vapor. We observed the same phenomenon previously for the red supergiants VY CMa (Wittkowski et al. \cite{Witt2012}), AH~Sco, UY~Sct, and KW~Sgr (Arroyo-Torres et al. (\cite{Arroyo2013}), as well as for the small-amplitude pulsating red giants RS~Cap (Marti-Vidal et al. \cite{Marti2011}), BK~Vir (Ohnaka et al. \cite{Ohnaka2012}), $\alpha$~Tau (Ohnaka \cite{Ohnaka2013_2}), and $\beta$~Peg (Arroyo-Torres et al. \cite{Arroyo2014}).

\begin{figure}
\centering
\includegraphics[width=0.95\hsize]{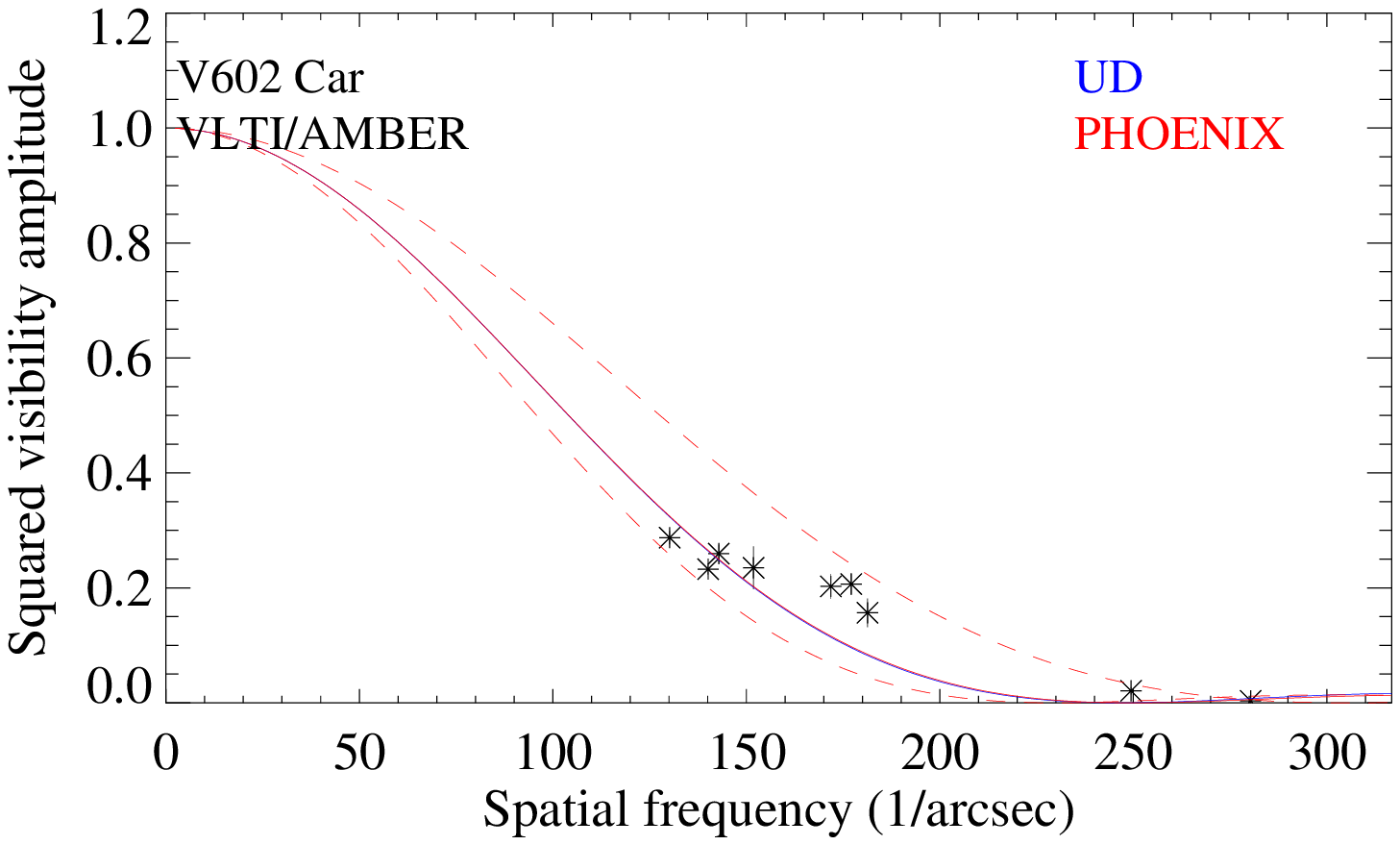}
\includegraphics[width=0.95\hsize]{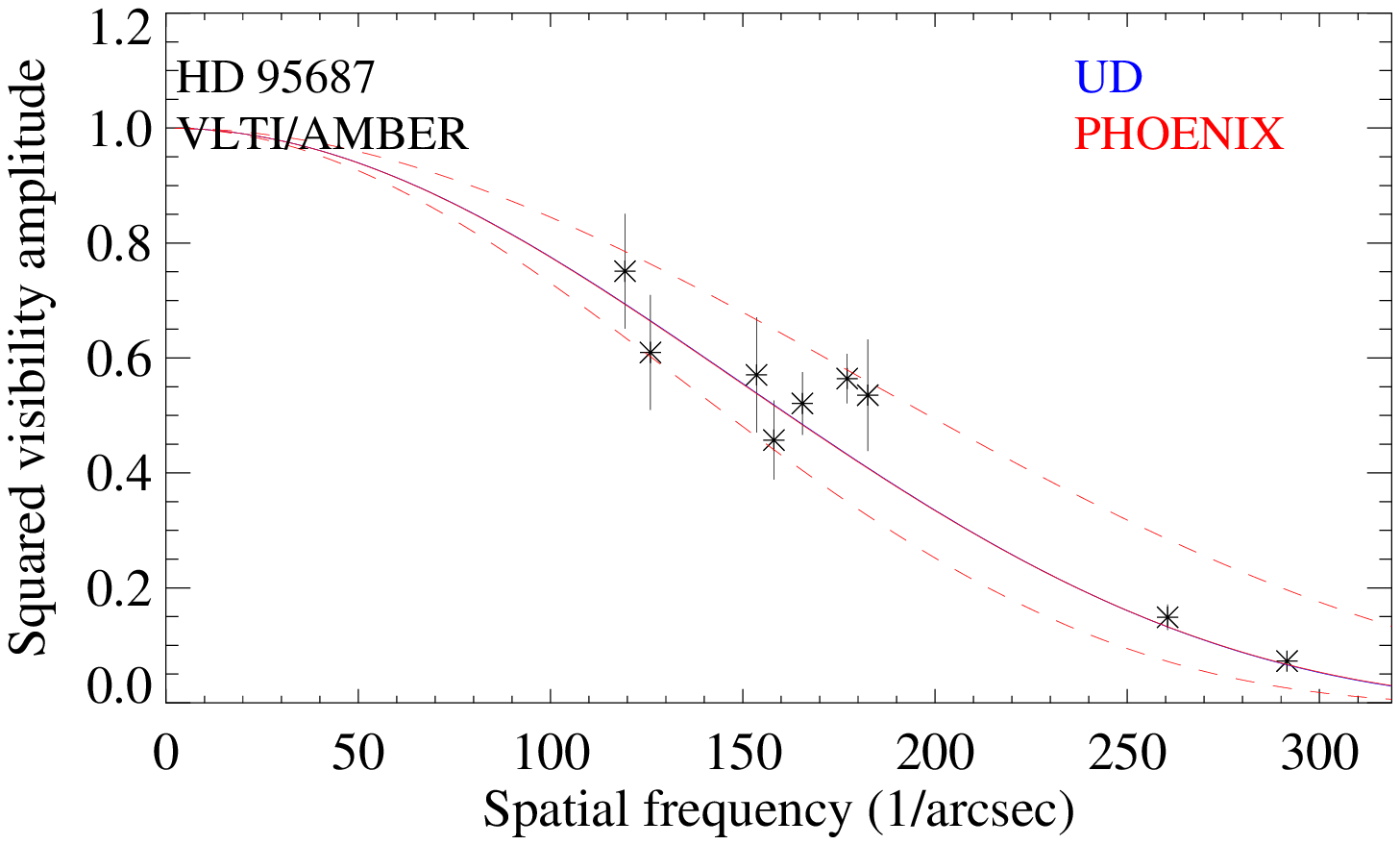}
\includegraphics[width=0.95\hsize]{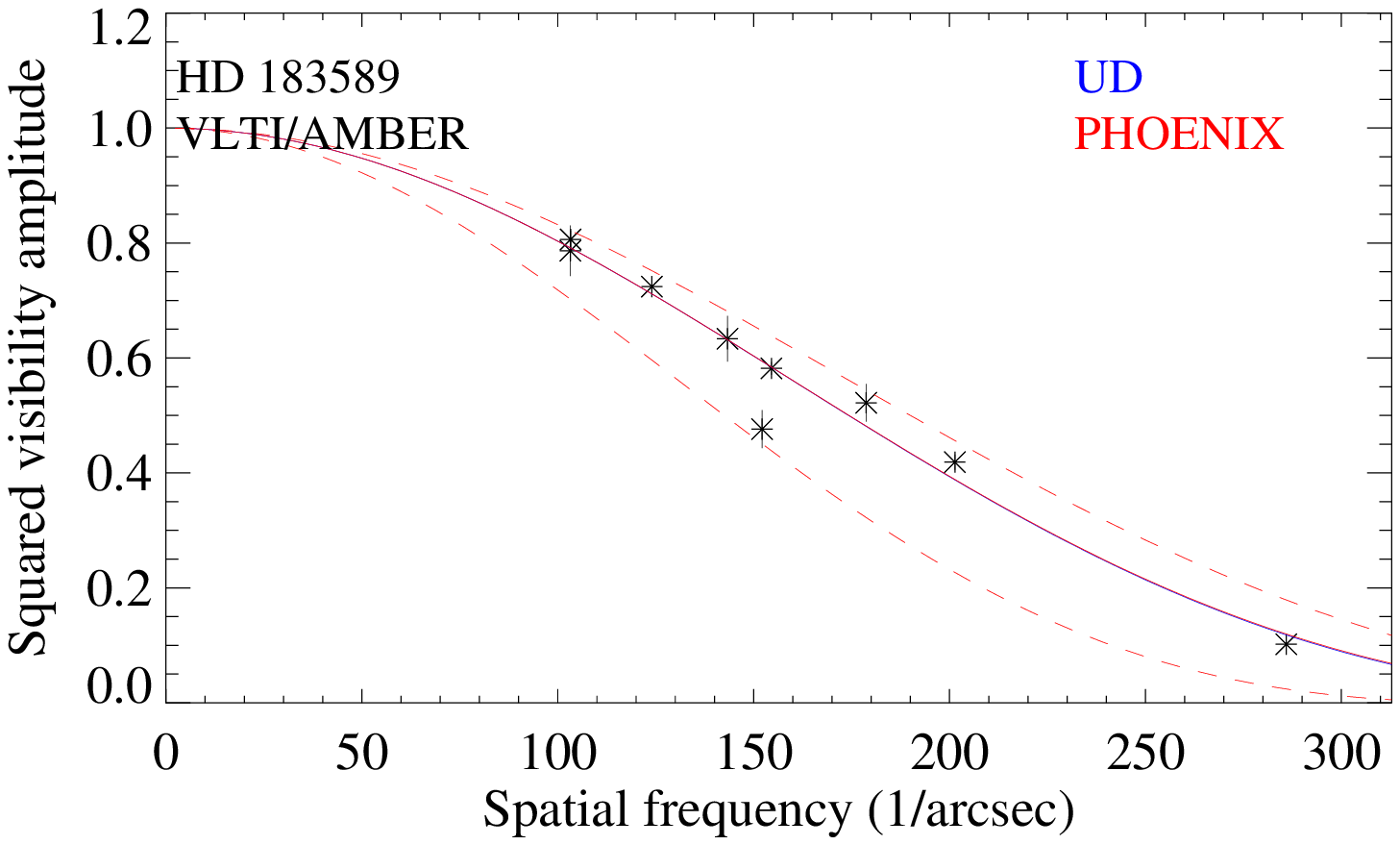}
\caption{Squared visibility amplitudes in the continuum bandpass for V602~Car, HD~95687, and HD~183589 (from top to bottom) as a function of spatial frequency. Each point represents an average of data points within  the continuum bandpass at 2.15--2.25\,$\mu$m.  Shown are data of all observing dates and both spectral setups. The red lines indicate the best-fit UD models and the blue lines (often indistinguishable from the red) the best-fit {\tt PHOENIX} models. The dashed lines indicate the maximum and minimum visibility curves, from which we estimated the angular diameter errors.}
\label{Vis_spacialFrec}
\end{figure}

The closure phase data of our sources in the 4th panels of Fig.~\ref{resul_V602Car_2604}  and Figs.~\ref{resul_V602Car_0404}--\ref{resul_HD183589_2907} show variations within the noise level, and are thus are not indicative of deviations from point symmetry. However, since our measurements lie in the first visibility lobe and the noise level is relatively high, we cannot exclude asymmetries on scales smaller than the observed stellar disk. We note that there are points in the observed closure phases whose deviation from zero is larger than the error bars. In general, small deviations from zero closure phases might indicate asymmetries in layers corresponding to certain atomic or molecular bands as previously observed for RSGs by e.g., Ohnaka et al. (\cite{Ohnaka2011}), Wittkowski et al. (\cite{Witt2012}), Ohnaka et al. (\cite{Ohnaka2013}). However, it is not clear whether in our case this deviation are real or whether they correspond to systematic uncertainties of the data reduction, as for instance due to the bad pixel mask.

\begin{table}
\caption{Calibration sources}
\centering
\begin{tabular}{lcccc}
\hline
\hline
 & Spectral type & Angular diameter (mas) \\
\hline
HR 4164 & K1 III & 1.64$\pm$0.12 \\
z Car & M6 & 1.54$\pm$0.11 \\
38 Aql & K3 III & 2.22$\pm$0.02 \\
HR 7404 & K2 & 1.17$\pm$0.08 \\
\hline
\end{tabular}
\label{calibrator}
\end{table}

\subsection{Estimate of the angular diameter}

The continuum band near 2.25\,$\mu$m appears to be largely free of contaminations by molecular layers. Thus, fits of {\tt PHOENIX} models to the continuum band allow us to estimate reliable angular diameters of our sources. The angular diameter, obtained in this way, corresponds to the size of the outermost model layer (0\% intensity radius). To estimate the Rosseland angular diameter (corresponding to the layer where the Rosseland optical depth equals 2/3), we multiplied our value of the angular diameter by the ratio between the Rosseland layer and the outermost model layer. This ratio was 0.92 for V602~Car, 0.95 for HD~95687, and 0.93 for HD~183589. The model fits used all available data taken during all nights and with any of the two spectral setups, as both setups include the continuum band near 2.25\,$\mu$m. Tab.~\ref{angular_diam} lists the resulting best-fit Rosseland angular diameters as well as the best-fit UD diameters. Fig.~\ref{Vis_spacialFrec} shows the continuum visibility data as a function of spatial frequency together with the best-fit {\tt PHOENIX} and UD models. The errors of the continuum visibilities data were computed as an average of the individual errors, whereas, the errors of the angular diameter are estimated from the differences between the  visibility curves lying at the maximum and minimum of our data as shown by the dashed lines in Fig.~\ref{Vis_spacialFrec}. Deviating visibility points are caused by remaining systematic uncertainties of the absolute visibility calibration. The data of V602~Car include two points near the first visibility null, which increases the precision of the best-fit angular diameter.

\begin{table}
\caption{Summary of estimated angular diameters}
\centering
\begin{tabular}{lccccc}
\hline
\hline
  & V602 Car & HD 95687 & HD 183589 \\
\hline
$\theta _\mathrm{UD}$ (mas)   & 4.94$\pm$0.75 & 3.17$\pm$0.50  & 2.95$\pm$0.50 \\ 
$\theta _\mathrm{Ross}$ (mas) & 5.08$\pm$0.75 & 3.26$\pm$0.50  & 3.04$\pm$0.50 \\
\hline
\end{tabular}
\label{angular_diam}
\end{table}
 
\subsection{Fundamental parameters}

\begin{figure}
\centering
\includegraphics[width=0.99\hsize]{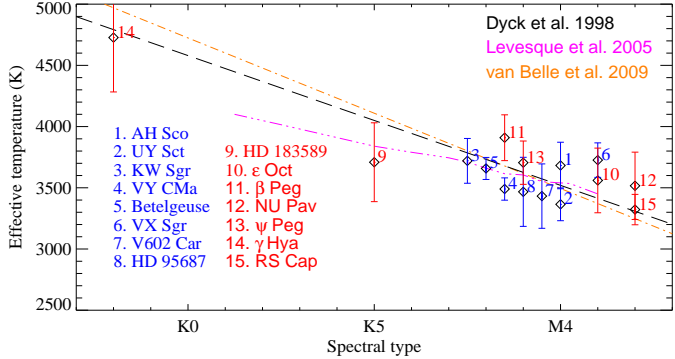}
\caption{Effective temperature versus spectral type of our sources, together with calibrations of the effective temperature scale by Dyck et al. (\cite{Dyck1998}), Levesque et al. (\cite{Levesque2005}), and van Belle et al. (\cite{Belle2009}). Also included are previous measurements of RSGs and red giants as listed in the main text. In blue are the RSGs and in red the red giants.}
\label{Teff_sp_R}
\end{figure}

\begin{figure}
\centering
\includegraphics[width=0.99\hsize]{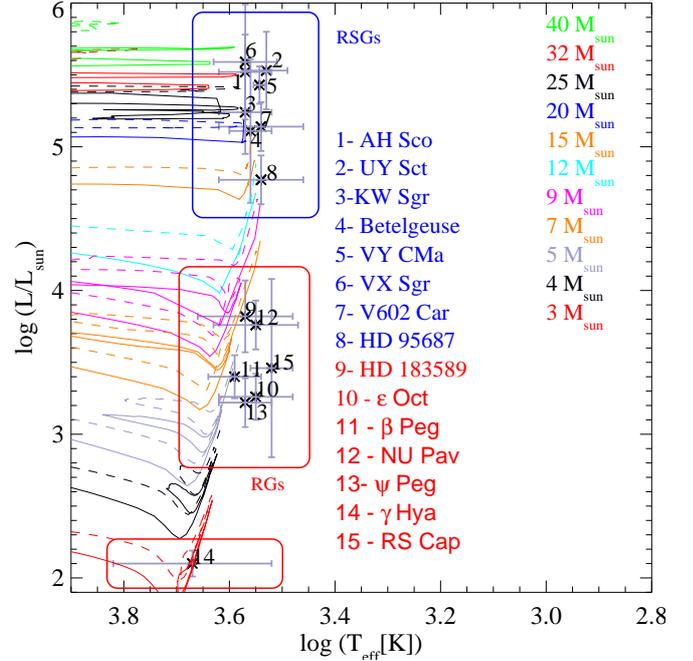}
\caption{Location of our sources in the HR diagram, compared to evolutionary tracks from Ekstr\"om et al. (\cite{Ekstrom2012}) for masses of 3\,$M_{\odot}$, 4\,$M_{\odot}$, 5\,$M_{\odot}$, 7\,$M_{\odot}$, 9\,$M_{\odot}$, 12\,$M_{\odot}$, 15\,$M_{\odot}$, 20\,$M_{\odot}$, 25\,$M_{\odot}$, 32\,$M_{\odot}$, and 40\,$M_{\odot}$. The solid lines indicate models without rotation, and the dashed lines with rotation. Also shown are previously measured sources as listed in the main text. In blue are the RSG stars, and in red the red giants.}
\label{diagram_HR}
\end{figure} 

We estimated the fundamental stellar parameters of our sources to place them on the HR diagram. In particular we calculated the effective temperature, the luminosity, the Rosseland-mean radius, the bolometric flux, and the distance in the same way as described in detail by Arroyo-Torres et al. (\cite{Arroyo2014}). We used the BVJHKs magnitudes from Kharchenko (\cite{Kharchenko2001}) and Cutri et al. (\cite{Cutri2003}) and the IRAS flux from IRAS (\cite{iras}). To convert the magnitudes into fluxes, we used the zero values from Johnson (\cite{Johnson1965}) and Cohen et al. (\cite{Cohen2003}). To deredden the flux values we used the color excess method applied to (V-K) and based on intrinsic colors from Ducati et al. (\cite{Ducati2001}), as described in Arroyo-Torres et al. (\cite{Arroyo2014}). 

V602~Car and HD~95687 belong to the cluster CAR~OB2 and we use the distance as determined by Humphreys et al. (\cite{Humphreys1978}). For HD~183589, we used the distance value from van Leeuwen (\cite{Leeuwen2007}). Lastly, the effective temperature is estimated from the angular diameter and the bolometric flux, the luminosity from the bolometric flux and the distance, and the Rosseland radius from the Rosseland angular diameter and the distance. We assumed a 15\% error in the flux, and a 10\% error in the distance for HD~183589. For V602~Car and HD~95687, we used the errors from Humphreys et al. (\cite{Humphreys1978}). The errors in the luminosity, effective temperature and radius were estimated by error propagation. The resulting fundamental parameters and their errors are listed in Table~\ref{fund_parameters}. 

\begin{table*}
\caption{Fundamental parameters of V602~Car, HD~95687, and HD~183589.}
\begin{center}
\begin{tabular}{ccccccc}
\hline
\hline
Parameter & V602 Car & HD 95687  & HD 183589 & Ref.   \\
\hline
$F_{bol}$ ($10^{-10}$ W$m^{-2}$) & 11.30$\pm$1.69 &  4.84$\pm$0.73  & 5.49$\pm$0.82 & 1  \\ 
d (pc) & 1977$\pm$75 &  1977$\pm$75  & 621$\pm$62  & 2 \\
L ($10^{31}$ W) & 5.28$\pm$0.89 &  2.26$\pm$0.38  & 0.25$\pm$0.06  & 1,2 \\
log(L/$L_{\odot}$) & 5.14$\pm$0.17 & 4.77$\pm$0.17  & 3.82$\pm$0.25 & - \\
$\theta _{Ross}$ (mas) & 5.08$\pm$0.75 & 3.17$\pm$0.50  & 2.95$\pm$0.50 & This work \\
R($R_\odot$) & 1050$\pm$165 & 674$\pm$109  & 197$\pm$39 & 2,4 \\
$T_\mathrm{eff}$ (K) & 3432$\pm$263 & 3467$\pm$282  & 3709$\pm$322 & 3,5 \\
log($T_\mathrm{eff}$) & 3.54$\pm$0.08 & 3.54$\pm$0.08  & 3.57$\pm$0.09 & -  \\
log($g$) & -0.30$\pm$0.16 & -0.14$\pm$0.14  & 0.80$\pm$0.17 & this work\\
M($M_\odot$) & 20-25 & 12-15 & 7-12 & 6 \\
\hline
\end{tabular}
\tablefoot{1: Kharchenko (\cite{Kharchenko2001}), Cutri et al. (\cite{Cutri2003}), IRAS (\cite{iras}); 2: Humphreys et al. (\cite{Humphreys1978}) - HD~95687, V602 Car (cluster CAR~OB2); van Leeuwen (\cite{Leeuwen2007}) - HD~183589; 6: Values obtain by the position of the stars in the HR diagram with the evolutionary tracks from Ekstr\"om et al. (\cite{Ekstrom2012}); We assumed a 15\% error in the flux. The distance error was based on the values from Humphreys et al. (\cite{Humphreys1978}) by V602~Car and HD~95687. From HD~183589, we assumed a 10\% error in the distance. The errors in the luminosity, effective temperature and radius were estimated by error propagation.}
\end{center}
\label{fund_parameters}
\end{table*}
\begin{figure}
\centering
\includegraphics[width=0.99\hsize]{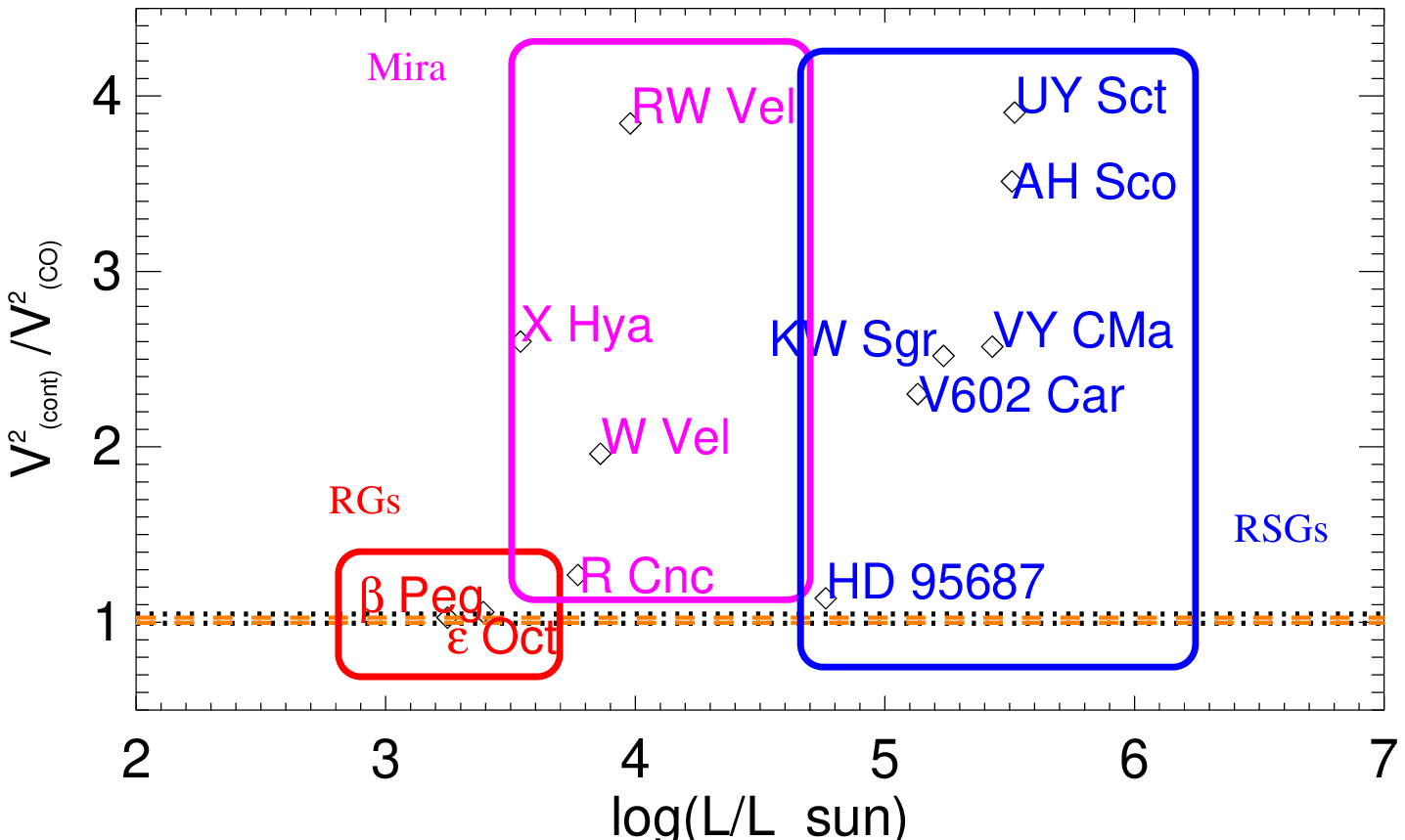}
\includegraphics[width=0.99\hsize]{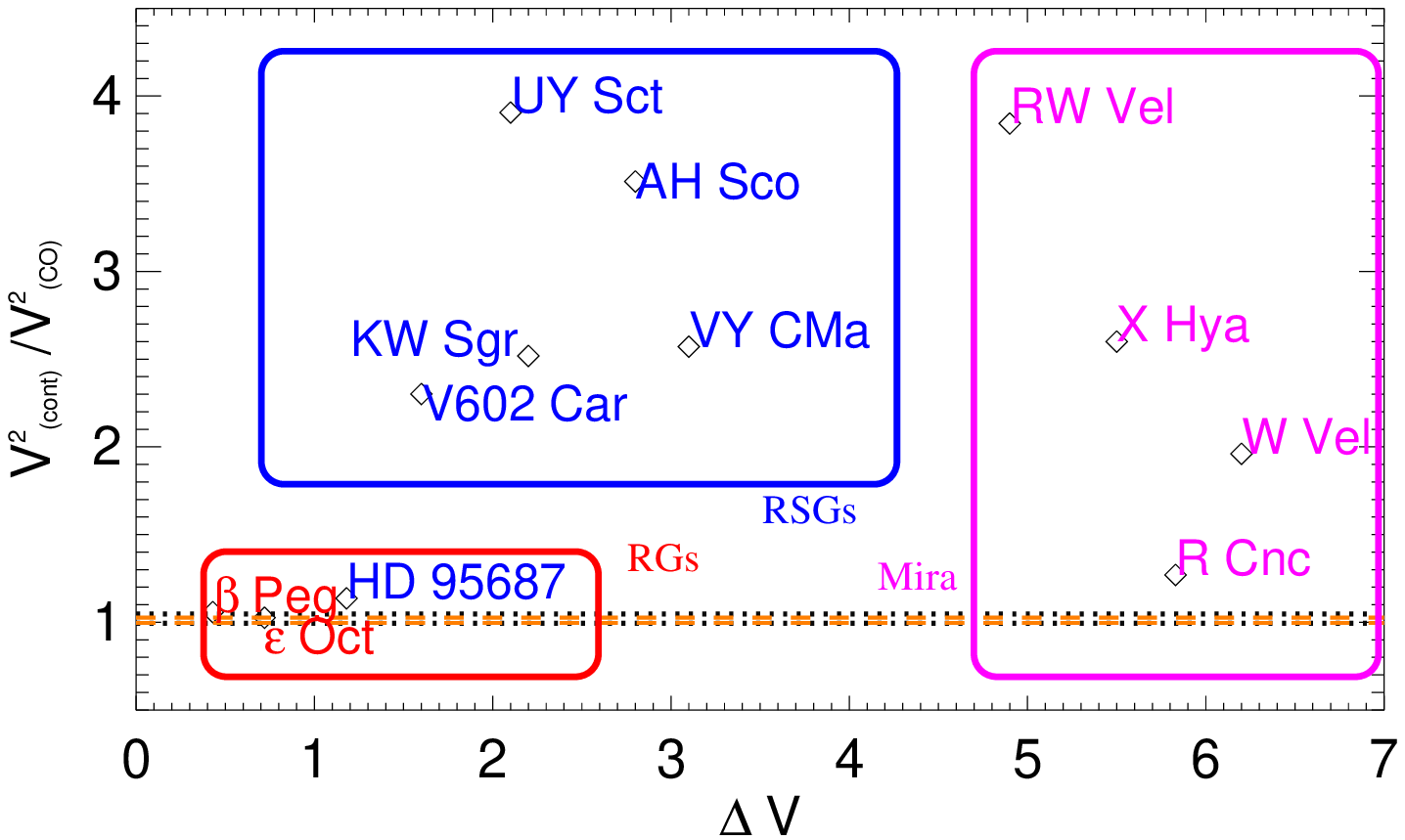}
\caption{Ratio between the square visibility in the continuum (average between 2.27 and 2.28\,$\mu$m) and the square visibility in the CO (2-0) line vs. log(L/L$_{\odot}$) (top) and the variability amplitude (bottom) for a sample of RSGs in blue (Arroyo-Torres et al. \cite{Arroyo2013}; Wittkowski et al. \cite{Witt2012}; this work), AGB stars in red (Arroyo-Torres et al. \cite{Arroyo2014}) and Mira variable stars in magenta (Wittkowski et al. \cite{Witt2011}) for visibilities in the continuum between 0.2 and 0.4. The dotted black lines show the range of predictions by the best-fit {\tt PHOENIX} model of these sources, the dashed orange the range of predictions by the best-fit RHD simulation. Both lines overlap.}
\label{fig:visratio}
\end{figure}

\begin{figure}
\centering
\includegraphics[width=0.99\hsize]{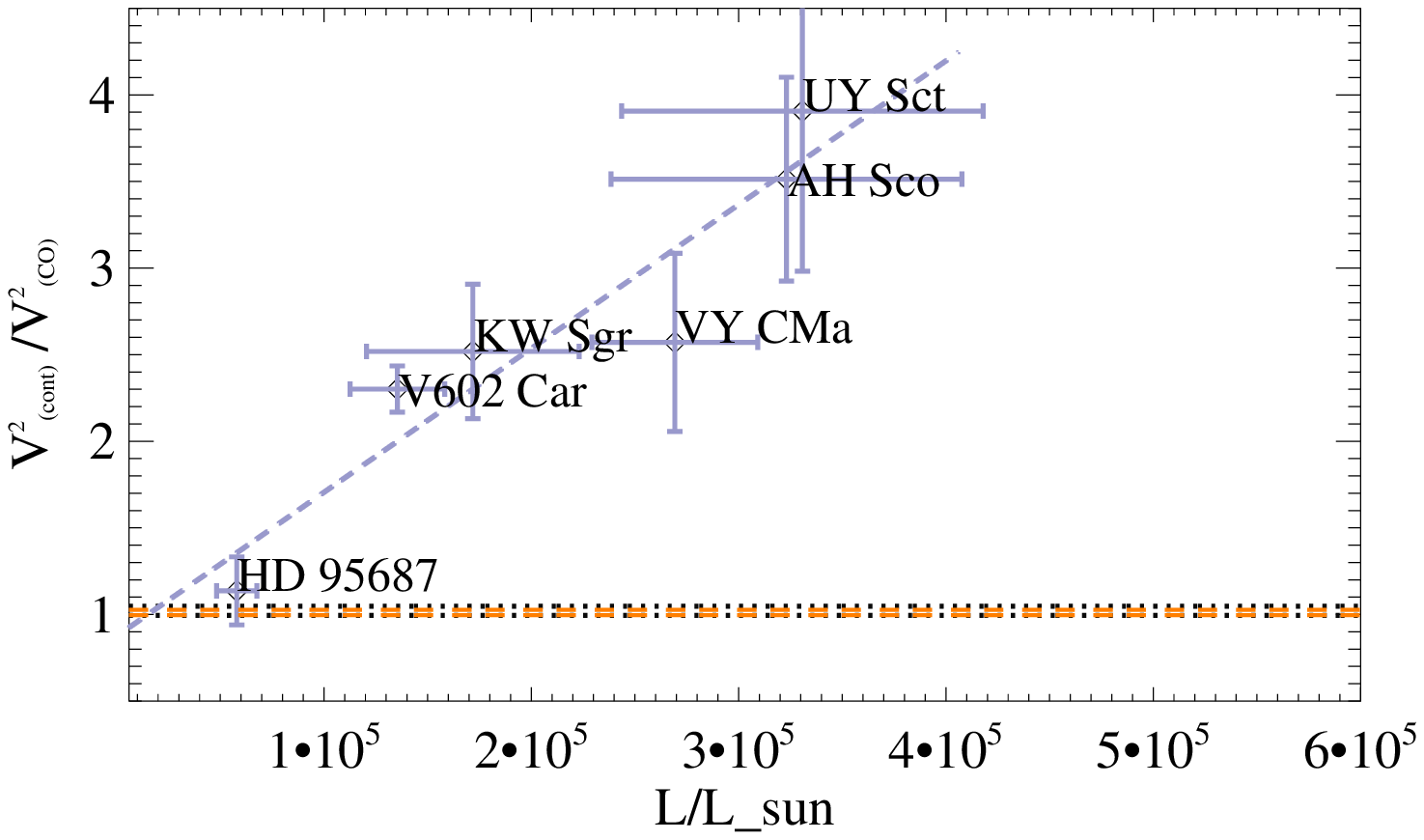}
\includegraphics[width=0.99\hsize]{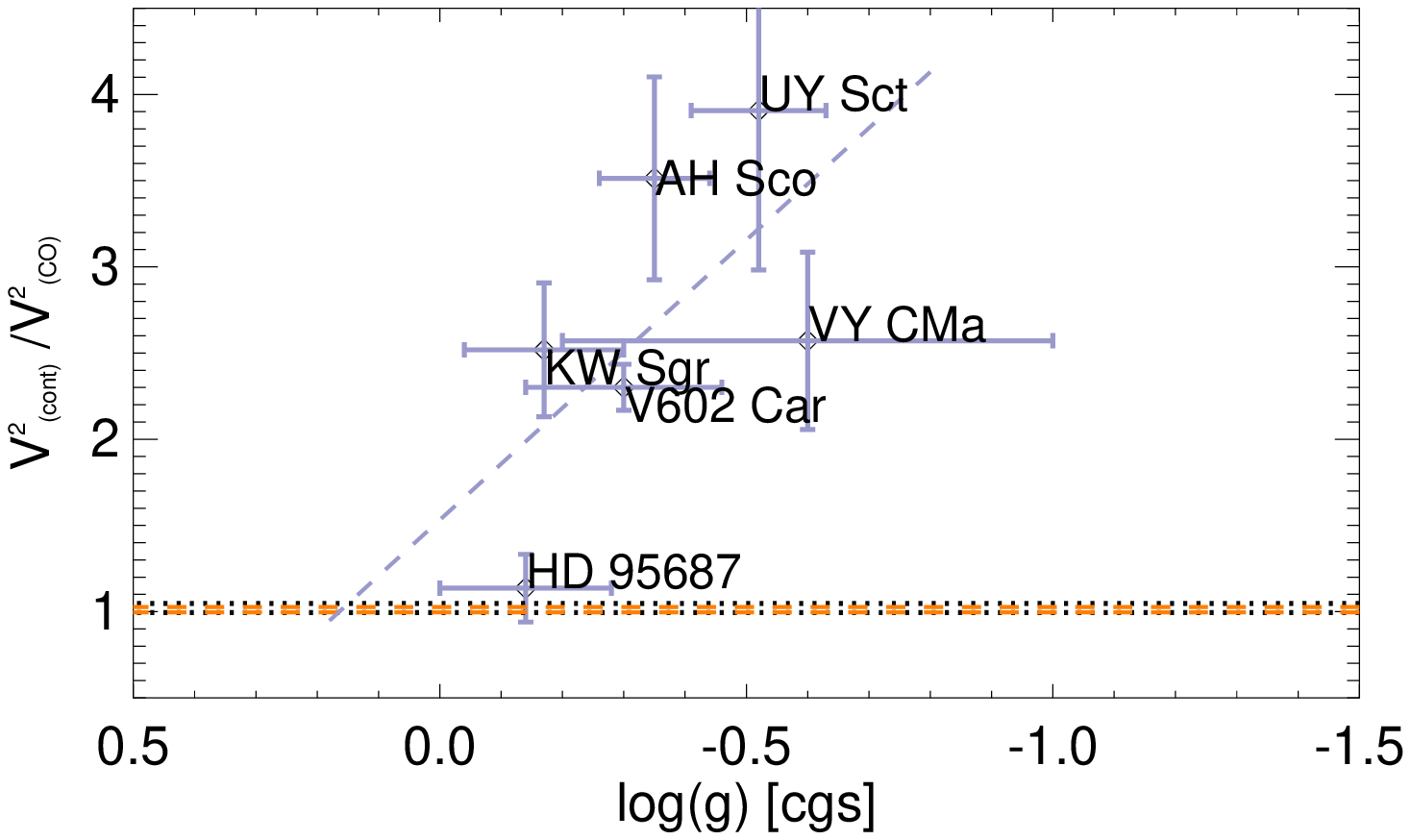}
\includegraphics[width=0.99\hsize]{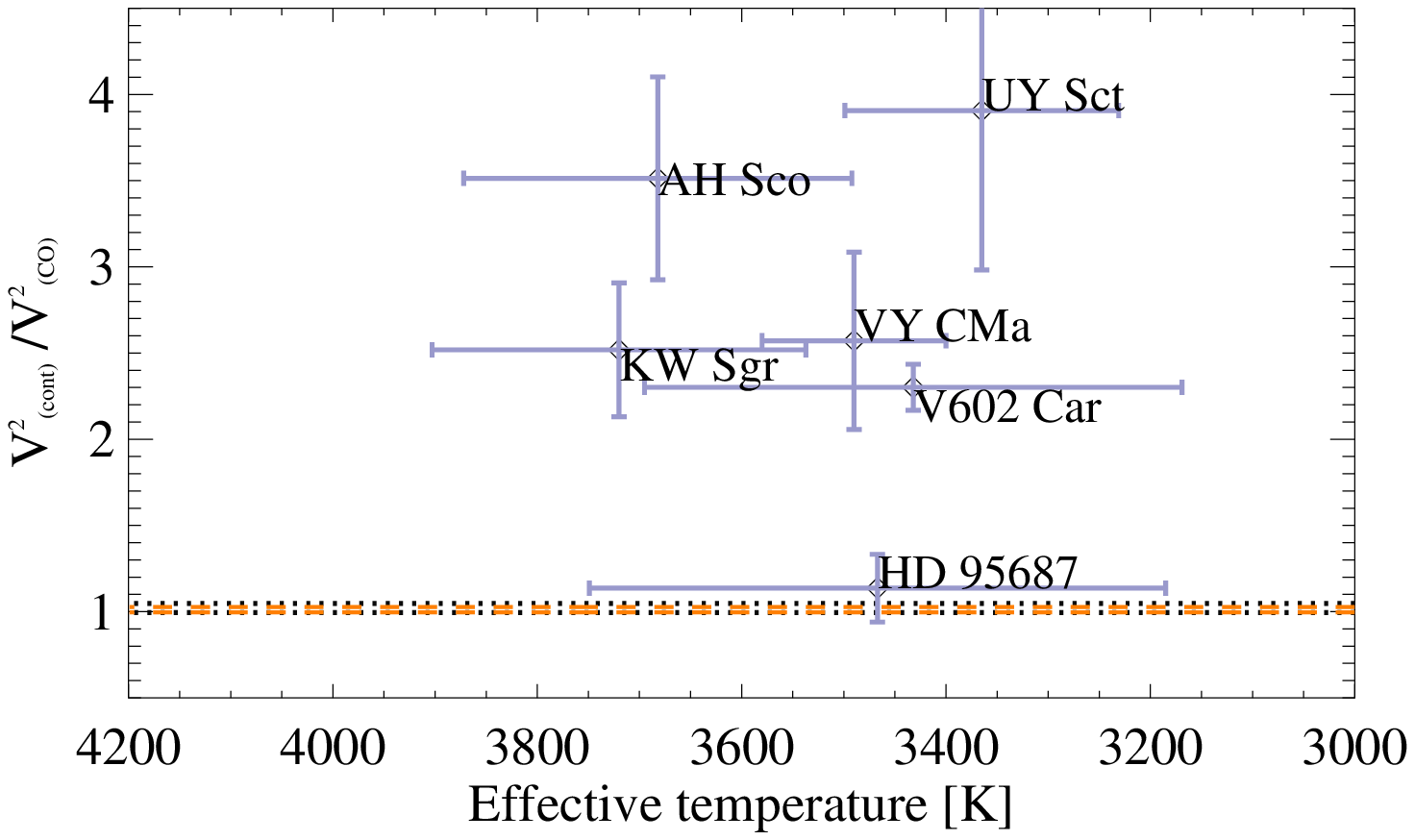}
\caption{Ratio between the square visibility in the continuum (average between 2.27 and 2.28\,$\mu$m) and the square visibility in the CO (2-0) line vs. log(L/L$_{\odot}$) (top), log(g) (middle), and the effective temperature (bottom) for the sample of RSGs (Arroyo-Torres et al. \cite{Arroyo2013}; Wittkowski et al. \cite{Witt2012}; this work). The dotted black lines show the range of predictions by the best-fit {\tt PHOENIX} model of these sources, and the dashed orange the range of predictions by the best-fit RHD simulation. Both lines are overlap. The dashed gray lines, in the top and middle, show linear fits.}
\label{fig:visratio_RSGs}
\end{figure}

The radius and luminosity of HD~183589 suggest that this is a source of lower luminosity and thus lower mass compared to the other observed RSG sources. It is at the limit between RSG and super-AGB stars. The visibility functions resemble those of the red giants as observed by Arroyo-Torres et al. (\cite{Arroyo2014}), which do not show indications of an extended molecular layer.

In Fig.~\ref{Teff_sp_R}, we plot the resulting effective temperatures vs. spectral types of our targets. For comparison, we include the calibrations of the effective temperature scale by Dyck et al. (\cite{Dyck1998}) for cool giants stars, by van Belle et al. (\cite{Belle2009}) for cool giants stars and RSG stars, and by Levesque et al. (\cite{Levesque2005}) for only RSGs. Fig.~\ref{diagram_HR} shows the position of our targets in the HR diagram, together with evolutionary tracks from Ekstr\"om et al. (\cite{Ekstrom2012}). Both figures also include the RSGs from our previous studies (VY~CMa from Wittkowski et al. \cite{Witt2012}; AH~Sco, UY~Sct, KW~Sgr from Arroyo-Torres et al. \cite{Arroyo2013}), as well as Betelgeuse based on the data by Ohnaka et al. (\cite{Ohnaka2011}) and VX~Sgr based on the distance by Chen et al. (\cite{Chen2007}) and the angular radius by Chiavassa et al. (\cite{Chiavassa2010}), analyzed by us in Arroyo-Torres et al. (\cite{Arroyo2013}). Finally, we included the red giant stars from  Mart{\'{\i}}-Vidal et al. (\cite{Marti2011}) and Arroyo-Torres et al. (\cite{Arroyo2014}). All sources are consistent within their errors with the different calibrations of the effective temperature scale and with the red limits of the evolutionary tracks. The positions on the HR diagram of our new sources are close to  evolutionary tracks corresponding to an initial mass of 20-25\,M$_{\odot}$ without rotation or 15-20\,M$_{\odot}$ with rotation (V602~Car), 12-15\,M$_{\odot}$ without rotation or 9-15\,M$_{\odot}$ with rotation (HD~95687), 5-12\,M$_{\odot}$ without rotation or 7-9\,M$_{\odot}$ with rotation (HD 183589). HD 183589 may thus also be a (super-)AGB star and not a RSG star (Siess \cite{Siess2010}). 

We note that the red giants with luminosities below $\log(L/L_\odot)\sim 4$ are located systematically to the right of the Ekstr\"om tracks, and that a better agreement for these sources can be found using the STAREVOL grid (Lagarde et al. \cite{Lagarde12}), which includes thermaline mixing unlike Ekstr\"om's grid (Arroyo-Torres et al. \cite{Arroyo2014}).

\section{Characterization of the extension of the molecular atmosphere}
\label{sec:charac}

We characterized the observed extensions of the molecular layers in order to better understand how the fundamental parameters affect them. We also want to compare the behavior of RSG stars and Mira stars, which also exhibit extended molecular layers (cf., e.g., Wittkowski et al. \cite{Witt2011}). We used the ratio of the observed visibilities in the continuum band (average between 2.27 and 2.28\,$\mu$m) and the first (2-0) CO line at 2.294\,$\mu$m as an observational indication of the contribution from extended atmospheric CO layers. Since this ratio depends on the value of the visibility in the continuum (V$^{2}_\mathrm{cont}$), i.e. on how well the source was resolved, we limited the study to continuum squared visibilities between 0.2 and 0.4, a range where the visibility function is nearly linear. Although this approach may be limited by the limited spectral resolution of our observations of $R\sim$\,1500 and by the low number of observations per source, it is appropriate for a first comparison of the extensions of RSG stars and other evolved stars.

Fig.~\ref{fig:visratio} shows the resulting ratios (V$^{2}_\mathrm{cont}$/V$^{2}_\mathrm{CO}$) for our sources vs. log(L/L$_{\odot}$) and $\Delta$V, considering the RSGs by Wittkowski et al. (\cite{Witt2012}), Arroyo-Torres et al. (\cite{Arroyo2013}) and this work (not represent HD~183589 because their visibilities in the continuum are greater than 0.4), as well as the giants from Arroyo-Torres et al. (\cite{Arroyo2014}). Results for Mira stars obtained by Wittkowski et al. (\cite{Witt2011}) are also shown for comparison. Fig.~\ref{fig:visratio_RSGs} shows the same visibility ratio but only for the RSG sample. In this case, we show the visibility ratio vs. L/L$_{\odot}$, log(g), and $T_\mathrm{eff}$. Also shown are the ranges of the predicted visibility ratios based on the {\tt PHOENIX} model atmospheres, as well as based on  3-D RHD simulations, which will be discussed in the next section.

Fig.~\ref{fig:visratio} shows that the giant stars (red) are consistent with the {\tt  PHOENIX} models and the convection models (RHD simulations). $\beta$~Peg shows an atmospheric extension larger than predicted by the PHOENIX models, however it is relatively small and it does not show up with any significance using this metric of plotting (V$^{2}_\mathrm{cont}$/V$^{2}_\mathrm{CO}$). The other giant stars do not show an extended CO band (the visibilities in the continuum are very similar to those in the CO (2-0) line). The RSG (blue) and Mira stars (magenta) show a more extended CO layer and therefore they are not consistent with hydrostatic models or convection models. The RSG and Mira stars show similar atmospheric extensions. 

Fig.~\ref{fig:visratio_RSGs} suggests a correlation between the visibility ratio of our RSGs and the luminosity and surface gravity, which is indicated by linear fits  (dashed gray lines). The correlations indicate an increasing atmospheric extension with increasing luminosity and with decreasing surface gravity. We do not observe a correlation between observed atmospheric extension and effective temperature or variability amplitude. In the case of Mira stars, we do not observe any correlation. The lack of a correlation between visibility ratio and the luminosity and/or surface gravity suggests that different processes may be responsible for extending the atmosphere in RSGs and Miras. On the other hand, we note that considerable atmospheric extensions for RSG stars are observed only for luminosities beyond $\sim1*10^5$~L$_{\odot}$ and for surface gravities below log(g)$\sim$0.

\section{Comparisons with convection and pulsation models}
\label{sec:comp}

In the following we discuss physical mechanisms that have been discussed as possible drivers of the observed large extensions of RSG atmospheres. In particular, we discuss convection models and pulsation models. Pulsation models have been successful to explain observed atmospheric extensions of Mira-variable AGB stars with shock fronts passing through the atmospheres.

\subsection{3-D RHD simulations}

The 3-D radiation-hydrodynamics simulations of red supergiant stars were computed with CO$^{5}$BOLD (Freytag et al. \cite{Freytag2012}). The code solves the coupled equations of compressible (magneto-)hydrodynamics and non-local radiative energy transport on a Cartesian grid. The "star-in-a-box" geometry is used and the computational domain is a cubic grid equidistant in all directions; the same open boundary condition is employed for all sides of the computational box. The tabulated equation of state takes the ionization of hydrogen and helium and the formation of H$_2$ molecules into account. The opacity tables are gray or use a frequency-binning scheme.

\begin{figure}
\centering
\includegraphics[width=0.99\hsize]{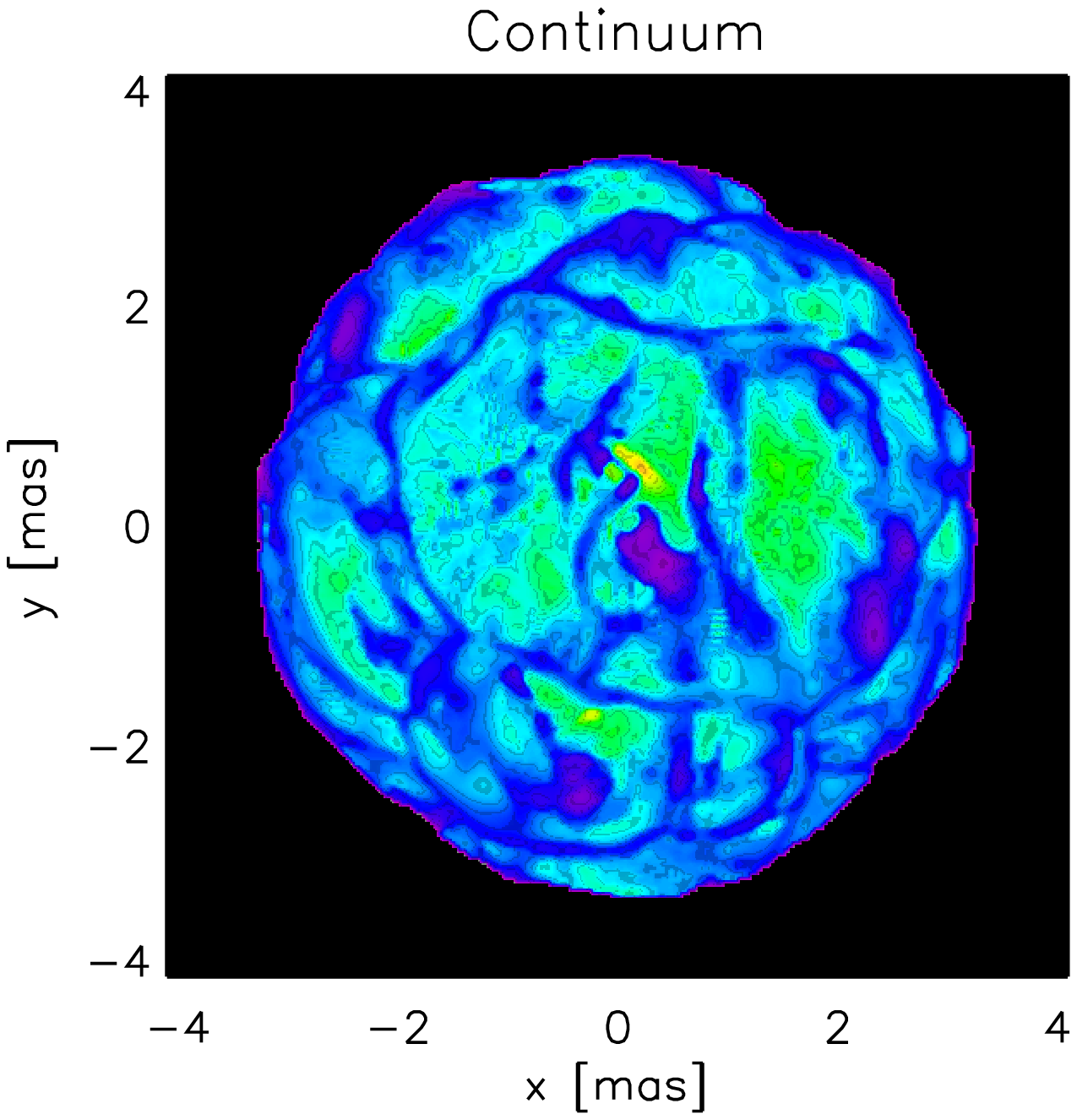}
\includegraphics[width=0.99\hsize]{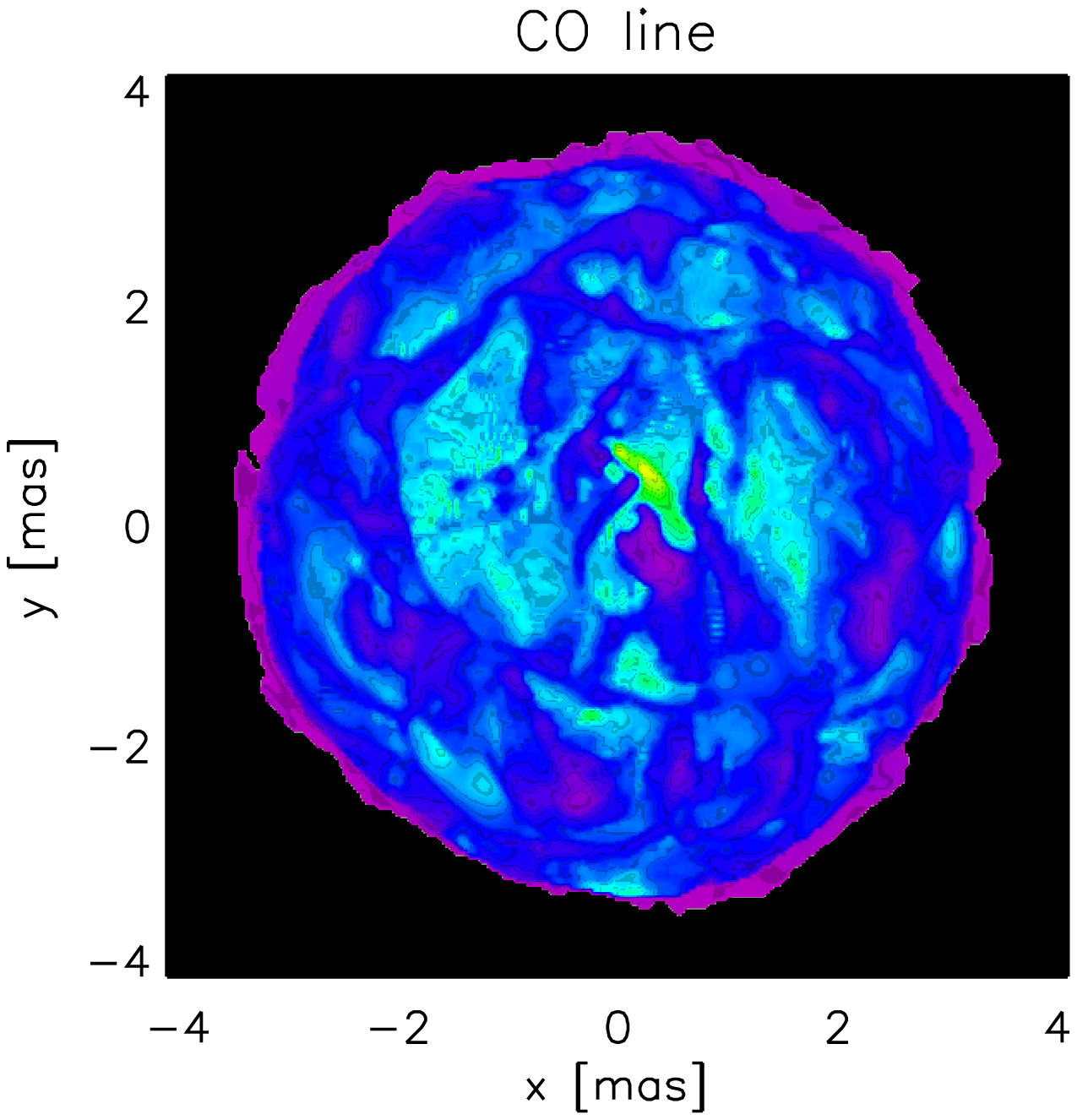}
\includegraphics[width=0.99\hsize]{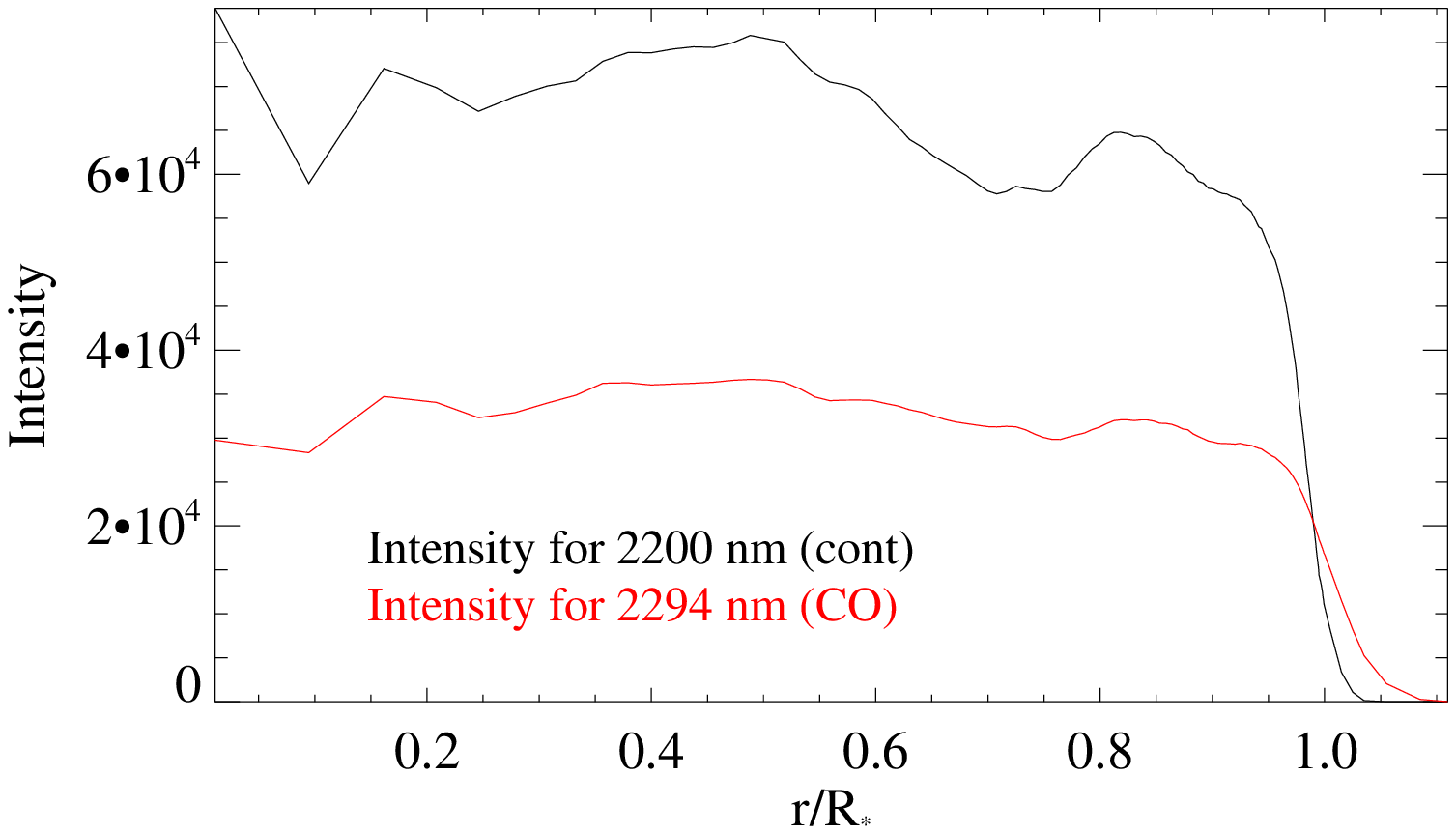}
\caption{3-D radiative-hydrodynamical (RHD) simulations of convection for one snapshot of model st35gm03n07 (see table \ref{simus}). (Top) Image of the intensity at a continuum wavelength of 2.20\,$\mu$m. (Middle) Image of the intensity at the CO (2-0) line at 2.294\,$\mu$m. (Bottom) Azimuthally averaged intensity profiles of both images, where the black line denotes the continuum wavelength and the red line denotes the CO line. The radii is defined by $r/R_{*}=\sqrt{1-\mu ^{2}}$ ($\mu$ is explained in the text).}
\label{Image}
\end{figure}

\begin{figure}
\centering
\includegraphics[width=0.95\hsize]{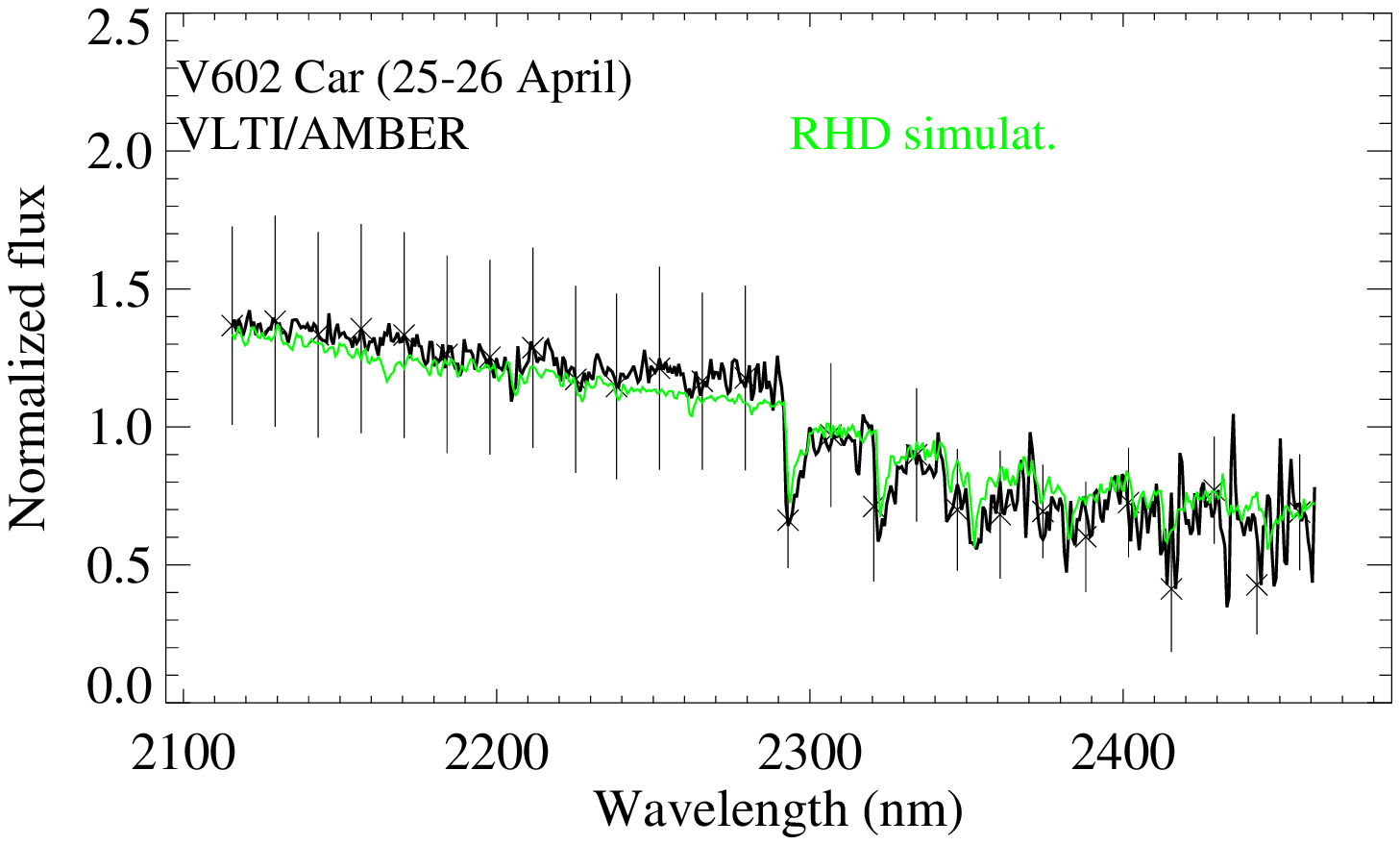}
\includegraphics[width=0.95\hsize]{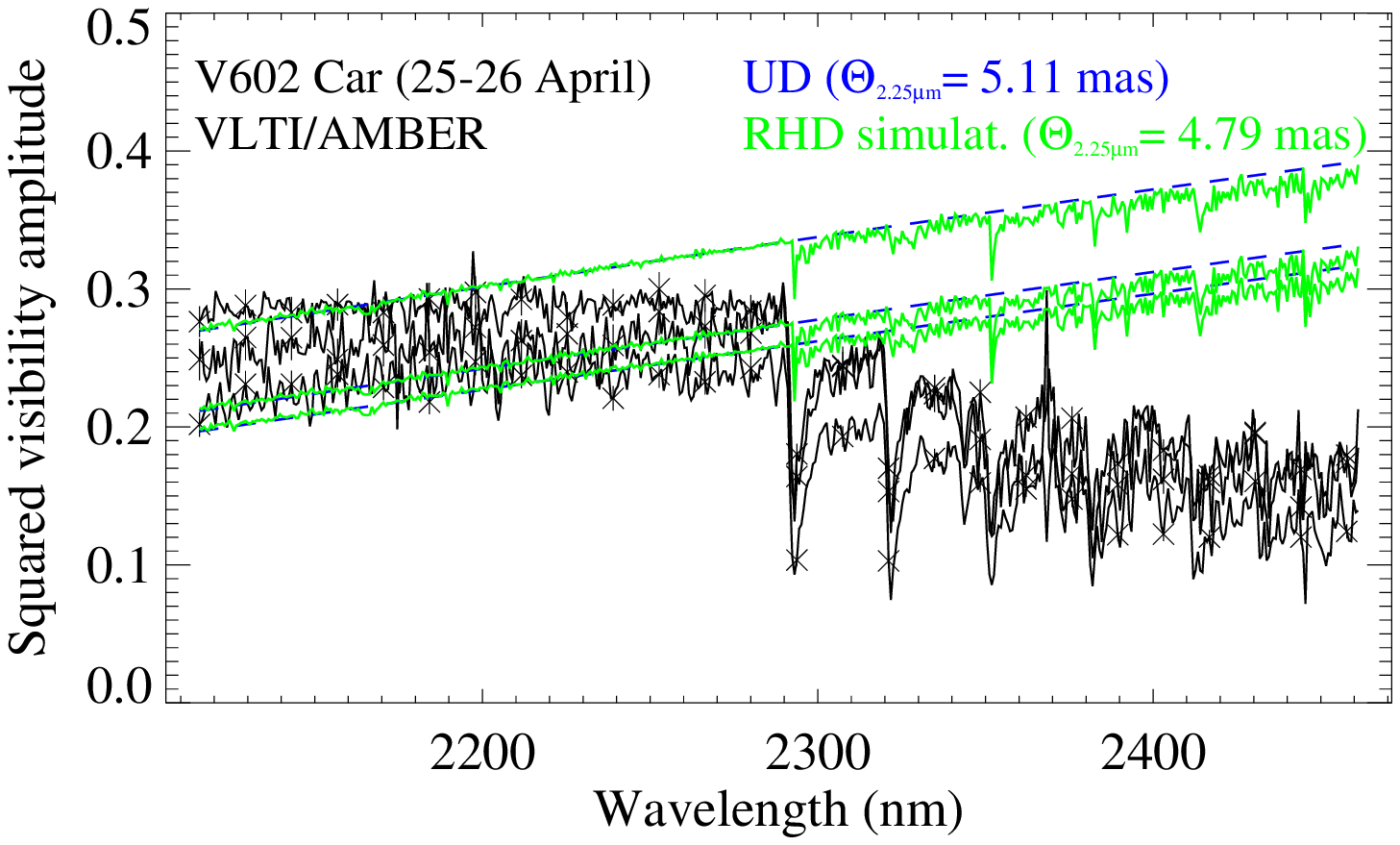}
\includegraphics[width=0.95\hsize]{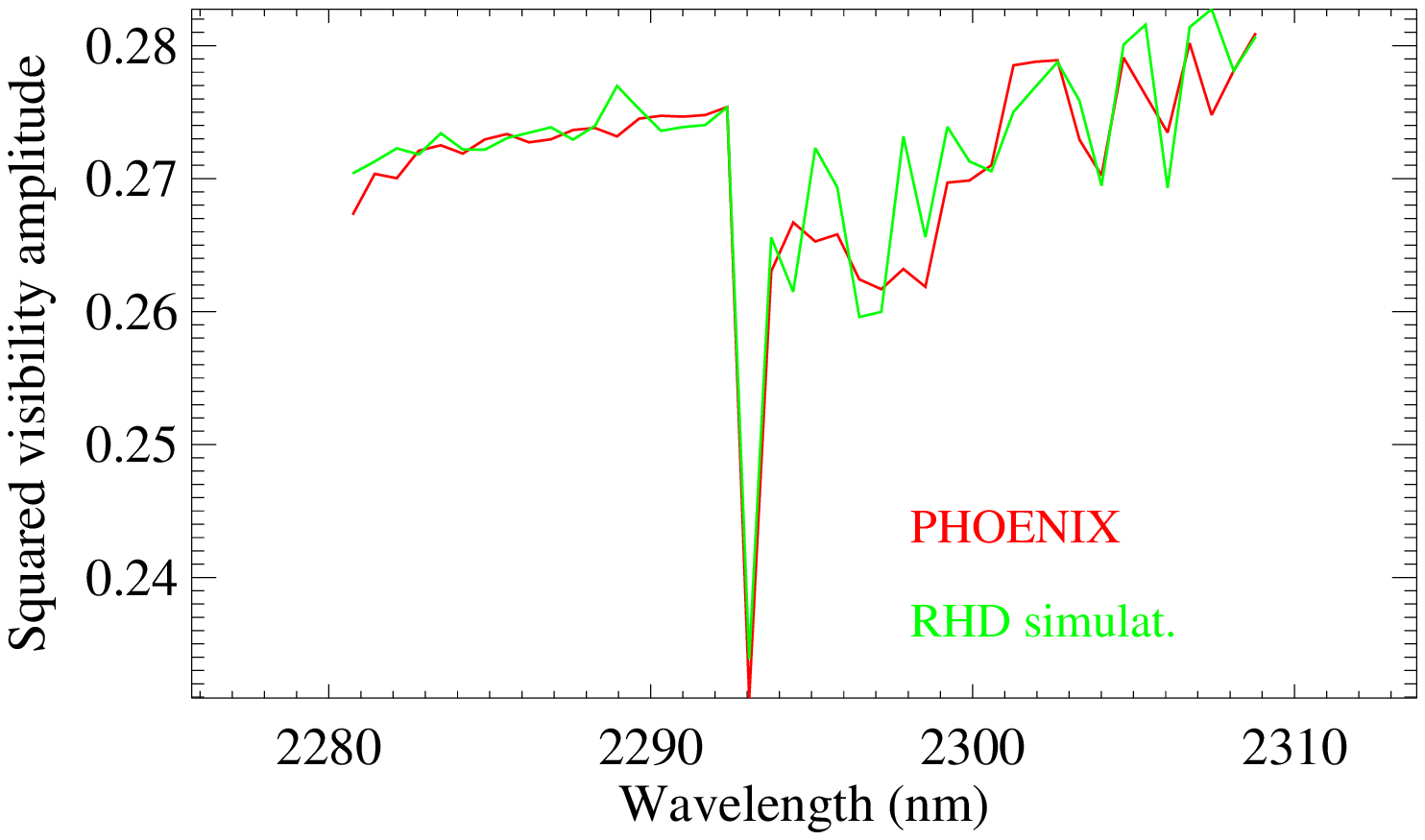}
\caption{Observed normalized flux (top), square visibility amplitudes (middle) of V602~Car as in Fig.~\ref{resul_V602Car_2604}, compared to the prediction by the 3-D RHD simulations of Fig.~\ref{Image}. The black curves are the observational data, the green curve the RHD simulation, and the blue curve the best-fit UD model (it is underplotted to the green curve).  Bottom: Enlargement of the synthetic squared visibility amplitudes of the CO (2-0) line (2.294\,$\mu$m) predicted by the {\tt PHOENIX} model atmosphere that best fits our V602~Car data (cf. Fig.~\ref{resul_V602Car_2604}) compared to the prediction by the RHD simulation of Fig.~\ref{Image}. The stellar parameters of the {\tt PHOENIX} mode are $T_\mathrm{eff}$=3400\,K, $\log(g)=$-0.5, $M$=20\,$M_\odot$. Those of the RHD simulation are $T_\mathrm{eff}$=3487\,K,; $\log(g)$=-0.335, $R$=830\,$R_\odot$, $M$=12\,$M_\odot$).}
\label{3D_PHOENIX}
\end{figure}

The main model parameters are (Chiavassa et al. \cite{Chiavassa2011}): the stellar mass (entering the gravitational potential), the input luminosity in the core, and the abundances that were used to construct the equation-of-state and opacity tables. Average values of stellar radius, effective temperature, and surface gravity have to be derived from a relaxed model (Chiavassa et al. \cite{Chiavassa2009,Chiavassa2011}). 

The models show large-scale convection cells (several in RSGs and only one to two for AGB stars) that span the entire convective envelope and have lifetimes of years. On the surface, there are smaller shorter-lived cells that resemble solar granules. According for instance to Samadi et al. (\cite{Samadi2001}), convective flows excite acoustic waves through turbulent Reynolds stresses (i.e., essentially velocities) and turbulent entropy fluctuations. In the part of the 3-D RHD simulations that comprises the stellar interior, both are largest in or close to the downdrafts. Accordingly, acoustic waves are produced mostly in the downdrafts, and particularly during merging of downdrafts. While the fast downdrafts themselves actually impede the outward propagation of the waves, the waves spread into the surrounding upflow regions in which they can easily reach the stellar surface. When these waves reach the thin and cold photospheric layers, they steepen into shocks that travel through the atmosphere with a decreasing density of the post-shock gas. Their induced dynamical pressure exceeds the thermal gas pressure leading to a significant increase in the density scale height.

We used the pure-LTE radiative transfer $\sc{Optim3D}$ (Chiavassa et al. \cite{Chiavassa2009}) to compute intensity maps from three of these RHD simulations. This code take the Doppler shifts occurring due to convective motions into account. The radiative transfer equation is solved monochromatically using pre-tabulated extinction coefficients as a function of temperature, density, and wavelength. The lookup tables are computed for the same chemical compositions as the RHD simulations using the same extensive atomic and molecular continuum and line opacity data as the latest generation of MARCS models (Gustafsson et al. \cite{Gustafsson2008}).

The three red supergiant (RSG) simulations are shown in the Table~\ref{simus}. On the other hand, we also tried the higher resolution simulation (401$^\mathrm{3}$ grid points) st36gm00n05 (also in table~\ref{simus}). We considered several snapshots for each simulation and computed intensity maps in the wavelength range 1.90-2.60\,$\mu$m with a constant spectral resolution of $R$=$\lambda / \delta_{\lambda} \sim$~20000. Moreover, for every wavelength, a top-hat filter including 5 wavelengths close by has been considered. In total, about 70000 images for every simulation snapshot have been computed to cover the wavelength range of the observations.

\begin{table*}
\caption{RHD simulations of red supergiant stars used in this work.}
\begin{center}
\begin{tabular}{c c c c c c c c c c c}
\hline
\hline 
model & grid  & $M_{\mathrm{pot}}$ &  $L$ &  $T_{\rm{eff}}$ & $R_{\star}$ &  $\log g$ \\
      &  [grid points] & [M$_\odot$] &  [L$_\odot$] & [K] & [R$_\odot$] &  [cgs]  & \\
	   &&&&&&&&\\
\hline
st35gm03n07$^a$ &235$^3$ & 12  & 91\ 932$\pm$1400 & 3487$\pm$12 & 830.0$\pm$2.0 & $-$0.335$\pm$0.002  \\
st35gm03n13$^b$ & 235$^3$ & 12  & 89\ 477$\pm$857 & 3430$\pm$8 & 846.0$\pm$1.1 & $-$0.354$\pm$0.001 \\
st36gm00n06 &    255$^3$  & 6 & 23\ 994$\pm$269& 3660$\pm$11& 391.3$\pm$1.5& 0.010$\pm$0.004 \\
st36gm00n05$^b$ &       401$^3$  & 6  & 24\ 233$\pm$535& 3710$\pm$20& 376.7$\pm$0.5& 0.047$\pm$0.001  \\
\hline
\hline 
\end{tabular}
\end{center}
\tablefoot{a) used in Chiavassa et al. \cite{Chiavassa2009, Chiavassa2010_bis, Chiavassa2011_bis}. b) Chiavassa et al. \cite{Chiavassa2011}}
\label{simus}
\end{table*}

Then, using the method explained in detail in Chiavassa et al. (\cite{Chiavassa2009}), we computed azimuthally averaged intensity profiles. These profiles  were constructed using rings regularly spaced in $\mu$ (where $\mu=cos(\theta$) with $\theta$ the angle between the line of sight and the radial direction).

Finally, we averaged the monochromatic intensity profiles to match the spectral channels of the individual observations (we reduced the spectral resolution to AMBER observations). Afterward, we estimated the synthetic visibility of each baseline using the Hankel transform in the same way as for the {\tt PHOENIX} models described above.

Fig.~\ref{Image} shows the intensity images at a continuum wavelength (2.20\,$\mu$m) and at the CO (2-0) line (2.294\,$\mu$m), together with the azimuthally averaged intensity profiles for the example of the snapshot of model st35gm03n07. The intensity images illustrate that the intensity in the CO line is lower by a factor of about 2 compared to the intensity in the continuum, which is consistent with observed flux spectra such as in  Fig.~\ref{resul_V602Car_2604} for the example of V602 Car. In the images, the linear intensity range is between 0 and 130000 erg/s/cm$^{2}$/A. Keeping the same scale seems to reduce the apparent luminosity of the CO image and thus also the contrast between bright and dark regions, however it is the opposite. The image of CO has a pseudo-continuum contribution due to CO opacities that increase the surface contrast and decrease the level of the structures, in particular at the limb. The CO line surface looks slightly more extended (purple color close to the stellar limb in the figure \ref{Image}, central panel) but only by a few percent ($\sim$7\%, estimated from Fig. \ref{Image} bottom panel  at the 0\% intensity)

Fig.~\ref{3D_PHOENIX} (top and middle panels) shows the example of our V602~Car data of Fig.~\ref{resul_V602Car_2604} compared with the 3-D RHD simulation of Fig.~\ref{Image}. In the bottom panel, we show the predicted squared visibility amplitudes of the best-fit {\tt PHOENIX} model to our data of V602~Car (cf. Fig.~\ref{resul_V602Car_2604}) compared to those of the RHD simulation of Fig.~\ref{Image}. The model-predicted visibility curves of the 3-D RHD simulation are very similar to the hydrostatic {\tt PHOENIX} model at the AMBER resolution and can thus not explain the large observed atmospheric extensions of RSG stars. 

\begin{figure}
\centering
\includegraphics[width=0.95\hsize]{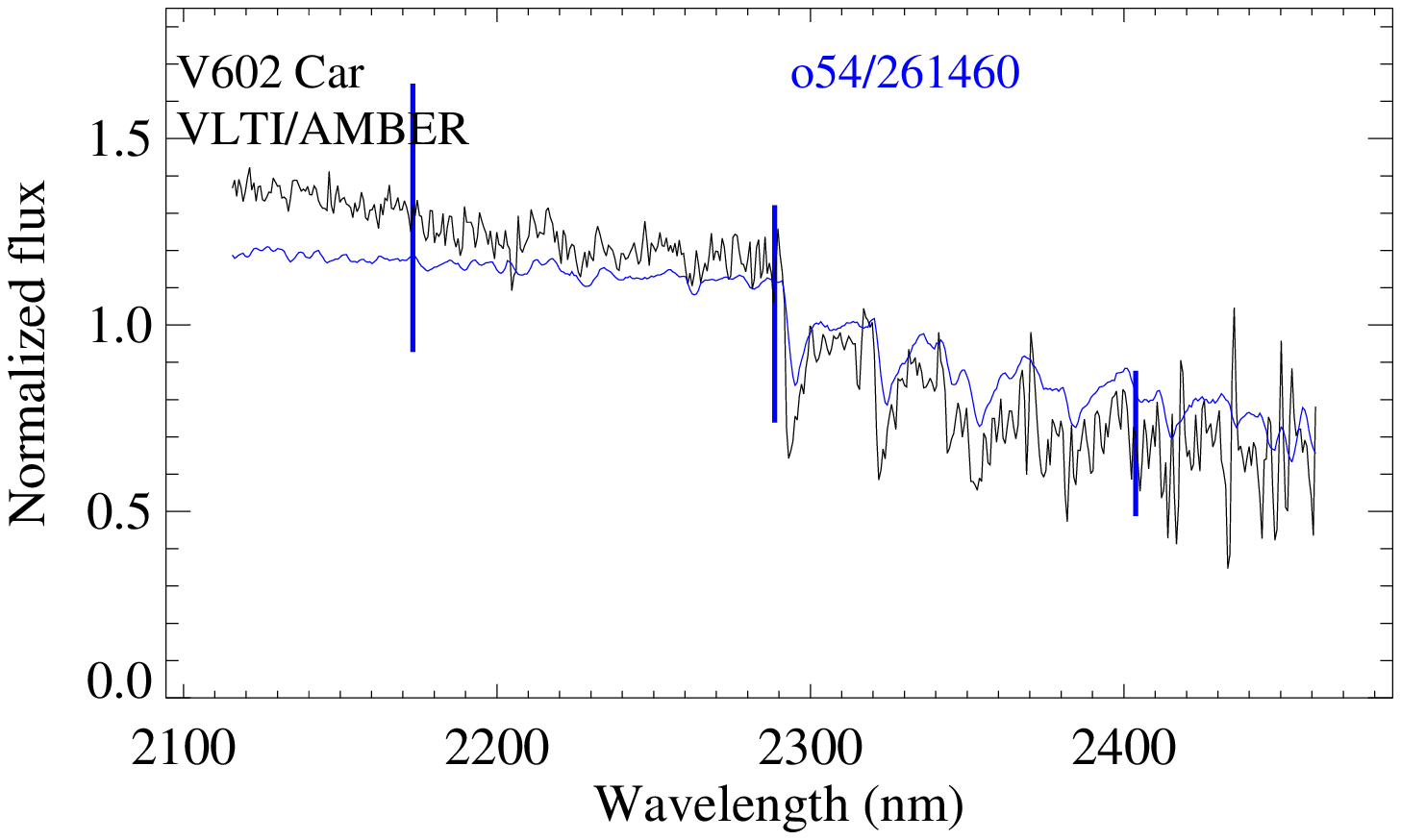}
\includegraphics[width=0.95\hsize]{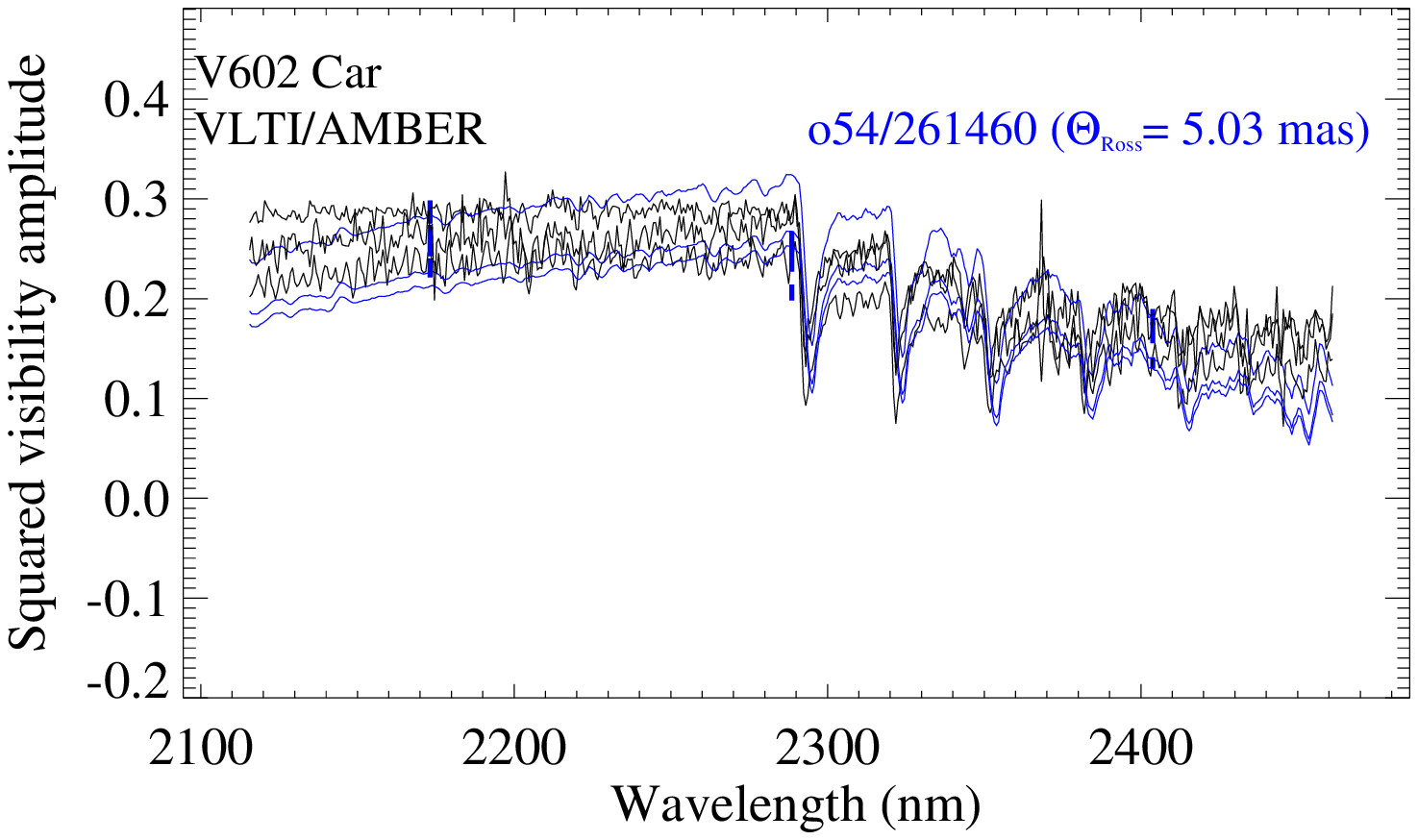}
\caption{Observed normalized flux (top) and squared visibility amplitudes (bottom) of our V602 Car from Fig.~\ref{resul_V602Car_2604}, compared to one of the best-fit {\tt CODEX} model atmospheres. The black lines denote the observational data, and the blue lines the model predictions for all three baselines (bottom). The model parameters are listed in the main text. The thick vertical lines
indicate the uncertainties at three wavelength intervals.}
\label{CODEX_model}
\end{figure}

\begin{figure}
\centering
\includegraphics[width=0.95\hsize]{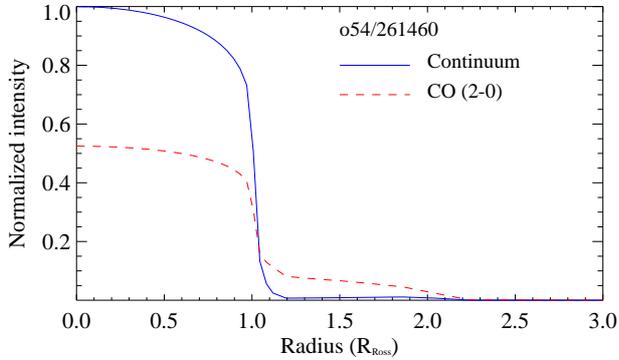}
\caption{Intensity profile of the {\tt CODEX} model o54/261460 in the continuum at 2.25\,$\mu$m (solid blue line) and at the CO (2-0) line at 2.294\,$\mu$m (dashed red line).}
\label{fig:CODEXprofile}
\end{figure}

\begin{figure}
\centering
\includegraphics[width=0.99\hsize]{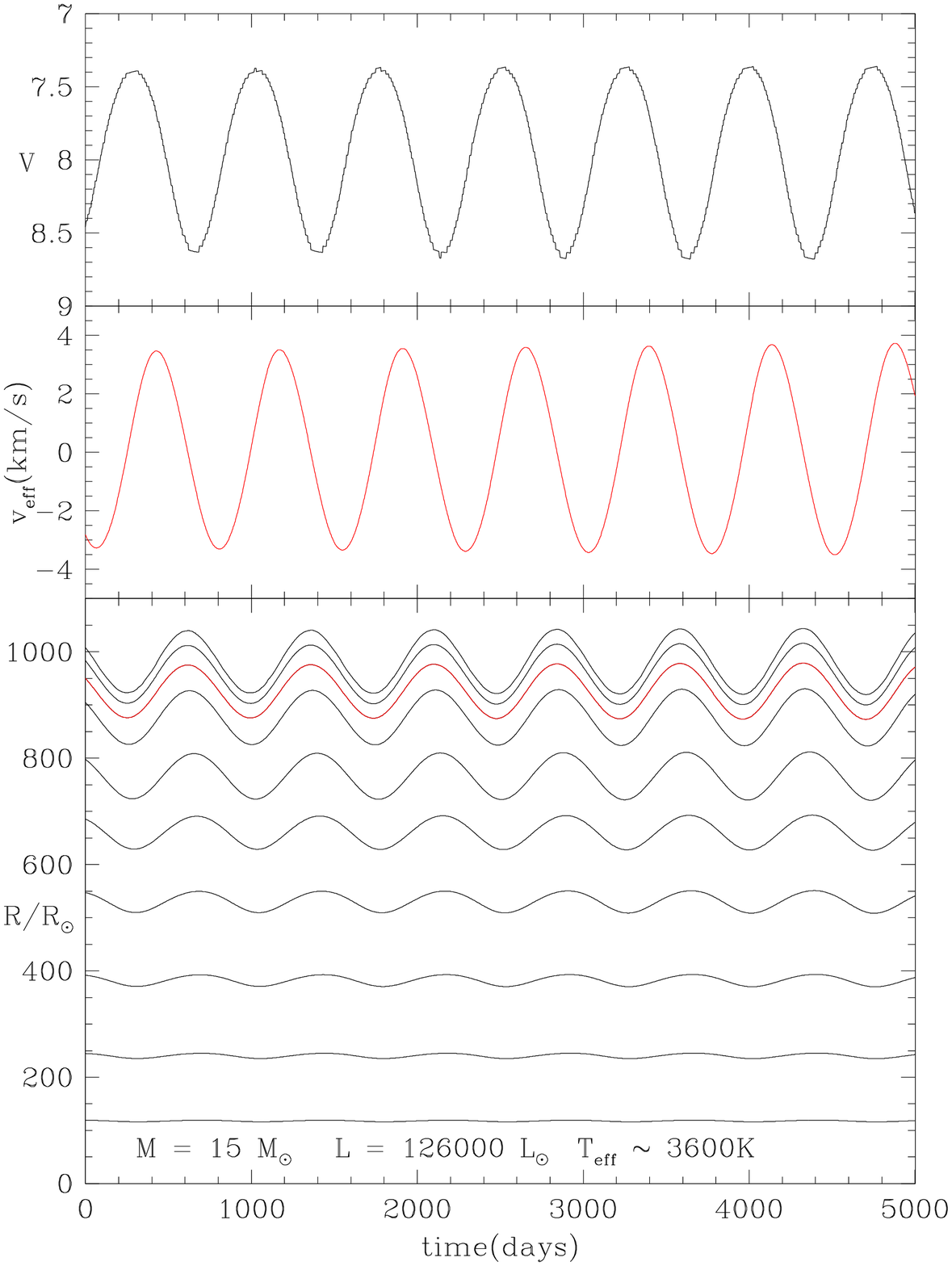}
\caption{Pulsation model of a RSG with M=15~M$_{\odot}$, L=126000~L$_{\odot}$, Teff$\sim$3600K. Bottom panel: Radius variation of selected mass zones in a pulsating supergiant model with M~=~15~M$_{\odot}$ and L~=~126000~L$_{\odot}$ (black curves). The red curve is the position of the photosphere (defined as  the layer where the Rosseland optical depth equals 2/3). Middle panel: The velocity at the photosphere. Top panel: The visual light curve of the model, where the bolometric correction is obtained from the tables in Houdashelt et al. (\cite{Houdashelt2000a}; \cite{Houdashelt2000b}).}
\label{V+veff+r_i}
\end{figure}

The results are similar for all the RHD simulations employed in this work. In summary, it appears from the visibility curves that the atmospheric extension of RHD simulations is comparable to typical hydrostatic {\tt PHOENIX} model within the spectral resolution of the actual observations. 

RHD simulations of RSGs do not lead to a considerable increase of the radius. However, the situation may be different for the AGB simulations (Freytag \& H\"ofner \cite{Freytag2008}), where the ratio of the pressure scale height to the stellar radius is larger, giving rise to relatively larger convective structures, to larger-scale waves and shocks, and finally to a noticeable increase of the stellar radius. Thus, shock waves in AGB simulations may explain the atmospheric extension we see in the observations of Mira stars.

\subsection{Pulsation models}
\label{sec:RSGpulsation}

Self-excited pulsation models of Mira-variable AGB stars (Ireland et al. \cite{Ireland2004a,Ireland2004b,Ireland2008,Ireland2011}; Scholz et al. \cite{Scholz2014}) have been successful to describe interferometric observations of these sources, including their extended atmospheric molecular layers (Woodruff et al. \cite{Woodruff2009}, Wittkowski et al. \cite{Witt2011}, Hillen et al. \cite{Hillen2012}). The observed visibility spectra of our RSG sources show similar features as those of Mira stars, in particular at the CO bandheads (see Sect. 3.3). We have thus as a first step investigated whether pulsating model atmospheres of Mira variables can provide a good fit to our RSG stars as well, although the stellar parameters, in particular mass and luminosity, are very different and variability amplitudes of RSGs are typically lower by a factor of 2-3 compared to Miras (Wood et al. \cite{Wood1983}). We used the recent {\tt CODEX} model series by Ireland et al. (\cite{Ireland2008}, \cite{Ireland2011}) and, as an example, have found best-fit models to the data of V602~Car as shown in Fig.~\ref{resul_V602Car_2604}. Fig.~\ref{CODEX_model} shows one of the best-fit {\tt CODEX} models compared to  these data. The shown model is model 261460 of the o54 series. The o54 series is designed to describe the prototype Mira variable omi~Ceti with a non-pulsating parent star of $M$=1.1\,$M_\odot$, $L$=5400\,$L_\odot$, $R$=216\,$R_\odot$, $P$=330\,days. Model 261460 is a model at phase 0.2 within a particularly extended cycle, with $L$=7420\,$L_\odot$ and $T_\mathrm{eff}$=3154\,K. The model does not show all observational details, in particular the detailed overall slope of the flux and visibility spectra. However, it does match the drops of the CO bandheads in the visibility spectrum. Fig.~\ref{fig:CODEXprofile} shows the intensity profile of this model, in the continuum (2.25\,$\mu$m) and in the CO (2-0) line (2.294\,$\mu$m), as an illustration of which kind of atmospheric extension is required to describe the observed visibility data. The intensity profile in the CO line extends to beyond 2 Rosseland radii, compared to an extension of a few percent as predicted by the 3-D RHD simulations of RSGs (Fig.~\ref{Image}).

As a next step, we calculated a new pulsation model based on stellar parameters that are typical for an RSG star: the non-pulsating parent star has $M$=15\,$M_{\odot}$, $L$=1.26$\cdot$10$^{5}$\,$L_{\odot}$, $R$=954\,$R_{\odot}$, and its effective temperature is about 3600\,K. The pulsation period of this model is about 750\,days. These parameters are close to the star V602~Car discussed above ($T_\mathrm{eff}$=3432$\pm$263\,K, $L$=(1.35$\pm$0.23)$\cdot$10$^{5}$\,$L_\odot$, $R$=1050$\pm$165\,$R_\odot$, $P$=645\,days). Fig. \ref{V+veff+r_i} shows the radius variation of selected mass zones of this pulsation model, the velocity at the photosphere, and the visual light curve. The amplitude of the photospheric radius variation is about 10\% with radial velocities of up to about 5\,km/sec. The model reproduces the amplitude of the visual light curve of V602~Car of about 1-1.5 mag. Whilst shock fronts enter the stellar atmosphere in a typical {\tt CODEX} model of a Mira variable at or below optical depth 1, leading to a geometric extension of the stellar atmosphere of the order a few Rosseland radii (e.g., Ireland et al. \cite{Ireland2011}), it turns out that no shock fronts reach at any phase the atmospheric layers in case of the RSG model. We also computed a higher amplitude version of this model with $V$ mag., velocity, and radius amplitudes that are larger by a factor of 2--3. Although the surface moves in and out more, there are still no shocks in the atmospheric layers. We note that the moving photosphere might induce shocks at higher radii, beyond the region of these models, where the density is very low. This, however, could also not explain the observed atmospheric extensions above the photospheric layers. In our test model, the ratio $(r(\tau_\mathrm{Ross}=10^{-4})-r(\tau_\mathrm{Ross}=1))/r(\tau_\mathrm{Ross}=1)$ which roughly measures the atmospheric extension is only 0.04 at four sample phases (-0.02, 0.27, 0.52, 0.76), and we see hardly depth-dependent outflow/infall velocities below 4\, km/s (Fig.~\ref{V+veff+r_i}). For comparison, the {\tt PHOENIX} model plotted in Fig. \ref{resul_V602Car_2604} has an extension of 0.04 as well. 

A few test computations with higher parent-star luminosity (3~10$^{5}$\,L$_{\odot}$) and higher mass (25 and 15\,M$_{\odot}$) yield similar results. This means that the hydrostatic approximation should be well suited for calculating the stratification of the RSG atmosphere. We note in this context that the atmospheric scale height, $H\propto T_\mathrm{eff}*R^{2}/M$, which would be an approximate measure of the atmospheric extension in case of a static stellar atmosphere, is of the same order of magnitude for typical Miras and for typical RSGs. For, e.g., the o54 model of the {\tt CODEX} series that has about o~Cet parameters and the above RSG model, one finds $H$(o54)/$H$(RSG) = 0.7 and very small atmospheric extensions $H$(o54)/$R$(o54)$\sim$0.023/$\mu$ and $H$(RSG)/$R$(RSG)$\sim$0.008/$\mu$. The mean molecular weight is $\mu \sim$1.3 for a normal solar element mixture with atomic hydrogen (molecular hydrogen is formed only in the high layers of the Mira models). We also note that H/R is approximately the ratio (gas particle thermal kinetic energy)/(gravitational potential at the stellar surface). In the RSG model, the typical gas particle has a speed of $\sim$8\,km/s in the atmosphere whereas pulsation velocities are only $\sim$3\,km/s. Hence, the pulsation energy is only $\sim$14\% of the particle thermal energy and, therefore, does not lead to significant atmospheric extension of the models.

Overall, the pulsation models for typical parameters of RSG stars lead to compact atmospheres with extensions similar to those of the {\tt PHOENIX} and RHD models discussed above, and can thus also not explain the observed extensions of the molecular layers.

\subsection{Discussion on alternative mechanisms}

We showed in Sects.~\ref{sec:results} and \ref{sec:charac} that our interferometric observations of RSGs provide evidence of extended molecular layers for a sample of RSGs with luminosities between about $1\times10^5$\,L$_\odot$ and $5\times10^5$\,L$_\odot$ and effective temperatures between about 3350\,K and 3750\,K. These extensions are comparable to those of Mira variable AGB stars, which typically reach to a few photospheric radii. Furthermore, our observations of RSGs indicate correlations of increasing atmospheric extension with increasing luminosity and decreasing surface gravity, where considerable extensions of the CO (2-0) line are observed for luminosities above $\sim1*10^5$~L$_{\odot}$ and for surface gravities below log(g)$\sim$0.

Comparisons of our interferometric data to available 1-D hydrostatic {\tt PHOENIX} model atmospheres, new 1-d pulsation models, and available 3-D RHD simulations show that all of these theoretical models result in very similar synthetic visibility values for the parameters of our observations. They indicate a compact atmospheric structure for all considered models, and none of these can currently explain the observed major extensions of the atmospheres of RSGs. 

A large number of spectroscopic and interferometric studies of RGs and RGSs show that classical models do not explain the extended atmospheres of these sources (Tsuji \cite{Tsuji2003, Tsuji2008}). Studies of resolved 12\,$\mu$m spectra of red giant and supergiant stars show strong absorption lines of OH and H$_{2}$O, which are even larger than expected from a classical photosphere without a \textit{MOLecular sphere} around the stars  (Ryde et al. \cite{Ryde2002, Ryde2003, Ryde2006a, Ryde2006b}).

Hereby, the new 1-D pulsation test models show velocity amplitudes below about 10\,km/sec. These velocity amplitudes are consistent with observed long-term average velocity curves of Betelgeuse of $\sim$\,9km/sec (Gray \cite{Gray2008}, Fig. 6) and of a sample of RSGs of $\le$\,10\,km/sec (Josselin \& Plez \cite{Josselin2007}). These 1-D pulsation models are also consistent with typical amplitudes of the long-period visual light curves of RSGs of 1--3\,mag.

In addition to these long-term average velocity curves, which we interpret as being caused by ordered pulsation motions and which are well explained by the new 1-D pulsation models, Gray (\cite{Gray2008}) and Josselin \& Plez (\cite{Josselin2007}) also report on higher velocity gradients and turbulent motions in the atmospheres on short time scales with velocities of up to 30\,km/sec. These were explained by granulation and giant convection cells accompanied by short-lived oscillations. In addition, Ohnaka et al.      (\cite{Ohnaka2011,Ohnaka2013}) reported on interferometric observations of Betelgeuse and Antares at two epochs each, which confirmed time variable atmospheric velocities of up to 20--30\,km/sec. They suggested that these motions are related to the wind driving mechanism. They also concluded that the density of the  extended outer atmospheres of Antares and Betelgeuse are significantly higher than predicted by current 3-D RHD simulations, so that convection alone can not explain the formation of the extended atmospheres. This conclusion is confirmed by our direct comparisons of interferometric data and synthetic visibilities based on current 3-D RHD simulations, and for a larger sample of 6 additional RSGs.

Current 3-D RHD simulations of RSGs are still limited in spatial resolution. However, we can compare older models of RSGs with low numerical resolution (235$^\mathrm{3}$ grid points) and new models with better resolution (401$^\mathrm{3}$ grid points). On the other hand, we also have better resolved models of AGB stars and even better resolved local models of solar-type stars -- with a range of resolutions. From these comparisons, we conclude that future RSG simulations with higher resolution are indeed important and desirable, but that we do not expect the atmospheric velocity fields due to convection and pulsations alone to grow enough to remotely reach the amplitude necessary to give a molsphere extension of a few stellar radii as observed.

Josselin \& Plez (\cite{Josselin2007}) suggested that high velocities and steep velocity gradients, possibly caused by convective motion, generate line asymmetries, that turbulent pressure decreases the effective gravity, and that this decrease combined with radiative pressure on lines initiates the mass loss. The spherically symmetric {\tt CODEX} Mira model atmospheres do, indeed, show noticeable radiative acceleration $a_\mathrm{rad}$ which significantly affects the effective gravity in the atmospheric layers (e.g., Fig. 6 with Eq.~1 in Ireland et al. \cite{Ireland2008}), but still without any net outward acceleration. One should realize that the pulsation-generated very irregular velocity stratification (resulting in irregular $\rho(r)$, e.g., Figs.~14/15/16 in Ireland et al. \cite{Ireland2011} and Fig.~2 in Scholz et al. \cite{Scholz2014}) favors large $a_\mathrm{rad}$ because of substantial line shifts.  Velocity (outward/inward) variations can be quite strong, up to a few km/s between shock fronts, apart from the big ``jumps'' at shock front positions. The new RSG pulsation models described in Sect.~\ref{sec:RSGpulsation} do not show this velocity stratification behavior. However, they are also solely based on long term velocity variations caused by pulsation motion. It is plausible that much steeper velocity gradients on shorter time scales, as reported by Gray (\cite{Gray2008}), and not included in these models, might generate accelerations on Doppler-shifted molecular lines that sufficiently extend the molecular atmosphere. Although already suggested in 2007, more detailed calculations of this effect for RSGs and its implementation in the 3-D RHD simulations are still pending and are also outside the scope of the present paper. Our observed correlation of increasing atmospheric extension with increasing luminosity and decreasing surface gravity  (cf. Fig.~\ref{fig:visratio_RSGs}) supports such a radiatively driven extension of the atmospheres of RSGs.

The proposed scenario of radiation pressure on Doppler-shifted lines is in a a way reminiscent of what happens in the winds of hot stars. However, it should be noted that --unlike for hot stars-- RSGs form dust at larger radii (typically $\sim$20 stellar radii, Danchi et al. \cite{Danchi1994}) so radiation pressure on dust cannot occur in the wind acceleration zone (Josselin \& Plez, \cite{Josselin2007}). In the case of RSGs, the proposed radiative pressure on molecular lines may only levitate the molecular atmosphere up to radii where dust can form (as suggested by Josselin \& Plez, \cite{Josselin2007}), analogous to shock fronts for Miras, albeit not lifting the material outside the gravitational potential. However, the detailed dust formation, condensation sequence, and mass-loss mechanisms are also not yet fully understood in the case of oxygen-rich AGB stars  (cf., e.g., Karovicova et al. \cite{Karovicova2013} for a recent discussion). Recent indications based on a polarimetric aperture masking technique (Norris et al. \cite{Norris2012}) and on mid-infrared interferometry (Wittkowski et al. \cite{Witt2007}; Karovicova et al. \cite{Karovicova2011, Karovicova2013}) point toward transparent dust grains forming already at relatively small radii of about 1.5 stellar radii. These may be grains of amorphous Al$_2$O$_3$ and/or magnesium-rich iron-free (``forsterite'') silicates, while iron-rich silicates form at larger radii. Verhoelst et al. (\cite{Verhoelst2006}) found a similar evidence for amorphous alumina in the extended atmosphere of the RSG Betelgeuse as close as $\sim$\,1.4 stellar radii, while iron-rich silicates indeed had an inner radius of 20 stellar radii in their model. However, Kami{\'n}ski et al. (\cite{Kaminski2013}) do not observe the presence of AlO in the inner outflow in the spectrum obtained with the Ultraviolet and Visual Echelle Spectrograph  (UVE) at the Very Large Telescope (VLT). On the other hand, they have detected AlO-bearing gas in the wind-acceleration zone, out to 20~R$_{*}$. If the presence of transparent grains is confirmed at small radii of $\sim$\,1.5 stellar radii, it would play a crucial role for the dust condensation sequence and the overall mass-loss process of both oxygen-rich AGB stars and RSGs.

Grunhut et al. (\cite{Grunhut2010}), Auri{\`e}re et al. (\cite{Auriere2010}), and Bedecarrax et al. (\cite{Bedecarrax2013}) detected weak ($<$1G) magnetic fields in one third of their sample of late type supergiants, with their sensitivity limit. They suggested that magnetic fields are generated by dynamo action, and that weak fields may be excited in all cool supergiants. It has been discussed that magnetic fields may also play a role in the mass-loss process of RSG stars alongside other processes as discussed above. Thirumalai \& Heyl (\cite{Thiru2012}) recently presented a magnetized hybrid wind model for Betelgeuse, combining a Weber-Davis stellar wind with dust grains, that was able to lift stellar material up from the photosphere and into the circumstellar envelope. One of us (BF) recently produced a set of 5-M$_\odot$-RSG RHD models without and with magnetic field. These first simulations with magnetic fields did not significantly alter the stratification of the model, however, more calculations are need to reach a real conclusion.  We also note that the good overall fit of the hydrostatic PHOENIX models in the continuum bands means that the temperature structure vs. optical depth is roughly correct, so that additional heating of the atmosphere is not a missing mechanism. In addition, rigid rotation or differential rotation (e.g., a rapidly rotating core) are processes that are not included in our models but may contribute to the atmospheric extension and to the mass-loss process.

\section{Summary and conclusions}
\label{sec:summ}

Our spectro-interferometric observations with AMBER are a good tool for studying  the continuum-forming layer and molecular layers separately and for constraining the atmospheric structure of RSGs. We used the continuum near 2.2\,$\mu$m, which is mostly free of molecular contamination, to estimate the angular diameters of the targets.  Together with estimates of the distances and the bolometric fluxes, we also derived fundamental stellar parameters such as the luminosity, the Rosseland radius, and the effective temperature. 

With the effective temperature and the luminosity, we located our targets in the HR diagram. Their locations are close to evolutionary tracks that correspond to initial masses of 20-25/15-20\,M$_{\odot}$(V602 Car), 12-15/9-15\,M$_{\odot}$ (HD 95687), and 5-12/7-9\,M$_{\odot}$ (HD 183589) with or without rotation. All target positions are consistent with the red limits of recent evolutionary tracks. HD~183589 shows a lower luminosity and thus lower mass compared to our other sources. It may more likely be a (super-)AGB star instead of a RSG star.

The near-infrared spectra of all our target are reproduced well by hydrostatic {\tt PHOENIX} model atmospheres, including the CO bands. We had observed the same behavior in our previous project (Wittkowski et al. \cite{Witt2012}, Arroyo-Torres et al. \cite{Arroyo2013}).

The observed visibility curves of our sample of RSGs show large drops in the CO (2.3--2.5\,$\mu$m) and partly in the H$_2$O (1.9--2.1\,$\mu$m) bands, indicating major extensions of the atmospheric molecular layers. As a first characterization of the extensions, we calculated the square visibility ratios between the nearby continuum and the CO (2-0) bandhead. More detailed observations using a larger number of data points, possibly imaging campaigns, and a higher spectral resolution to better isolate the CO bandhead are desirable for a more accurate description of the extensions. Nevertheless, we observed a linear correlation between the visibility  ratios of our RSGs and the luminosity and surface gravity, indicating an increasing atmospheric extension with increasing luminosity and decreasing surface gravity. With that, we observed considerable atmospheric  extensions of RSGs only for luminosities beyond $\sim$\,10$^5$\,L$_\odot$ and for surface gravities below $\log(g)\sim$0. We did not observe a correlation with effective temperature or variability amplitude. Compared to Mira stars, we noticed comparable extensions between RSGs and Mira stars, which extend to a few stellar radii. Mira stars did not exhibit the correlations with luminosity and surface gravity that we observed for RSGs. This suggests that the extended atmospheric structure may be generated by different processes for each type of star. For Miras, the extensions are believed to be triggered by shocks that are generated by pulsations and that enter the atmospheric layers.

The synthetic visibility amplitudes of hydrostatic {\tt PHOENIX} models did not predict the strong visibility drops in the molecular bands, and can thus not explain their observed major extensions. To further constrain processes that were discussed as possible drivers of the extensions, we compared our data to available 3-D RHD simulations of convection and to new 1-D pulsation models for typical parameters of our RSGs. Both models resulted in a compact atmospheric structure as well, produced similar observable synthetic visibility values as the {\tt PHOENIX} models, and could therefore not explain the major extensions. We note that the actual atmospheric extension of 3-D RHD simulations is not enough to take the contribution of the molecular extended layers into account. Improvements in this direction may alter the appearance of the stellar surface with respect to typical 3-D RHD simulations of the photosphere as shown in Fig. \ref{Image}. This may be particularly important for interferometric observations taken with broad bandpasses, which are contaminated by molecular bands. 

Our observed correlation of increasing atmospheric extension with increasing luminosity and decreasing surface gravity supports a scenario of a radiatively driven extension caused by radiation pressure on Doppler-shifted molecular lines. We speculate that another ingredient of the mass-loss process could be acceleration on dust grains that may form already at a few stellar radii, a process which may possibly be further supported by magnetic fields or, differential rotation (e.g., a rapidly rotating core). In this case, the radiative acceleration on molecular lines would only be needed to levitate the atmosphere up to the point where the dust grains are formed.

\begin{acknowledgements}
This research has made use of the AMBER data reduction package of the Jean-Marie Mariotti Center, and the SIMBAD database operated at CDS, Strasbourg, France. We gratefully acknowledge support from the PSMN (P{\^o}le Scientifique de Mod{\'e}lisation Num{\'e}rique) computing center of ENS de Lyon. We also thank the CINES (Centre Informatique National de l'Enseignement Sup{\'e}rieur) for providing some of the computational resources necessary for this work. BAT and JMM acknowledge support by the Spanish Ministry of Science and Innovation though the grants AYA2009-13036-C02-02 and AYA2012-38491-C02-01. 
\end{acknowledgements}

\end{document}